\documentclass[a4paper,12pt]{article}
\usepackage[utf8]{inputenc}
\usepackage{amsfonts}
\usepackage{amsmath}
\usepackage[english]{babel}
\usepackage{amssymb}
\usepackage{amsthm}
\usepackage{booktabs}
\usepackage{bbm}
\usepackage{float}
\usepackage{multirow}
\usepackage{graphicx} 
\usepackage{graphics}
\usepackage[comma,authoryear,sort]{natbib}
\usepackage{longtable}
\usepackage[]{threeparttable}
\usepackage{rotating}
\usepackage{xcolor}
\usepackage{comment}
\usepackage{tikz}
\usepackage{eurosym}
\usepackage{textcomp}
\usepackage{makecell}
\usepackage{url}
\usepackage{hyperref}

\usepackage{algorithm}	% those two packages are helpful if you use a lot of algorithms in your thesis
\usepackage{algorithmic}

\usepackage{bibunits}

\defaultbibliography{SACM}
\defaultbibliographystyle{chicago}

\def\tr{{\rm T}}

\DeclareMathOperator{\rvech}{rvech}

\DeclareMathOperator*{\argmaxB}{argmax}

\def\diag{{\rm diag}}
\def\diagv{{\rm diagv}}
\def\tr{{\rm tr}}

\def\bsS{{\boldsymbol \Sigma}}
\def\bsEta{{\boldsymbol \eta}}
\def\bsMu{{\boldsymbol \mu}}
\def\bsLambda{{\boldsymbol \lambda}}

\def\bsBeta{{\boldsymbol \beta}}

\newcommand{\overbar}[1]{\mkern 1.5mu\overline{\mkern-1.5mu#1\mkern-1.5mu}\mkern 1.5mu}

\newtheorem{remark}{Remark}

\title{Scalable Fitting Methods for Multivariate Gaussian Additive Models with Covariate-dependent Covariance Matrices}
\author{ V. Gioia$\,^1$, M. Fasiolo$\,^2$, S. N.  Wood$\,^3$,   R. Bellio$\,^1$\\
$\,^1\,$\small University of Udine, Department of Economics and Statistics \\
$\,^2\,$\small University of Bristol, School of Mathematics\\
$\,^3\,$\small University of Edinburgh, School of Mathematics \\
\small gioia.vincenzo@spes.uniud.it
}
\date{}

\def\spacingset#1{\renewcommand{\baselinestretch}%
{#1}\small\normalsize} \spacingset{1}

\newcommand{\blind}{0}

\begin{document}
\begin{bibunit}

\if0\blind
{
 \title{\bf Scalable Fitting Methods for Multivariate Gaussian Additive Models with Covariate-dependent Covariance Matrices}
  \author{Vincenzo Gioia%\thanks{
    %The authors gratefully acknowledge \textit{please remember to list all relevant funding sources in the unblinded version}}
    \hspace{.2cm}\\
    Department of Economic, Business, Mathematical and Statistical Sciences, \\
    University of Trieste\\
    and \\
    Matteo Fasiolo \\
    School of Mathematics, University of Bristol\\
    and \\
    Ruggero Bellio\\
    Department of Economics and Statistics, University of Udine\\
    and \\
    Simon N. Wood\\
    School of Mathematics, University of Edinburgh\\
    }
  \maketitle
}
\if1\blind
{
  \bigskip
  \bigskip
  \bigskip
  \begin{center}
    {\LARGE\bf Scalable Generalized Additive Covariance Matrix Modelling}
\end{center}
  \medskip
} \fi

\bigskip

\begin{abstract}
\noindent
We propose efficient computational methods to fit multivariate Gaussian additive models, where the mean vector and the covariance matrix are allowed to vary with covariates, in an empirical Bayes framework. To guarantee the positive-definiteness of the covariance matrix, we model the elements of an unconstrained parametrisation matrix, focussing particularly on the modified Cholesky decomposition and the matrix logarithm. A key computational challenge arises from the fact that, for the model class considered here, the number of parameters increases quadratically with the dimension of the response vector. Hence, here we discuss how to achieve fast computation and low memory footprint in moderately high dimensions, by exploiting parsimonious model structures, sparse derivative systems and by employing block-oriented computational methods. Methods for building and fitting multivariate Gaussian additive models are provided by the \verb|SCM| \verb|R| package, available at \verb|https://github.com/VinGioia90/SCM|, while the code for reproducing the results in this paper is available at \verb|https://github.com/VinGioia90/SACM|.
\noindent
\end{abstract}

\noindent
%3 to 6 keywords, that do not appear in the title
\vfill

\newpage
\spacingset{1} % DON'T change the spacing!

\section{Introduction}
\label{chap2:intro}

This paper proposes computationally efficient methods for fitting multivariate Gaussian regression models in which both the mean vector, $\boldsymbol\mu$, and the covariance matrix, $\boldsymbol\Sigma$, are allowed to vary with the covariates via additive models, based on spline basis expansions. A key challenge in this context is to ensure the positive definiteness of $\boldsymbol\Sigma$, which is achieved here by modelling the elements of an unconstrained parametrisation matrix, $\boldsymbol \Theta$. While several alternative parametrisations are discussed in \citet{pinheiro1996}, we focus on the matrix logarithm (logM) of \citep{chiu1996} and the modified Cholesky decomposition (MCD) of $\boldsymbol\Sigma^{-1}$ \citep{pourahmadi1999}, which are widely used in applications.

Model fitting is performed in an empirical Bayesian framework, where maximum a posteriori (MAP) estimates of the regression coefficient are obtained via Newton optimisation methods as in \cite{wood2016}, while prior hyperparameters are selected by maximising a Laplace approximation to the marginal likelihood (LAML) via the generalised Fellner-Schall (FS) iteration of \cite{woodfasiolo2017}. Such methods require the first- and second-order derivatives of the log-likelihood w.r.t. the elements of $\boldsymbol\mu$ and $\boldsymbol \Theta$, hence the first contribution of this work is to provide efficient methods for evaluating them. In particular, we show that this can be achieved by exploiting sparsity, in the MCD case, and, under the logM model, by developing highly optimised derivative expressions. Note that the latter are more efficient than those provided by \cite{chiu1996} and are useful beyond the additive modelling context considered here. For example, the optimisation and sampling methods used by \cite{kawakatsu2006matrix} and \cite{ishihara2016matrix} to fit, respectively, GARCH and multivariate volatility models, require evaluating the derivatives of the logM parametrisation. The derivatives of the logM parametrisation are also used to fit the multivariate covariance GLMs via the software package discussed in \cite{bonat2018} and to aid model interpretability in the multivariate volatility models considered by \cite{bauer2011forecasting}. Interestingly, \cite{Deng2013} proposed methods to fit logM-based covariance matrix models by maximum penalised likelihood methods, but avoided derivative computation by approximating the log-likelihood.

The only model-specific input of the fitting framework outlined above are the derivatives of the log-likelihood w.r.t. the response distributions parameters (here $\boldsymbol\mu$ and $\boldsymbol \Theta$) so, in principle, no further modification is needed to fit the models considered here. However, given that the number of elements of $\boldsymbol \Theta$ grows quadratically with the dimension $d$ of the response vectors and that each element can be modelled via linear predictors containing several regression coefficients, the memory requirement can be substantial even for moderate $d$. Hence, the second contribution of this paper is to propose a block-oriented computational framework that limits memory requirements, while not affecting the leading cost of computation. We also show how to further speed up computation in cases where many linear predictors are left unmodelled, with a realistic example of such scenarios being provided in the application to electricity load modelling considered here. 

As discussed in \cite{woodfasiolo2017}, the FS update is guaranteed to maximise the LAML only under linear-Gaussian models, which is not the case here due to the non-linear relation between $\boldsymbol\Sigma$ and $\boldsymbol \Theta$. Hence, the third contribution of this work is to consider an ``exact'' version of the FS update, which is guaranteed to provide an ascent direction on the LAML objective. Further, we provide methods for efficiently computing the update that are applicable to any non-linear or non-Gaussian additive model.

At the time of writing, we are not aware of other work focussing on the computational challenges that must be addressed when fitting regression models based on either the MCD or the logM parametrisation. However, the models considered here are closely related to the MCD-based multivariate Gaussian models considered by \citet{muschinski2024} and to the regression copula models of \cite{klein2025bayesian}, which exploit the logM parametrisation for correlation matrices proposed by \cite{archakov2021}. In the former, the covariance matrix is allowed to vary with covariates, as is the case in the unconstrained models considered by \cite{tastu2015space} and \cite{bonat2016}. Related models are also the fixed covariance, MCD-based regression models of \cite{cottet2003bayesian} and the multivariate functional additive mixed models of \cite{volkmann2023multivariate}.

The rest of the paper is structured as follows. In Section \ref{sec:model} we describe the model class considered here and the proposed fitting framework. In Section \ref{sec:effComp} we explain how to efficiently compute the log-likelihood derivative w.r.t. the elements of $\boldsymbol\mu$ and $\boldsymbol \Theta$ under the MCD and logM parametrisations, comparing the proposed methods with automatic differentiation and assessing the relative performance of models based on either parametrisation. We discuss methods for limiting memory usage and for efficiently computing the exact FS update in, respectively, Section \ref{sec:blocked_D_beta} and \ref{sec:effic_fellner}, while in Section \ref{sec:App} we consider an application to probabilistic electricity demand modelling.

\section{Multivariate Gaussian Additive Models} \label{sec:model} 
This section defines additive mean and covariance matrix models, and details the corresponding fitting methods.  Henceforth, $\text{diag}(\cdot)$ is the vector-to-matrix diagonal operator, while   $\text{tr}(\mathbf X)$ and $|\mathbf X|$ denote the trace and the determinant of a matrix $\mathbf X$, respectively.

\subsection{Model Structure} \label{sec:model_gaussian}
Consider $d$-dimensional, independent random vectors $\mathbf{y}_i=(y_{i1}, \ldots, y_{id})^\top$, $i=1, \ldots, n$, normally distributed with mean vector $\bsMu_i$ and covariance matrix $\bsS_i$, that is $\mathbf y_i\sim \mathcal{N}(\bsMu_i, \bsS_i)$.  Let $\bsEta$ be the $n \times q$ matrix of linear predictors, where $q=d+d(d+1)/2$ is the number of distributional parameters controlling the mean and covariance matrix, and indicate its $i$-th row and $j$-th column with, respectively, $\bsEta_{i\cdot}$ and  $\bsEta_{\cdot j}$. The mean vector $\boldsymbol \mu_i$ is modelled via $\mu_{ij}=\eta_{ij}$, $j=1, \ldots, d$, while $\eta_{ij}$, $j=d+1, \ldots, q$, model the unconstrained entries of a covariance matrix reparametrisation. By omitting the subscript $i$, the latter can be formalised via a symmetric $d \times d$ parametrisation matrix  

\begin{equation}\label{mat:theta}
\boldsymbol \Theta=\begin{pmatrix} \eta_{d+1} & \eta_{2d+1} & \eta_{2d+2} & \cdots & \cdots & \eta_{q-d+2}\\
\eta_{2d+1} & \eta_{d+2} & \eta_{2d+3}   & \cdots & \cdots & \eta_{q-d+3} \\
\eta_{2d+2} & \eta_{2d+3} & \eta_{d+3}&   \cdots  & \cdots & \vdots \\                               \vdots & \vdots & \vdots & \ddots   & \cdots & \vdots \\
\vdots & \vdots & \vdots & \vdots   &  \ddots&\eta_{q}\\
\eta_{q-d+2} & \eta_{q-d+3} & \cdots &  \cdots & \eta_{q} & \eta_{2d}
\end{pmatrix}. \end{equation}\\
%defined via $\diag(\boldsymbol \Theta)=(\eta_{d+1}, \ldots, \eta_{2d})^\top$ and $\rvechs(\boldsymbol \Theta)= (\eta_{2d+1}, \ldots,\eta_q)^\top$, where $\rvechs(\cdot)$ is the strict row-wise half-vectorisation operator that for the $d \times d$ matrix $\boldsymbol \Theta$  corresponds to $\rvechs(\mathbf Z) = (\Theta_{21}, \Theta_{31}, \Theta_{32},  \dots, \Theta_{(d-1)(d-1)})^\top$.
The elements of $\boldsymbol \Theta$ are related to specific entries of the logM- and MCD-based covariance matrix regression models, as will be clarified in Section \ref{sec:effComp}. Although such models can be defined without introducing $\boldsymbol \Theta$, using this auxiliary matrix eases the notation. 

In an additive modelling context, the elements of $\boldsymbol \eta_{i\cdot}$ are modelled by  
\begin{equation} \label{eq:lpi}
    \eta_{ij}=\sum_{h}^{} f_{hj}(x^h_i)\,\,, \;\;\;  j = 1, \ldots, q, 
\end{equation}
where the $f_{hj}(\cdot)$'s can be linear or smooth effects of $x^h_i$, that is the $h$-th element of an $s$-dimensional vector of covariates $\boldsymbol x_i$. %Here, we restrict the formulation to univariate effects, although the r.h.s. of (\ref{eq:lpi}) could account for bivariate smooth effects and smooth factor interactions 
However, the r.h.s. of (2) can be straightforwardly generalised beyond univariate effects.  Assume that the smooth functions appearing in (\ref{eq:lpi}) are built using linear spline basis expansions, so that the linear predictors can be expressed as $\bsEta_{\cdot j} = \mathbf X^j \bsBeta_j$, $j=1, \ldots, q$, where $\mathbf X^j$ is an $n \times p_j$ model matrix and $\bsBeta_j$ includes the regression coefficients of both linear and smooth effects. Let $\bsBeta = ({\bsBeta_1}^\top, \dots, {\bsBeta_q}^\top)^\top$ be the $p$-dimensional vector of coefficients belonging to all the linear predictors. To avoid overfitting, assume that the effects' complexity is controlled via an improper Gaussian prior, under which $ {\boldsymbol \beta} \sim \mathcal{N}(\mathbf 0,  {\bf S}^{\boldsymbol \lambda} {}^{-})$. Here ${{\bf S}}^{\boldsymbol \lambda} {}^{-}$ is a Moore-Penrose pseudoinverse of ${{\bf S}}^{\boldsymbol \lambda}=\sum_{u=1}^{U}\lambda_u{{\bf S}}^u$, where $\boldsymbol \lambda = (\lambda_1, \ldots, \lambda_U)^\top$ is the vector of positive smoothing parameters, controlling %the strength of the prior penalties, 
the prior precision, and the ${{\bf S}}^u$'s are positive semi-definite matrices. See \citet{Wood2017} for a detailed introduction to smoothing spline bases and penalties.

\subsection{Model Fitting} \label{sec:GAM_gen}

Let $\log p(\bsBeta|\mathbf y, \boldsymbol \lambda)$ be the posterior distribution of $\bsBeta$, conditional on $\boldsymbol \lambda$. In this work, we consider maximum a posteriori (MAP) estimation of $\boldsymbol \beta$, that is 
\begin{equation} \label{eq:log_posterior}
\hat{\bsBeta} = \underset{\bsBeta}{\argmaxB}\, \mathcal{L}(\bsBeta), \;\;\; \text{where} \;\;\; \ \mathcal{L}(\bsBeta) = \log p(\bsBeta|\mathbf y, \boldsymbol \lambda) \propto \bar\ell - \frac{1}{2} \bsBeta^\top {\bf S}^{\boldsymbol \lambda} \bsBeta \,\,,\\
\end{equation}
$\bar\ell = \sum_{i=1}^n\ell_i = \sum_{i=1}^n \log p(\mathbf y_i|\bsBeta)$, with $\ell_i$ the $i$-th log-likelihood contribution, and $ \bsBeta^\top {\bf S}^{\boldsymbol \lambda}\bsBeta/2 \propto \log p(\bsBeta|\boldsymbol \lambda)$ the prior log density.   
To do so, we use Newton's algorithm, which requires the gradient and Hessian of $\mathcal{L}(\bsBeta)$ w.r.t. $\bsBeta$. These are respectively
$$
\nabla_{\bsBeta} \, \mathcal{L} (\bsBeta)  = \mathcal L^{\bsBeta} (\bsBeta) = \sum_{i=1}^n \ell_i^{\bsBeta} - {{\mathbf S}}^{\bsLambda} \bsBeta \,\, , \quad \text{and} \quad
    \nabla_{\bsBeta}^\top \nabla_{\bsBeta} \, \mathcal{L} (\bsBeta)  = \mathcal L^{\bsBeta \bsBeta} (\bsBeta)
    =
    \sum_{i=1}^n \ell_i^{\bsBeta \bsBeta} - { {\mathbf S}}^{\bsLambda}\,\,,
$$
where $\big(\ell_i^{\bsBeta}\big)_j = \partial \ell_i^{\bsBeta} / \partial \beta_j$ and $\big(\ell_i^{\bsBeta\bsBeta}\big)_{jk} = \partial^2 \ell_i^{\bsBeta} / \partial \beta_j\partial \beta_k$. Gradient and Hessian are formed by blocks depending on the parametrisation-specific log-likelihood derivatives w.r.t. $\bsEta$, that is 
\begin{equation} \label{eq:gr_hess_block2cap1}
\bar\ell^{\bsBeta_j} = {\mathbf {X}^j}^\top \ell^{\bsEta_{j}}\, , \quad \text{and} \quad \bar\ell^{\bsBeta_j \bsBeta_k} = {\mathbf {X}^k}^\top \text{diag}(\ell^{\bsEta_j \bsEta_k}) {\mathbf {X}^j}\,\,,
\end{equation}
where $(\ell^{\bsEta_j})_i = \partial \ell_i / \partial \eta_{ij}$ and  $(\ell^{\bsEta_j\bsEta_k})_i = \partial^2 \ell_i / \partial \eta_{ij} \partial \eta_{ik}$. For fixed $\bsLambda$, the efficient computation of the log-likelihood derivatives w.r.t. $\bsBeta$ is essential for fast model fitting. Hence, Section \ref{sec:effComp} focuses on computing $\ell^{\bsEta_{j}}$ and $\ell^{\bsEta_j \bsEta_k}$ efficiently for both logM- and MCD-based models, while Section \ref{sec:blocked_D_beta} proposes a bounded-memory strategy to compute %for computing 
the Hessian blocks $\bar \ell^{\bsBeta_j \bsBeta_k}$.

However, the real challenge when fitting the GAMs is selecting the smoothing parameters. 
Here, we select $\boldsymbol \lambda$ by maximising a Laplace approximation to the marginal log-likelihood (LAML), $\mathcal{V}(\boldsymbol \lambda) = \log \int p(\mathbf y| \bsBeta)p(\bsBeta|\boldsymbol \lambda) d \bsBeta$, that is  
\begin{equation} \label{eq:LAML}
\tilde{\mathcal{V}}(\bsLambda) =  \mathcal{L}(\hat{\bsBeta})+\frac{1}{2}\log|{\bf S}^{\bsLambda}|_{+}-\frac{1}{2}\log|\boldsymbol {\mathcal{H}}|+\frac{M_{p}}{2}\log(2\pi)\,\,,
\end{equation}
where $M_{p}$ is the dimension of the null space of ${\bf S}^{\bsLambda}$, $|{\bf S}^{\bsLambda}|_{+}$ is the product of its positive eigenvalues,   $\hat{\bsBeta}$ is the maximiser of $\mathcal{L}(\bsBeta)$, and $\boldsymbol{\mathcal{H}}=- \mathcal L^{\bsBeta \bsBeta} (\bsBeta)$  % its negative Hessian, 
evaluated at $\hat{\bsBeta}$. To maximise $\tilde{\mathcal{V}}(\bsLambda)$ efficiently, we consider the generalised Fellner-Schall iteration proposed by \cite{woodfasiolo2017}. This starts by considering the first derivatives of $\tilde{\mathcal{V}}(\bsLambda)$ w.r.t. $ \lambda_u$
%
%\begin{equation} \label{eq:LAML_score}
%\frac{\partial \tilde{\mathcal{V}}(\bsLambda)}{\partial \lambda_u} = \frac{1}{2} \left\{ -\hat  \bsBeta^\top {\mathbf S}^u \hat  \bsBeta + {\rm tr}({\mathbf S^{\boldsymbol \lambda}}^{-}\mathbf S^{u}) - \tr \left(\boldsymbol{ \mathcal{H}}^{-1} {\mathbf S}^u  \right) - \tr \left(\boldsymbol{ \mathcal{H}}^{-1}{\bar\ell}^{\hat \bsBeta \hat \bsBeta\rho_u}  \right)  \right\} = -a + b - c \,,
%\end{equation}
%
%for $u=1, \ldots, U$, where ${\bar\ell}^{\hat \bsBeta \hat \bsBeta\rho_u}=\partial \bar \ell^{\hat \bsBeta \hat \bsBeta}/\partial \lambda_u$,
%
\begin{equation} \label{eq:LAML_score}
\frac{\partial \tilde{\mathcal{V}}(\bsLambda)}{\partial \lambda_u} = \frac{1}{2} \left\{ -\hat  \bsBeta^\top {\mathbf S}^u \hat  \bsBeta + {\rm tr}({\mathbf S^{\boldsymbol \lambda}}^{-}\mathbf S^{u}) - \tr \left(\boldsymbol{ \mathcal{H}}^{-1} {\mathbf S}^u  \right) - \tr \left(\boldsymbol{ \mathcal{H}}^{-1}{\bar\ell}^{\hat \bsBeta \hat \bsBeta\lambda_u}  \right)  \right\} = \frac{1}{2}(-a + b - c) \,,
\end{equation}
for $u=1, \ldots, U$, where ${\bar\ell}^{\hat \bsBeta \hat \bsBeta\lambda_u}=\partial \bar \ell^{\hat \bsBeta \hat \bsBeta}/\partial \lambda_u$, $b = {\rm tr}({\mathbf S^{\boldsymbol \lambda}}^{-}\mathbf S^{u}) - \tr \left(\boldsymbol{ \mathcal{H}}^{-1} {\mathbf S}^u  \right)$, and with terms $a$ and $c$ being defined accordingly. \cite{woodfasiolo2017} show that $b \geq 0$, provided that $\boldsymbol{ \mathcal{H}}$ is positive semi-definite (if it is not, then an appropriately perturbed or expected version can be used instead). Under the further assumption that $c = 0$, it is easy to see that the update $\lambda_u^* = \lambda_u b/a$ (henceforth the Fellner-Schall update, FS) will increase (decrease) $\lambda_u$ when $\partial \tilde{\mathcal{V}}(\bsLambda)/\partial \lambda_u > 0$ ($< 0$), thus providing a LAML ascent direction. 

From a computational perspective, the FS update is attractive because it avoids the computation of the term $c$ which, as explained in Section \ref{sec:effic_fellner}, requires the third-order derivatives of the log-likelihood w.r.t. $\bsEta$. However, the assumption $c = 0$ does not hold under any of the models considered here, hence the FS iteration is not guaranteed to reach the LAML maximiser. Nonetheless, under the logM parametrisation the third derivatives are expensive to compute, hence the FS update is the most attractive option for scalable computation. Instead, under the MCD parametrisation such derivatives can be computed cheaply, which enables the adoption of the following iteration \citep{Wood25GAM}
\begin{equation} \label{eq:EFS_exact} 
\lambda_u^* = \begin{cases} \lambda_u(b - c)/a & c \leq 0 \\  \lambda_u b/(a+c) & c > 0 \end{cases}\,.
\end{equation} 
We call such update the ``exact'' FS (EFS) update because its direction agrees with the sign of the corresponding LAML gradient element, while avoiding the $c = 0$ assumption.  

The next section focusses on the efficient computation of the log-likelihood derivatives w.r.t. $\bsEta$, under either the logM or MCD parametrisation. %Then Section \ref{sec:blocked_D_beta} considers the scalable computation of the log-likelihood derivatives w.r.t. $\bsBeta$, while Section \ref{sec:effic_fellner} provides details on the efficient computation of the terms appearing in (\ref{eq:LAML_score}), which are needed for the FS and EFS updates, and compares the computational performance of the FS, EFS and other fitting methods under the MCD parametrisation.

\section{Efficient Computation of Parametrisation-Specific log-Likelihood Derivatives} \label{sec:effComp}

%Let us introduce the following notation.  %We distinguish $\diag(\cdot)$ as the vector-to-matrix diagonal operator, from  $\diagv(\cdot)$ as the matrix-to-vector diagonal operator. 
 %, and indicate with $\mathbf X_{r\cdot}$ and $\mathbf X_{\cdot s}$ the $r$-th row and the $s$-th column of a matrix $\mathbf X$, respectively. 
%This section details the efficient computation of the log-likelihood derivatives w.r.t. $\boldsymbol \eta$ under the logM and  MCD parametrisations. %, which are, respectively, introduced in a joint mean and covariance matrix regression context by \cite{chiu1996} and \citet{pourahmadi1999}. 
If we define $\mathbf r_i=\mathbf y_i - \boldsymbol \mu_i$, then the $i$-th log-likelihood contribution is 
\begin{equation} \label{eq:loglik}  \ell_i = -\frac{1}{2}\log |\boldsymbol \Sigma_i|-\frac{1}{2}\mathbf r^\top_i \boldsymbol \Sigma^{-1}_i \mathbf r_i \,\, , \end{equation}
up to an additive constant.
In the following, we omit the subscript $i$ to lighten the notation (e.g., $\boldsymbol \eta_{i\cdot}$ is simply $\boldsymbol \eta$) and denote the Hadamard product with $\odot$, while $\mathbf X_{q\cdot}$ and $\mathbf X_{\cdot r}$ represent, respectively, the $q$-th row and the $r$-th column of a matrix $\mathbf X$. 
 
%\newpage 

\subsection{Log-Likelihood Derivatives under the logM Parametrisation} \label{sec:logM}

\cite{chiu1996} consider an unconstrained parametrisation matrix $\boldsymbol \Theta$ defined by $\boldsymbol \Sigma = \exp(\boldsymbol \Theta)$, where $\exp(\cdot)$ is the matrix exponential such that, if $\boldsymbol \Theta = \mathbf U \boldsymbol \Gamma \mathbf U^\top$ is the eigendecomposition of $\boldsymbol \Theta$, then $\boldsymbol \Sigma = \mathbf U\exp(\boldsymbol \Gamma)\mathbf U^\top$. Hence, $\boldsymbol \Theta$ is the matrix logarithm of $\boldsymbol \Sigma$, and the positive-definiteness of the latter is guaranteed. Given that $\log |\boldsymbol \Sigma|=\tr(\boldsymbol \Theta)$, the log-likelihood given in (\ref{eq:loglik}) is now 
\begin{equation} \label{eq:loglik_logM}
 \ell= -\frac{1}{2} \left\{\tr(\boldsymbol \Theta) + \mathbf r^\top\exp(-\boldsymbol \Theta)\mathbf r\right\} = -\frac{1}{2} \sum_{j=d+1}^{2d}\eta_j- \frac{1}{2}\mathbf s^\top \mathbf L  \mathbf s\,\, ,
\end{equation}
where $\mathbf s=\mathbf U^\top \mathbf r$ and $\mathbf L = \exp (-\boldsymbol \Gamma)$. 
Although the cost of evaluating the log-likelihood by eigen-decomposing $\boldsymbol \Theta$ is $\mathcal{O}(d^3)$, the main bottleneck during model fitting is the computation of the log-likelihood derivatives w.r.t. $\bsEta$ (i.e., the elements of $\boldsymbol \Theta$), which are required by the methods described in Section \ref{sec:GAM_gen}. 
% Then it is immediate that $\Sigma^{-1} = U \exp(-\Gamma)U^T$ and easy to get to $\log|\Sigma| = tr(\Omega)$.

% Let $\mathbf {U}$  and ${\boldsymbol \Lambda}$ be, respectively, the orthonormal matrix of eigenvectors and the diagonal matrix containing the eigenvalues resulting from eigen-decomposing $\boldsymbol \Sigma$. Let ${\boldsymbol \Gamma}$ be a diagonal matrix with non-zero elements $\gamma_j = \Gamma_{jj} = \log\Lambda_{jj}$. Then, the logM parametrisation of $\boldsymbol \Sigma$ is
% \begin{equation}\label{eq:log_cov}
% \log \boldsymbol \Sigma =\mathbf U {{\boldsymbol \Gamma}} \mathbf U^\top\,\,.
% \end{equation}

% The non-redundant entries of $\log \boldsymbol \Sigma$ are constraint-free and can be covariate-dependent, %modelled using covariates, 
% as the positive definiteness of $\boldsymbol \Sigma$ is ensured by construction \citep[see][for more  details]{chiu1996}. 

% In terms of the generic parametrisation matrix defined in (\ref{mat:theta}), we simply have $\boldsymbol \Theta = \log \boldsymbol \Sigma$. 

To the best of our knowledge, analytic expressions for such derivatives can be obtained in two main ways. The first relies on the chain rule for the matrix exponential \cite[see e.g.][]{mathias1996chain}, where the derivatives are obtained as functions of augmented matrices. This approach requires computing the exponential of block-triangular matrices via a scaling and squaring algorithm based on a Pad\'e approximation \cite[see e.g.][Section 10.3]{higham2008functions}, which is expensive.
For example, obtaining
$\ell^{\eta_j \eta_k}$, for $l, k \in \{d+1, \ldots, q\}$, involves building a matrix $4d \times 4d$ and then computing its matrix exponential. The second approach involves expressing the directional derivatives of the matrix exponential via an integral and solving it using the spectral decomposition \cite[see e.g.][]{najfeld1995}. This is the approach taken by \cite{chiu1996}, which we improve below to obtain derivative expressions that are much more computationally efficient. 

Let $\mathbf Z$ and $\mathbf W$ be two $(d-1)\times(d-1)$ lower triangular matrices such that ${\rm Z}_{jk}=k\mathbbm{1}_{\{k\leq j\}}$ and ${\rm W}_{jk}=(j+1)\mathbbm{1}_{\{k\leq j\}}$. Denoting with $\rvech(\cdot)$ the row-wise half-vectorisation operator, that is $\rvech(\mathbf Z) = (Z_{11}, Z_{21}, Z_{22}, Z_{31}, Z_{32}, Z_{33}, \dots, Z_{(d-1)(d-1)})^\top$, we define $\mathbf z=\rvech( \mathbf Z) $ and  $\mathbf w=\rvech( \mathbf W)$.  
Then, log-likelihood gradient, $\ell^{\bsEta} = (\ell^{\eta_1}, \dots, \ell^{\eta_q})^\top$, is 
\begin{align}
\ell^{\eta_l}&=\big[\mathbf U \mathbf L \mathbf s\big]_l \,\,,  \;\;\;\,  \quad \quad \quad \quad \quad  l = 1, \dots, d, \label{eq:grad_logm0}  \\
\ell^{\eta_l}&=\frac{1}{2} [\boldsymbol \Xi]_{l-d,l-d} -\frac{1}{2}\,\, , \;\;\; \quad  \,\, \, \,  l =d+1,\ldots, 2d, \label{eq:grad_logm1} \\
\ell^{\eta_l}& =[\boldsymbol \Xi]_{z_{l-2d},w_{l-2d}}\,\, , \;\;\; \quad \quad \quad  l =2d+1,\ldots, q, \label{eq:grad_logm2}
\end{align}
where $
\boldsymbol \Xi =  \mathbf F \boldsymbol \Delta \mathbf F^\top
$
with $\mathbf F =  \mathbf U \diag(\mathbf s)$ and  $\boldsymbol \Delta$ is a $d \times d$ matrix such that  $\Delta_{jj} =e^{-\gamma_j}$ and  $\Delta_{jk}= \big(e^{-\gamma_k}-e^{-\gamma_j}\big)/(\gamma_j  - \gamma_k)$,   with $\gamma_j = \Gamma_{jj}$. See SM \ref{app:Dllk_Deta_logm} for the derivation of (\ref{eq:grad_logm0} - \ref{eq:grad_logm2}).

\begin{remark} Computing $\ell^{\eta_l}$, for $l = 1, \dots, d$, requires two $\mathcal{O}(d^2)$ matrix-vector multiplications, while computing $\ell^{\eta_l}$, for  $l = d+ 1, \dots, q$,  requires computing $\boldsymbol \Xi$, which is of order $\mathcal{O}(d^3)$, and then simply accessing its elements. Hence, the total cost is $\mathcal{O}(d^3)$, which is of the same order as computing the log-likelihood itself by eigen-decomposing $\boldsymbol \Theta$. In contrast, the first-order derivative formulas provided by \cite{chiu1996}, have an $\mathcal{O}(d^4)$ cost because they involve computing, for each of the $\mathcal{O}(d^2)$ elements of the gradient, a sum over $\mathcal{O}(d^2)$ terms (see formulas (16) and (A.1) therein).

%have the same $\mathcal{O}(d^5)$ cost as using directly the expressions (\ref{eq:naive1}) and (\ref{eq:naive2}) in SM \ref{app:Dllk_Deta_logm}, due to the computation, for each $j= d+1, \ldots, q$, of $\mathbf U^\top {\mathbf V}^j \mathbf U$, where the ${\mathbf V}^j$'s are basis matrices ${\mathbf V}^j$ resulting from $\boldsymbol \Theta= \sum_{j=1}^{d(d+1)/2}{\eta_{j+d}  \mathbf V^j}$. 

\end{remark}%See SM \ref{app:Dllk_Deta_logm} for an example of the matrices ${\mathbf V}^j$.
%have the same $\mathcal{O}(d^5)$ cost as using directly the expressions (\ref{eq:naive1}) and (\ref{eq:naive2}) above, due to the computation of $\mathbf U^\top {\mathbf V}^j \mathbf U$, for each $j= d+1, \ldots, q$. 
%While exploiting the sparsity of ${\mathbf V}^j$ would reduce the cost to $\mathcal{O}(d^4)$, this is still $d$ times slower than the proposed approach.
 
The first $d$ rows of the Hessian matrix are given by
\begin{align}
\ell^{\eta_l\eta_m} &= -\big[\mathbf U \mathbf L \mathbf U^\top\big]_{lm}\,\, ,  \;\;\;  \quad \quad \quad \quad \quad \quad  \,\, m = l, \ldots, d,  \label{eq:Hess_1_1} \\
\ell^{\eta_l\eta_m} &=   -\mathbf U_{ l\cdot} \big(\boldsymbol \Pi_{(m-d)\cdot} \odot \mathbf U_{(m-d)\cdot}\big)^\top\,\,, \;\;\; \,\,\,  m=d+ 1, \ldots, 2d,  \label{eq:Hess_1_2} \\
\ell^{\eta_l\eta_m} &= -\mathbf U_{l\cdot}   {\tilde {\mathbf u}^{m-2d}}\,\,, \;\;\;  \quad \quad \quad \quad \quad \quad \quad \, m=2d+ 1, \ldots, q, \label{eq:Hess_1_3}
\end{align}
for $l=1,\ldots,d$. Here $\boldsymbol \Pi=\mathbf F \mathbf \Delta$ and the $j$-th entry of  ${\tilde {\mathbf u}^{m-2d}}$, for $j=1, \ldots, d$, is given by
\begin{equation*}  
{\tilde{\mathbf  u}^{m-2d}_j} =\Pi_{ z_{m-2d}j}{\rm U}_{w_{m-2d}j } +\Pi_{ w_{m-2d}j} {\rm U}_{ z_{m-2d}j}\,\, .
\end{equation*} 
We provide the details for deriving (\ref{eq:Hess_1_1} - \ref{eq:Hess_1_3}) in SM \ref{app:Dllk_Deta_logm}.

\begin{remark} While (\ref{eq:Hess_1_1}) requires no extra computation, if the eigen-decomposition of $\boldsymbol \Theta$ is available, (\ref{eq:Hess_1_2}) and (\ref{eq:Hess_1_3}) use the matrix $\boldsymbol \Pi=\mathbf F \mathbf \Delta$, which implies a $\mathcal{O}(d^3)$ cost. Further, for each $m$ and $l$, (\ref{eq:Hess_1_2}) and (\ref{eq:Hess_1_3}) require $\mathcal{O}(d)$ vector-vector operations, leading to an $\mathcal{O}(d^4)$ total cost. Note that the mixed derivatives blocks in (\ref{eq:Hess_1_2}) and (\ref{eq:Hess_1_3}) are not required by the fitting procedure of \citet{chiu1996}, which updates the linear predictors controlling $\boldsymbol \mu$ and $\log \boldsymbol \Sigma$ in separate steps, rather than via a joint
Newton update. Further, mixed derivatives are required here for smoothing parameter selection via the FS and EFS updates.  
\end{remark} 

%Further, as explained in Section \ref{sec:GAM_gen}, such derivatives are needed for smoothing parameter selection.

%\begin{prop} \label{prop3}
Denoting with $l' = l-d$, $l'' = l-2d$, $m'=m-d$ and $m''=m-2d$, the elements of the last $q-d$ rows of the Hessian matrix are 
\begin{align} 
\ell^{\eta_l\eta_m} &=  -\sum_{j=1}^{d} {\rm U}_{l'j}{\rm U}_{m'j} {\rm K}_{l'm'j}\,\,, \,\,\quad \quad \quad \quad \quad  \quad \quad \quad \,\,\, \,\,\,\,\,\, m=l, \ldots, 2d, \label{eq:Hess_2_2}\\
\ell^{\eta_l\eta_m} &=  -\sum_{j=1}^{d} {\rm U}_{l'j}\big({\rm U}_{w_{m''}j}{\rm K}_{l'z_{m''}j} + {\rm U}_{z_{m''}j}{\rm K}_{l'w_{m''}j}  \big)\,\,, \,\,    \,\,\,  m=2d+1, \ldots, q, \label{eq:Hess_2_3}
\end{align}
for $l=d+1, \ldots, 2d$, while for $l=2d+1, \ldots, q$, and $m=l, \ldots, q$, they take the form
\begin{align}
\ell^{\eta_l\eta_m} =  -\sum_{j=1}^{d} \big\{&{\rm U}_{w_{l''}j}\big({\rm U}_{w_{m''}j}{\rm K}_{z_{l''}z_{m''}j} + {\rm U}_{z_{m''}j}{\rm K}_{z_{l''}w_{m''}j}  \big)  + \nonumber \\ & {\rm U}_{z_{l''}j}\big({\rm U}_{w_{m''}j}{\rm K}_{w_{l''}z_{m''}j} + {\rm U}_{z_{m''}j}{\rm K}_{w_{l''}w_{m''}j}  \big)\big\}\,\,, \label{eq:Hess_3_3}
\end{align}
where $\mathbf K$ is a $d\times d \times d$ array, with elements
$${\rm K}_{rst} =  {\rm F}_{rt}{\rm F}_{st} \frac {\Delta_{tt}}{2} + {\rm F}_{rt} {\rm A}_{st} + {\rm F}_{st} {\rm A}_{rt} + {\rm A}'_{rst} + {\rm A}''_{rst}\,\,.$$
Here, $\mathbf A$ is a $d\times d$ matrix given by $\mathbf A = \mathbf F \tilde {\boldsymbol \Delta}^T$, where $\tilde {\boldsymbol \Delta}$ is such that $\tilde \Delta_{jk}=(\Delta_{jk}-\Delta_{kk})/(\gamma_j-\gamma_k)$ if $j\neq k$ and 0 otherwise; $\mathbf A'$ is a $d\times d\times d$ array given by $\mathbf A'_{\cdot\cdot j} = \mathbf F\, \big(\mathbf F \odot \overbar{\boldsymbol{\Delta}}_{j}\big)^\top$, where $\overbar{\boldsymbol{\Delta}}_{j}$ is a matrix whose $k$-th row, $k=1, \ldots,d$, is given by $\tilde {\boldsymbol \Delta}_{\cdot j}^\top$; 
and $\mathbf A''$ is a $d\times d\times d$ array given by $\mathbf A''_{\cdot\cdot j} = \mathbf F\, \big(\mathbf F \boldsymbol \Delta^*_{\cdot\cdot j}\big)^\top$, where $\Delta^*_{kk'j}=(\Delta_{k'j} - \Delta_{kk'})/(\gamma_k-\gamma_{j})$ if $j\neq k $ and $k'\neq j, k$, and 0 otherwise. 
%\end{prop}
SM \ref{app:Dllk_Deta_logm} provides details that lead to (\ref{eq:Hess_2_2} %), (\ref{eq:Hess_2_3}) and (
- \ref{eq:Hess_3_3}).
 
\begin{remark} Computing $\ell^{\eta_l\eta_m}$, for $l = d+1, \dots, q$, and $m=l, \ldots, q$, requires computing the $d \times d \times d$ arrays ${\mathbf K}$, $\mathbf A'$ and $\mathbf A''$ at cost $\mathcal{O}(d^4)$. Then, for each $l$ and $m$, we need to compute the sum of $\mathcal{O}(d)$ terms involving the elements of ${\mathbf K}$ and $\mathbf U$, which implies an $\mathcal{O}(d^5)$ cost. Instead, the expressions in \cite{chiu1996} (see formulas (17) and (A.2) therein), involve a sum over $\mathcal{O}(d^3)$ terms for each $l$ and $m$, leading to a total cost of $\mathcal{O}(d^7)$.
\end{remark} 

In summary, the expressions for the first and second log-likelihood derivatives w.r.t. $\bsEta$ provided above save, respectively, $\mathcal{O}(d)$ and $\mathcal{O}(d^2)$ computation, relative to the formulas provided by \cite{chiu1996}. Recall that such derivatives are required by the model fitting methods of Section \ref{sec:GAM_gen} and that the EFS iteration requires also the third-order derivatives.  However, as demonstrated by SM \ref{app:Dllk_Deta_logm}, obtaining efficient formulas for the third-order derivatives would be very tedious. Further, the results discussed in Section \ref{NumExp_eta} show that they would be expensive to compute due to the lack of sparsity, which makes the EFS iteration unattractive under the logM parametrisation. As explained in Section \ref{sec:blocked_D_beta} the lack of sparsity can lead to a large memory footprint even when third-order derivatives are not computed, but the block-based strategy proposed therein alleviates the issue.

\subsection{Log-Likelihood Derivatives under the MCD Parametrisation}\label{sec:MCD}
The MCD parametrisation of \cite{pourahmadi1999} is defined as follows. Let $\mathbf C$ be the lower-triangular Cholesky factor of $\boldsymbol \Sigma$, such that $\boldsymbol \Sigma=\mathbf C\mathbf C^\top$. Then the MCD of $\boldsymbol \Sigma^{-1}$ is
\begin{equation}\label{mcd_prec}
\boldsymbol{\Sigma}^{-1}=\mathbf{T}^\top\mathbf{D}^{-2}\mathbf{T} \, ,
\end{equation}
where $\mathbf D$ is a diagonal matrix such that  ${\rm D}_{jj}={\rm C}_{jj}$, $j=1, \ldots d$, and  $\mathbf T= \mathbf D \mathbf C^{-1}$.  %Under the MCD parametrisation, the linear predictors $\eta_j$, $j=d+1, \ldots, q$, model the elements of $\log \mathbf D^2$ and $\mathbf T$. 
Note that the non-zero elements of $\log \mathbf D^2$ and $\mathbf T$ are unconstrained, hence they can be modelled via the linear predictors $\eta_j$, $j = d+1, \ldots, q$. In terms of the general parametrisation matrix $\boldsymbol \Theta$, defined in (\ref{mat:theta}), we have $\Theta_{kk}= \log {\rm D}^2_{kk}, k= 1, \ldots, d,$ and $\Theta_{kl}= {\rm  T}_{kl}, 1 \leq l < k \leq d$. 
%\cite{pourahmadi1999, pourahmadi2000maximum} introduce such a covariance matrix regression model, adopting the maximum likelihood estimation approach and studying its asymptotic properties. However, the author reduces model complexity by imposing a strong parametric structure for modelling $\log \mathbf D^2$ and $\mathbf T$.

An important advantage of the MCD parametrisation is the interpretability of the decomposition elements, which can be obtained by regressing the response vector elements on their predecessors. In particular, \cite{pourahmadi1999} shows that a random vector $\mathbf y$ such that $\text{cov}(\mathbf y) = \boldsymbol{\Sigma}$ can be generated by the process
$$y_1 = \epsilon_1,\;\;\; y_{l} = -\sum_{k=1}^{l-1}T_{lk}y_{k} + \epsilon_{l}\,\,, \quad  l = 2, \dots, d, $$
where $\mathbbm{E}(\epsilon_{k}) = 0$, $\text{var}(\epsilon_{k}) = D_{kk}^2$ for $k=1,\dots, d$, and $\text{cov}(\epsilon_l, \epsilon_k) = 0$, for $l\neq k$. Hence, the lower-triangular elements of $\mathbf T$ can be interpreted as the negative regression coefficients of an element of $\mathbf y$ on its predecessors, while the elements of $\mathbf D^2$ are related to the variance of the regression residuals. 

Note the interpretation of the MCD elements depends on the order of the response vector elements, which can aid model selection when the elements of $\mathbf y$ have some natural ordering. In particular, see \cite{pourahmadi1999} and \cite{gioia2022additive} for examples where the response elements have, respectively, a temporal and a spatial order.
%this aspect is particularly appealing to facilitate model specification and to reduce model complexity, for instance by setting to zero some sub-diagonals of ${\rm \mathbf  T}$, thus recovering the antedependence model (see \cite{pourahmadi1999} and reference therein), %of \cite{gabriel1962ante}, 
%or modelling them only via intercepts. %, as done by \cite{gioia2022additive}. 
When the elements of $\mathbf y$ do not have a natural ordering, the dependence of the MCD on the ordering is undesirable, hence \cite{kang2020variable} has proposed a strategy to mitigate the ordering effect. %\citep[see e.g.][]{kang2020variable, barratt2022covariance}, 
However, note that \citet{muschinski2024} and  \citet{barratt2022covariance} found the covariance model fit to be insensitive to random permutations of the variables, which suggests that ordering should be checked but might be not a problem in practice.  

From a computational perspective, fitting MCD-based multivariate  Gaussian additive models is very efficient, relative to logM. Indeed, by using the property $\log |\boldsymbol \Sigma| = \tr(\log \mathbf D^2)$ within (\ref{eq:loglik}), we obtain that a log-likelihood contribution is
\begin{equation} \label{eq:loglik_MCD}
 \ell= -\frac{1}{2} \left\{\tr(\log \mathbf D^2)+\mathbf r^\top\mathbf{T}^\top\mathbf{D}^{-2}\mathbf{T}\mathbf r\right\}\,\, .
\end{equation}
which, in contrast with the logM case, can be directly expressed in terms of the linear predictors, thus facilitating the model fitting procedures (see SM \ref{app:Dllk_Deta_mcd} for details). 
Evaluating (\ref{eq:loglik_MCD}) takes $\mathcal{O}(d^2)$ operations, lower than the $\mathcal{O}(d^3)$ required under the logM parametrisation. Further, the derivatives of  (\ref{eq:loglik_MCD}) w.r.t. $\bsEta$ are easier to obtain, relative to the logM case, hence they are reported in SM \ref{app:Dllk_Deta_mcd}. More importantly, they become increasingly sparse as the order of differentiation increases, and this greatly speeds up computation.

\begin{remark} The log-likelihood gradient w.r.t. $\boldsymbol \eta$ is dense and requires $\mathcal{O}\big(d^3\big)$ computations, as for the logM parametrisation. The Hessian is sparse, with a ratio of non-zero to total unique elements of $\mathcal{O}(d^{-1})$. Most of these $\mathcal{O}\big(d^3\big)$  elements cost $\mathcal{O}(d)$, leading to a total cost of $\mathcal{O}(d^4)$, which should be compared with $\mathcal{O}\big(d^5\big)$ under the logM model. The third-order derivatives are even sparser, the ratio of non-zero to the total unique elements being $\mathcal{O}\big(d^{-3}\big)$. Most of these $\mathcal{O}\big(d^3\big)$ elements cost $\mathcal{O}(d)$, leading to an $\mathcal{O}\big(d^4\big)$ total cost. See SM \ref{app:Dllk_Deta_mcd} for MCD derivative expressions and SM \ref{app:sparsity} for an analysis of their sparsity patterns.
\end{remark}

Recall from Section \ref{sec:GAM_gen} that the first two log-likelihood derivatives w.r.t. $\bsEta$ are needed for model fitting, while the third-order derivatives are required for smoothing parameter optimisation via the EFS iteration. The sparsity of the third-order derivatives makes the EFS iteration a feasible option under the MCD parametrisation, hence Section \ref{sec:effic_fellner} compares the FS and the EFS iteration in a simulation-based scenario.

%For clarity of presentation, the first three derivatives are reported in SM \ref{app:Dllk_Deta_mcd}, either in matrix and indicial form, the latter being computationally convenient because leverages the diagonal and triangular structure, respectively, of the matrices  $\mathbf D$ and $\mathbf T$.

%The $\bsEta$ vector can be divided into the sub-vectors with elements controlling respectively, $\bsMu$, $\log \mathbf{D}^2$ and $\mathbf T$. Hence, the gradient vector can be divided into three corresponding sub-vectors, the Hessian into six unique matrix blocks and the third-order derivatives array into ten unique arrays.

%Note that, while the sparsity of the log-likelihood derivatives directly reduces the computational effort required to fit the MCD-based multivariate Gaussian additive model, its impact on memory storage requirements is more complex to quantify. In particular, the memory needed to store the evaluated derivatives depends on the model structure and the computational details of the chosen fitting framework. Section  \ref{sec:blocked_D_beta} %\ref{sec:BMCS} 
%focuses on such issues.

\subsection{Comparison between the logM and MCD Parametrisations} \label{NumExp_eta}

The analytical results provided so far show that the computational cost for evaluating $\ell^{\eta_l}$, $l=1, \ldots,q$, is of the same order for the logM and MCD parametrisations, while evaluating $\ell^{\eta_l\eta_m}$, $l,m=1, \ldots,q$, under the logM parametrisation cost $\mathcal{O}(d)$ times more than for the MCD. Here we empirically assess the computational time needed to evaluate the second-order derivatives under each model,  
by considering $n=10^3$ and $d \in \{5, 10, \dots, 50\}$.
To provide a benchmark, we compare the timings obtained using the derivative expressions provided here (henceforth called \texttt{Efficient}) with those obtained by using automatic differentiation (henceforth \texttt{AD}) via \textbf{TMB} package \citep {tmbPackage}.

\begin{figure}[htbp]
\centering
\includegraphics[scale=0.54]{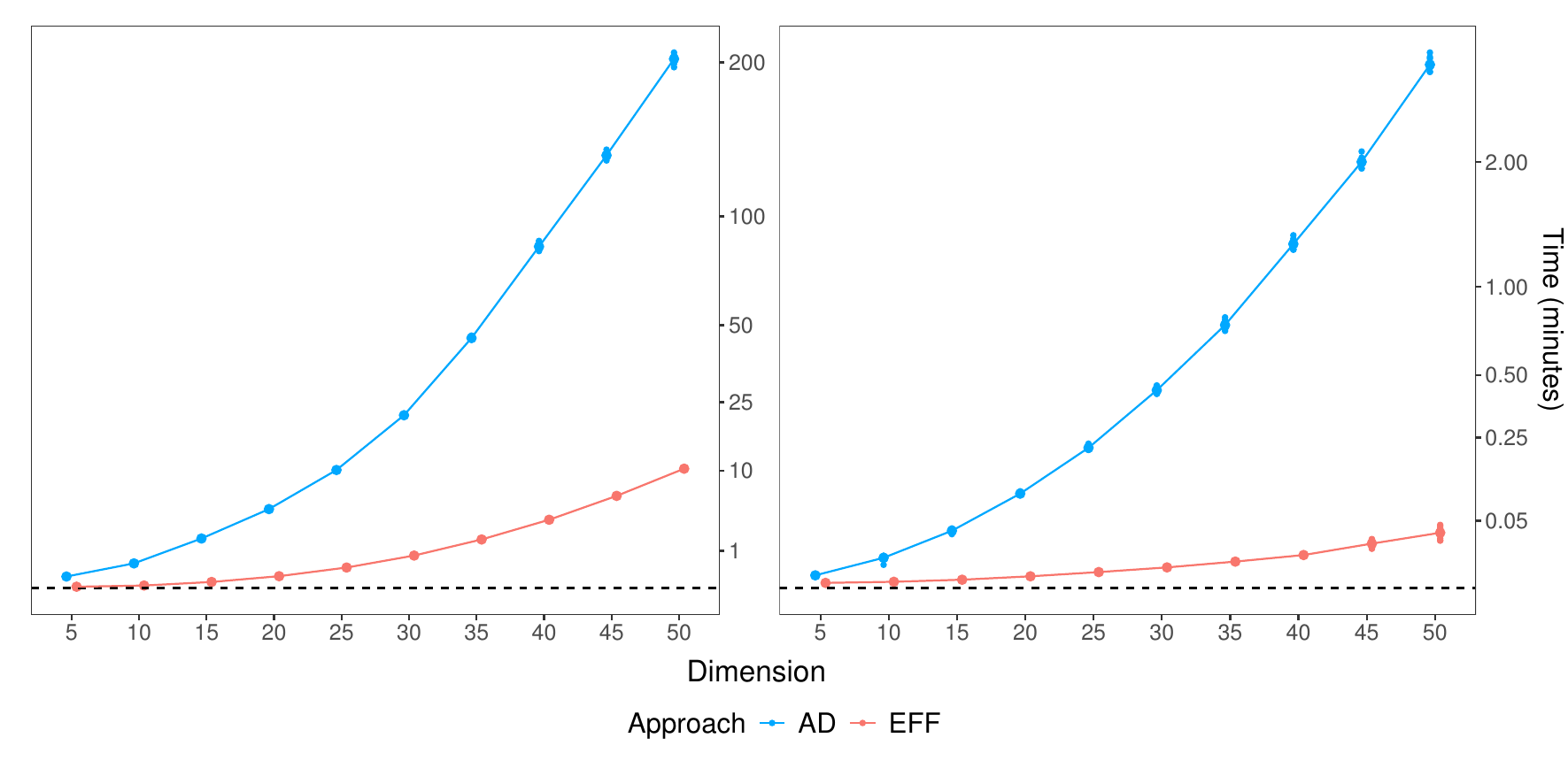}
\caption{Time (in minutes) for evaluating $\ell^{\eta_l\eta_m}$, $l,m=1, \ldots,q$, under the logM (left) and the MCD (right) parametrisations. The times for the \texttt{Efficient} (EFF) and \texttt{AD} are reported in red and blue, respectively. The connected bigger dots represent the mean times, while smaller dots depict the ten observed times. The $y$-axes are on a square root scale.}%The black dashed line report the reference null time.}
\label{fig:score_hessian}
\end{figure}

Figure \ref{fig:score_hessian} shows the results obtained over ten runs. By comparing the $y$-axes scales it is clear that the MCD is more scalable than the logM model as $d$ grows. Further, the relative performance of \texttt{AD} deteriorates as $d$ increases, particularly under the MCD model. 
A tentative explanation is that, under the logM parametrisation, the evaluation of the derivatives via \texttt{AD} is less optimised than the expressions provided in Section \ref{sec:logM}, which required considerable analytical work aimed at efficient evaluation (see SM \ref{app:Dllk_Deta_logm}). Under the MCD model, AD is not, to our best knowledge, automatically detecting the sparsity pattern of the Hessian, hence it is computing more elements than necessary.

We also compare the logM and the MCD parametrisations in the context of additive covariance matrix model fitting, using the FS iteration described in Section \ref{sec:GAM_gen}. To do this, we generate the data from $\mathbf y_i \sim \mathcal{N}(\boldsymbol \mu_i, \bsS_i)$, $i=1, \ldots, n$, where $\bsS_i$ is obtained from  $\boldsymbol \Theta_i$ via either the MCD or the logM parametrisation. In particular, the true mean vector and the lower triangular entries of $\boldsymbol \Theta_i$, are simulated according to  
\begin{align*}
\eta_{ij} = &\,\,  m_{1j} \sin(\pi x_{i1})\, + \exp(m_{2j} x_{i2}) \, +   \\
&\,\, m_{3j}  x^{11}_{i3} \{m_{4j}  (1 - x_{i3})\}^6 \,+\, m_{5j} (m_{6j} x_{i3})^3  (1 - x_{i3})^{10} \,\, ,\,\, \quad \quad \quad \,\, j=1, \ldots, d, \nonumber \\
\eta_{ij} = &\,\, m_{7j} + m_{8j}  \sin\{2 \pi (x_{i1} + m_{9j})\}\, + m_{10j} \cos\{2 \pi (x_{i1}+m_{9j})\}\,+ \\ 
&\,\, m_{11j}  \sin\{2 \pi (x_{i2} + m_{12j})\} + m_{13j} \cos\{2 \pi (x_{i2}+m_{12j})\} \,\,, \,\,\,\,\,\,\, \quad  j=d+1, \ldots, q,\nonumber
\end{align*}
where $x_{i1}$, $x_{i2}$ and $x_{i3}$ are generated from the standard uniform $\mathcal{U}(0,1)$,  
$m_{1j}, m_{2j} \sim \mathcal{U}(1,3)$, $m_{3j} \sim \mathcal{U}(0,0.5)$, $m_{4j}, m_{5j},m_{6j} \sim \mathcal{U}(9,11)$, 
$m_{7j},m_{10j}\sim \mathcal{U}(-0.25,0.25)$, $m_{8j}, m_{9j},m_{13j} \sim \mathcal{U}(-0.5,0.5)$, and $m_{11j},m_{12j} \sim \mathcal{U}(-1,1)$.

We fit the model $\eta_{ij} = f^{10}_{j1}(x_{i1})+f^{10}_{j2}(x_{i2}) + f^{10}_{j3}(x_{i3})$, for $j=1, \ldots, d$, and $\eta_{ij} = f^{10}_{j1}(x_{i1})+f^{10}_{j2}(x_{i2})$, for $j=d+1, \ldots, q$, where the $f^{k}$'s are smooth functions built using thin-plate spline bases of dimension $k$, under both the MCD and logM parametrisation. The simulation study is replicated ten times, with $n=10^4$ and over the grid $d\in\{5, 10,15,20\}$.

%\begin{align}
%\eta_{ij} &= f^{10}_{j1}(x_{i1})\,\,, \quad j=1, \ldots, d, %\nonumber\\
%\eta_{ij} &= f^{10}_{j1}(x_{i1})+f^{10}_{j2}(x_{i2})\,\,, \quad j=d+1, \ldots, q, \nonumber
%\end{align}
%where the superscript denote the number of  basis function used. Hence, $\dim(\bsBeta)= 10 \times d + 19 \times d(d+1)/2$. 

%Hence, $\dim(\bsBeta)= 29 \times d + 19 \times d(d+1)/2$. 

\begin{figure}%[htbp]
\centering
\includegraphics[scale=0.54]{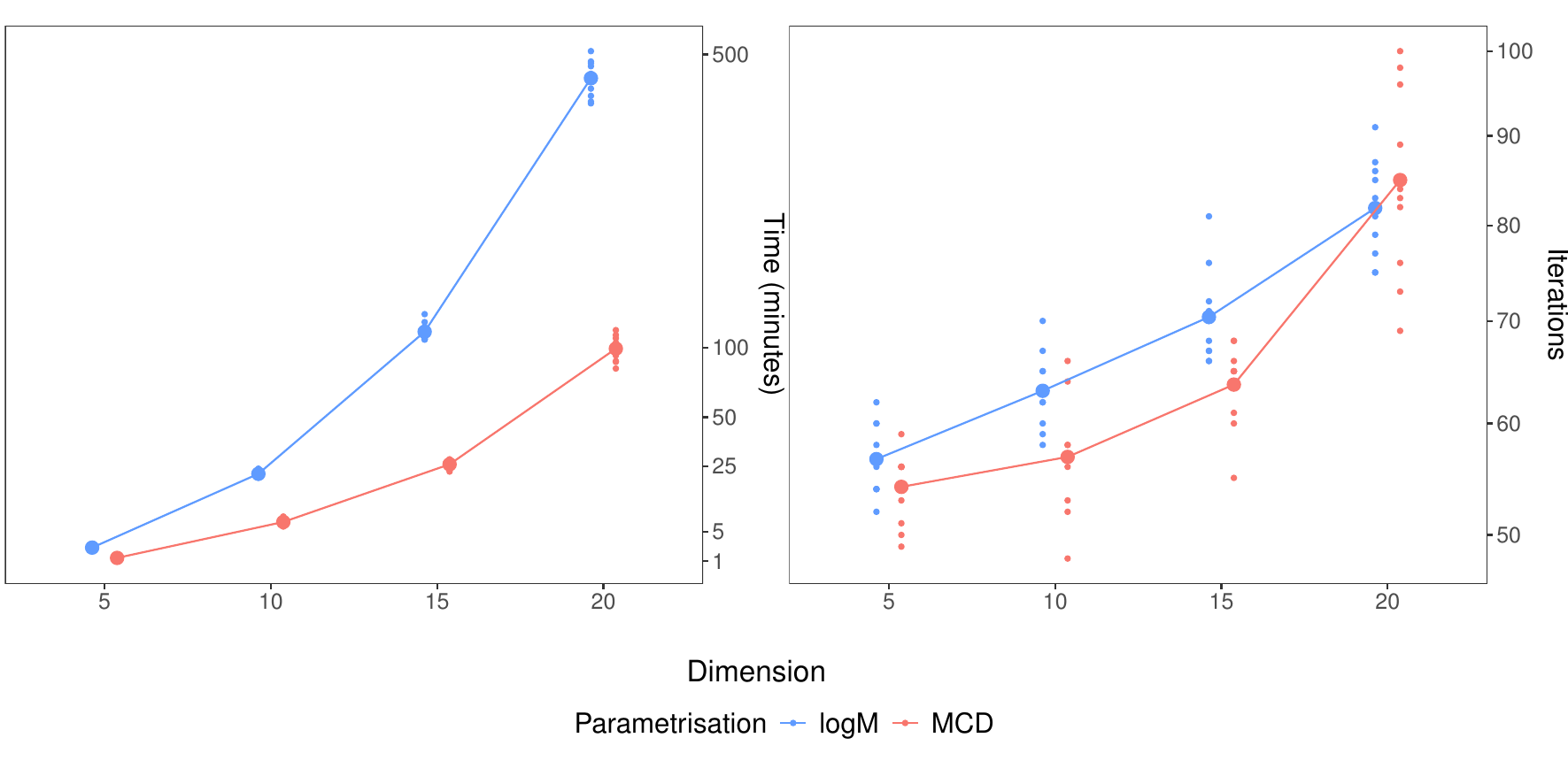}
\caption{Model fitting results when the true $\boldsymbol \Sigma$ is generated via the MCD parametrisation. Left: fitting times (in minutes); Right: number of inner Newton-Raphson iterations. The connected bigger dots are the mean values (times and iterations), the smaller dots are the ten observed values. 
%Blue refers to the logM-based model and red to the MCD one. 
The $y$-axes are on a square root scale.} 
\label{fig:performance_mcd_logm_genMCD} 
\end{figure}

Figure \ref{fig:performance_mcd_logm_genMCD} shows the model fitting times and the total number of Newton's iterations used to obtain $\hat{\boldsymbol \beta}$ (a MAP estimate of $\boldsymbol \beta$ must be computed at each FS optimisation step), for both the logM- and MCD-based models, when the true covariance matrix $\bsS_i$ is related to $\boldsymbol \Theta_i$ via the MCD model. 
The results confirm that the MCD is more scalable than the logM parametrisation as $d$ increases. The plot on the right shows that the number of iterations is similar between the two parametrisations, hence the lower fitting time of the MCD-based model is not due to faster convergence. Instead, the MCD model is faster to fit because computing the Hessian w.r.t. $\bsBeta$, which is the dominant computation cost of model fitting via the FS iteration, is more scalable under this model. This is due to the second-order derivatives w.r.t. $\bsEta$ being increasingly sparse as $d$ increases which %, following formula (\ref{eq:gr_hess_block2cap1}), 
leads to a sparse Hessian w.r.t. $\bsBeta$. Similar fitting times are obtained when the true covariance $\bsS_i$ of the simulated responses is related to $\boldsymbol \Theta_i$ via the logM parametrisation; see SM \ref{Sec:sim1} for details.

\section{Limiting Memory Requirements} \label{sec:blocked_D_beta}%\label{sec:BMCS}

The previous section focussed on the computation of the log-likelihood derivatives w.r.t. $\bsEta$, under either the MCD or the logM parametrisation. This section is concerned with the memory footprint required during model fitting and proposes strategies to mitigate its size.

\subsection{Block-based Derivatives Computation} \label{sec:blocked_computation}

Fitting multivariate Gaussian additive models using the methods described in Section \ref{sec:GAM_gen} can require a considerable amount of memory. In particular, recall that obtaining MAP estimates, $\hat \bsBeta$, via Newton's methods requires the gradient and Hessian of the log-posterior w.r.t. $\bsBeta$ and assume, for simplicity, that all the model matrices $\mathbf {X}^j$, $j = 1, \dots, q$, have $p$ columns.
Then, computing the gradient and the Hessian via (\ref{eq:gr_hess_block2cap1}) requires $\mathcal{O}(npq)$ storage for the design matrices. However, the memory footprint can be reduced by an $\mathcal{O}\big(q)$ factor if $\mathbf {X}^j$ and $\mathbf {X}^k$ are discarded immediately after being used to compute $\bar\ell^{\bsBeta_j}$ and $\bar\ell^{\bsBeta_j \bsBeta_k}$. Doing so for each $j$ and $k$ has an $\mathcal{O}(nq^2p)$ cost, but the matrix-matrix products in (\ref{eq:gr_hess_block2cap1}) cost $\mathcal{O}\big(nq^2p^2\big)$ operations, hence the leading cost of computation is not altered. Under the MCD parametrisation, the cost of computing all the $\bar\ell^{\bsBeta_j \bsBeta_k}$ blocks is reduced by an $O(d)$ factor, due to the sparsity of $\ell^{\bsEta_j \bsEta_k}$, but this does not affect the memory required to store the model matrices, hence discarding and re-evaluating them can still be advantageous.

In contexts where $n$ is large and $p$ is moderate, storing even a single model matrix $\mathbf {X}^j$ can still require a considerable amount of memory. However, in such scenarios the memory footprint can be limited by adopting block-oriented methods for computing (\ref{eq:gr_hess_block2cap1}). In particular, assume that the $n$ observations have been divided into $B$ blocks and that, for simplicity, $n$ is a multiple of $B$. Indicate with $n_B = n/B$ the number of observations in each block, and with $\mathbf {X}_b^j$ the $b$-th row-wise block of $\mathbf {X}^j$. Analogously, let $\ell^{\bsEta_{j}}_b$ and $\ell^{\bsEta_j \bsEta_k}_b$, for $b=1, \ldots, B$, be sub-vectors of $\ell^{\bsEta_{j}}$ and $\ell^{\bsEta_j \bsEta_k}$, respectively.
Then, (\ref{eq:gr_hess_block2cap1}) %the first two derivatives of the log-likelihood w.r.t. $\bsBeta_j$ 
can be written as
\begin{equation} \label{eq:gr_hess_block}
\bar\ell^{\bsBeta_j} = \sum_{b=1}^{B} {\mathbf {X}_b^j}^\top \ell^{\bsEta_{j}}_b\, , \quad \text{and} \quad \bar\ell^{\bsBeta_j \bsBeta_k} = \sum_{b=1}^{B} {\mathbf {X}^k_b}^\top \text{diag}\big(\ell^{\bsEta_j \bsEta_k}_b\big) {\mathbf {X}^j_b}\,\,.
\end{equation}
If $\mathbf {X}^j_b$ and $\mathbf {X}^k_b$ are discarded after being used to compute a term in (\ref{eq:gr_hess_block}), then memory requirements are reduced by an $\mathcal{O}(B)$ factor, relative to stardard evaluation via (\ref{eq:gr_hess_block2cap1}).

Block-oriented GAM fitting methods that do not require storing the full model matrix have been considered by \cite{wood2015generalized}, while \citet{wood2017generalized} focus on speeding up GAM computation via a covariate discretisation scheme. However, such methods apply to GAMs with only one linear predictor (i.e., $q = 1$). Here $q$ is $\mathcal{O}\big(d^2\big)$, which leads to additional computational challenges. In particular, storing $\ell^{\bsEta_{j}}$ and $\ell^{\bsEta_j \bsEta_k}$ for $j = 1, \dots, q$, and $k = j, \dots, q$, requires, respectively, $\mathcal{O}\big(nq\big)$  and $\mathcal{O}\big(nq^2\big)$ memory, which can be considerable even for moderate $d$. For example, \cite{gioia2022additive} consider a regional electricity net-demand data set where $n \approx 10^5$ and $d = 14$, which they model using a multivariate Gaussian additive model comprising $q = 119$ linear predictors. Under the logM parametrisation, storing $\ell^{\bsEta_j \bsEta_k}$ requires nearly 6~GBytes of memory, whereas under the MCD this is reduced to approximately 1~GB due to sparsity. By computing $\ell^{\bsEta_{j}}$ and $\ell^{\bsEta_j \bsEta_k}$ and discarding them as soon as the $b$-th accumulation term in (\ref{eq:gr_hess_block}) has been computed, the memory requirements are reduced to $\mathcal{O}\big(n_Bq^2\big)$, without incurring any additional computational cost.

%One might argue that each $\ell^{\bsEta_j \bsEta_k}$ vector is used solely to compute the corresponding $\bar\ell^{\bsBeta_j \bsBeta_k}$ block and that, having computed the latter, it could be discarded immediately, without entailing any additional computational cost. That is, there seems to be no need to store $\ell^{\bsEta_j \bsEta_k}$ for more than a single pair of indices $j$ and $k$ at the same time, hence the memory cost of such derivatives should be just $\mathcal{O}\big(n\big)$. However, this ignores the fact that the computation of the log-likelihood derivatives w.r.t. $\bsEta$ is better performed jointly for all required values of $j$ and $k$. In particular, under both the MCD and the logM parametrisation, the computation of $\ell^{\bsEta_j \bsEta_k}$ often involves the same terms across different values of the indices $j$ and $k$, and such terms would need to be computed repeatedly if the derivatives w.r.t. $\bsEta_j$ and $\bsEta_k$ where evaluated separately (i.e., within different function calls), depending on the value of $j$ and $k$. Hence, it is more efficient to break the computation by creating the sub-vectors $\ell^{\bsEta_j \bsEta_k}_b$, for $b \in \{1, \dots, B\}$, which will be computed and stored jointly, for $j = 1, \dots, q$ and $k = j, \dots, q$. Such sub-vectors can be discarded as soon as the $b$-th accumulation term in (\ref{eq:gr_hess_block}) has been computed, which limits the memory requirements to $\mathcal{O}\big(n_Bq^2\big)$ and does not entail any extra computational cost.

\subsection{Exploiting Parsimonious Model Structures} \label{sec:parsimonious}

In principle, the block-oriented strategy just described solves any memory requirement issue when computing $\bar\ell^{\bsBeta_j \bsBeta_k}$ because, given $d$, we can choose $B$ to obtain blocks of the desired size. But, as $d$ increases, limiting the memory needed to store the vectors $\ell^{\bsEta_j \bsEta_k}$, requires 
$B$ to increase at an $\mathcal{O}\big(d^4\big)$ rate. Hence, the number of rows, $n_B$, of ${\mathbf {X}_b^j}$ decreases rapidly with $d$, which is detrimental from a computational point of view. In particular, the matrix multiplications required to compute $\bar \ell^{\bsBeta_j \bsBeta_k}$ are level-3 (i.e. matrix-matrix) BLAS routines, %\citep{BLASsite}
optimized by numerical linear algebra libraries such as ATLAS or OpenBLAS. Such libraries provide implementations of level-3 routines designed to speed up computation by making optimal use of cache memory and other processor-dependent features, provided that they are multiplying matrices of sufficiently large size. Hence, reducing the size of the matrix blocks that are being passed to such routines, can lead to slower computation of $\bar \ell^{\bsBeta_j \bsBeta_k}$. 

%The impact of using small blocks on the overall model fitting time is significant as computing $\bar \ell^{\bsBeta_j \bsBeta_k}$, for each $j$ and $k$, requires  $\mathcal{O}\big(nd^4p^2\big)$ operations, under the logM parametrisation and assuming that each linear predictor is controlled via $p$ parameter. Hence, when $p^2 > d$, the Hessian w.r.t. $\bsBeta$ determines the leading cost of computation, as computing $\ell^{\bsEta_j \bsEta_k}$ involves $\mathcal{O}\big(nd^5\big)$ operations. Under the MCD model, both derivatives cost $\mathcal{O}\big(d\big)$ less, due to sparsity, but their relative cost is unchanged. However, the number of blocks required to keep memory usage constant as $d$ increases is $\mathcal{O}\big(d^3\big)$, rather than $\mathcal{O}\big(d^4\big)$, hence the computational slowdown due to small blocks is less severe than under the logM model.

While, under a simple setting where ${\mathbf {X}^j}$ is an $n \times p$ matrix for every $j$, keeping a small memory footprint as $d$ increases requires smaller and smaller blocks, there are practically relevant modelling settings that lead to more efficient computation. Here we focus on \emph{parsimoniuous} scenarios, where a significant proportion of the linear predictors do not depend on the covariates, but are fixed to an intercept. Examples are provided by \cite{gioia2022additive}, who consider models where over 60\% of the $q = 105$ linear predictors controlling an MCD-based covariance matrix parametrisation are modelled only via intercepts. More generally, in cases where the elements of the response variable are related in space and/or time, as in the example considered in Section \ref{sec:App}, one might expect parsimonious model structures to emerge as $d$ increases. For instance, as explained in Section \ref{sec:MCD}, under the MCD model the elements of the $\mathbf{T}$ factor can interpreted as regression coefficients between elements of $\mathbf y$ and, if the distance between such elements increases with $d$, it seems sensible to keep constant the elements of $\mathbf{T}$ that quantify the dependence between distant responses. 

When dealing with parsimonious model structures, it is possible to rearrange the computation of $\bar\ell^{\bsBeta \bsBeta}$ to use larger matrix blocks, which leads to a more efficient use of level-3 BLAS routines. Recall that the derivatives $\ell^{\bsEta_j \bsEta_k}_b$ should be computed and stored jointly, for all values of $j$ and $k$, so that they can be used to compute the $b$-th term in the accumulation appearing in (\ref{eq:gr_hess_block}). However, when both $\bsEta_j$ and $\bsEta_k$ are modelled only via intercepts, the corresponding design matrices are simply vectors of ones and $\bar\ell^{\bsBeta_j \bsBeta_k}$ is a scalar equal to the sum of the elements of $\ell^{\bsEta_j \bsEta_k}_b$. Hence, the elements of the latter can be accumulated as soon as they are computed,  %, rather than computed and stored as is the case for the blocks where $\bsEta_j$ and/or $\bsEta_k$ depend on the covariates. 
as detailed in Algorithm \ref{algo:parsim_deriv}.% details the steps needed to compute $\bar\ell^{\bsBeta \bsBeta}$ as just described. %An analogous approach can be used to compute the $\bar\ell^{\bsBeta}$.

\begin{figure}%[htbp]
\centering
\includegraphics[scale = 0.52]{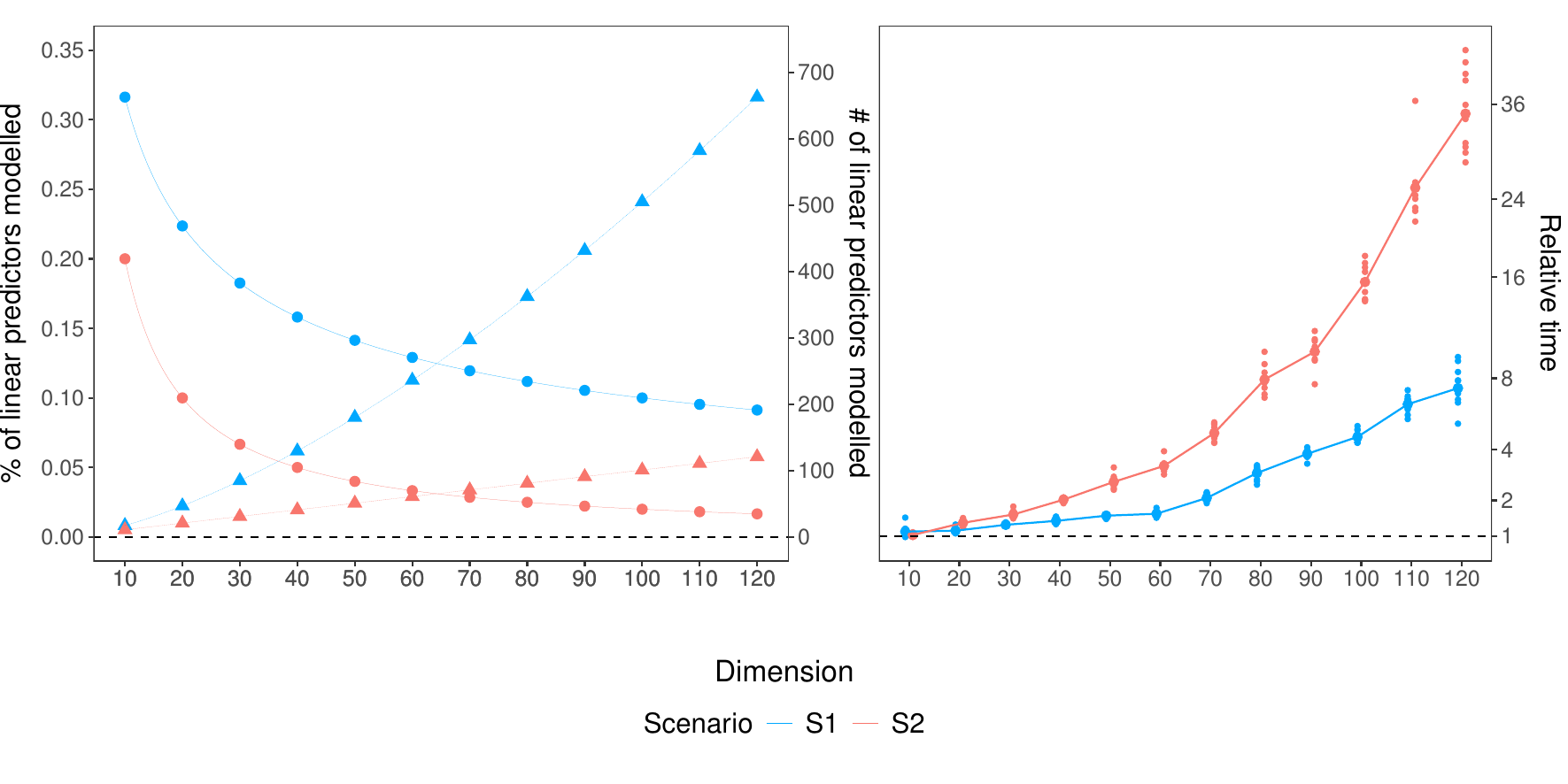}
\caption{Parsimonious modelling results for the logM model. Left: Total (triangles) and relative (circles) number of linear predictors modelled under the scenarios defined in (\ref{eq:pars}).  Right: Ratio of the time needed to compute $\bar\ell^{\bsBeta \bsBeta}$ under the \texttt{Standard} and the \texttt{Parsimonious} approach.  The connected dots are the ratios of the mean times, while the small dots are the ratios of ten observed times. The y-axis on the right plot is on a square root scale.} 
\label{fig:hessian_beta}
\end{figure}

To quantify the effect of parsimony on the size of the matrix blocks, assume that a fraction $\alpha \in [0, 1]$ of the $q$ linear predictors is controlled only via an intercept. Then, using Algorithm \ref{algo:parsim_deriv}, the storage required by  $\ell^{\bsEta \bsEta}_b$ is $\mathcal{O}\big(n_b(1-\alpha)^2q^2\big)$. Hence, taking into account the structure of parsimonious models permits expanding $n_b$ by an $(1-\alpha)^{-2}$ factor, while fulfilling the same memory requirements. To assess how larger blocks translate into more efficiency during level-3 BLAS computations, we consider a numerical study where we simulate $n=10^3$ multivariate Gaussian responses in $d = 5, 10, \dots, 100$ dimensions, from a model where the mean vector elements are $\eta_{ij} =  \sum_{k=1}^{10} {\rm X}_{ik} \beta_{jk}$, $j = 1, \ldots, d$,
%constant, that is $\eta_{ij} =  \beta_{1j}$ for $j=1, \ldots, d$, 
while the linear predictors controlling the covariance matrix are
\begin{equation} \label{eq:pars}
\eta_{ij} =  \beta_{1j} + (1-c_j) \sum_{k=2}^{10} {\rm X}_{ik} \beta_{jk}\, , \quad  c_j \sim {\rm Bernoulli}(\alpha_{s}), \quad j = d+ 1, \ldots, q,  
\end{equation}
%\begin{equation*}
%\eta_{ij} = \begin{cases} \beta_{1j} \quad \quad \quad \quad \quad \text{if} \quad c_j=1\,, \\
% \sum_{k=1}^{10} {\rm X}_{ik} \beta_{jk}\quad \text{if} \quad c_j=0\,,
%\end{cases} \quad \quad j = d+ 1, \ldots, q, 
%\end{equation*} 
where $s=1,2$, indicates one of two scenarios. In particular $\alpha_1 = 1 - d^{-1/2}$ and $\alpha_2 = 1 - 2d^{-1}$ which, considering that $q = \mathcal{O}\big(d^2\big)$, implies that the number of modelled (i.e., non-constant) linear predictors grows, respectively, at a $\mathcal{O}\big(d^{3/2}\big)$ and  $\mathcal{O}\big(d\big)$ rate. These parsimonious regimes are shown on the left of Figure \ref{fig:hessian_beta}. Note that the second scenario is highly parsimonious but still realistic as it includes, for example, MCD-based banded models where only the elements of $\mathbf D^2$ and of the first few sub-diagonals of $\mathbf T$ depend on covariates. Under both scenarios, we evaluate $\bar \ell^{\bsBeta \bsBeta}$ using either the standard block-oriented computation as in (\ref{eq:gr_hess_block}) (\texttt{Standard}) or Algorithm \ref{algo:parsim_deriv} (\texttt{Parsimonious}). In both cases, the number of blocks, $B$, is such that storing $\bar \ell^{\bsBeta \bsBeta}$ requires at most 1 GByte. 

%although such a strategy becomes effective only with  $d>25$ under the logM parametrisation.

%does not activate in low and moderate dimensional settings, while it is adopted considering $d>25$ for the logM parametrisation. %Indeed, with $d=50$ the storage of $\ell^{\bsEta_j \bsEta_k}$ requires 7  GBytes of RAM. %Further, note that the computational time gains of adopting the \texttt{Parsimonious} approach becomes noteworthy for moderate and large dimensions of the outcome vector taking into account the whole model fitting procedure and with sample sizes more representative than $n=1000$. 
%\cite{}

The plot on the right of Figure \ref{fig:hessian_beta} shows the time required to evaluate $\bar\ell^{\bsBeta \bsBeta}$ using the \texttt{Standard} approach relative to the \texttt{Parsimonious} method, based on ten runs under the logM parametrisation. Clearly, the performance gains of the \texttt{Parsimonious} approach are proportional to $d$, particularly in the second scenario. Under the MCD model, the gains are marginal due to the sparsity of the derivatives w.r.t. $\boldsymbol \eta$. See SM \ref{Sec:sim2} for further details.

\begin{algorithm}[t]%[htbp] 
\caption{Computing $\bar\ell^{\bsBeta \bsBeta}$ in a parsimonious modelling setting}
\label{algo:parsim_deriv}
\label{alCross}
%\setstretch{1.15}
Let $\mathcal{S}_j \subset \{j, \dots, q\}$, for $j \in \{1, \dots, q\}$, be sets of indices such that $\ell^{\bsEta_j \bsEta_k} \neq \mathbf 0$ if $k \in \mathcal{S}_j$, and $\ell^{\bsEta_j \bsEta_k} = \mathbf 0$ otherwise. Let $\mathcal{F} \subset \{1, \dots, q\}$ be the indices of the fixed linear predictors, that is $\mathbf {X}^j = {\boldsymbol 1}$ for $j \in \mathcal{F}$, where ${\boldsymbol 1}$ is a vector of ones. Define $\mathcal{D}_j = \{k : k \in \mathcal{S}_j \cap \mathcal{F}\}$ if $j \in \mathcal{F}$ and $\mathcal{D}_j = \emptyset$ otherwise. Let $\mathcal{I}_b \subset \{1, \dots, n\}$ be the indices of the $n_b$ observations belonging to the $b$-th block. Let $\ell^{\bsEta_j \bsEta_k}$ be either scalars, initialised a zero if $\{j,k\} \subset \mathcal{F}$, or vectors with $n_b$ elements, otherwise. 
%Initialise at zero the elements of the vector $\ell^{\bsEta_j \bsEta_k}_\mathcal{D}$ of length $\sum_{j = 1}^q \text{Card}(\mathcal{D}_j)$ and of the matrix $\ell^{\bsEta_j \bsEta_k}_{\mathcal{D}^c}$, with $n_b$ rows and $\sum_{j = 1}^q \text{Card}(\mathcal{D}_j^c)$ columns. 
Finally, let $\bar\ell^{\bsBeta_j \bsBeta_k}$ be matrices of appropriate size, with elements initialised at zero.  
%\begin{algorithmic}[1]
For $b = 1, \dots, B$:
\begin{enumerate}
\item For $i \in \mathcal{I}_b$, $j=1, \ldots, q$ and $k \in \mathcal{S}_j$, compute $\ell^{\eta_j \eta_k}_i$ and then either store it or accumulate it, that is
\begin{align*}
\big(\ell^{\bsEta_j \bsEta_k}_{\mathcal{D}^c}\big)_i &\leftarrow \ell^{\eta_j \eta_k}_i  \;\;\; \quad \quad \quad \quad \, \text{if} \;\;\; k \in \mathcal{D}^c_j,\\
\ell^{\eta_j \eta_k}_\mathcal{D} &\leftarrow \ell^{\eta_j \eta_k}_\mathcal{D} + \ell^{\eta_j \eta_k}_i  \quad \;\;\; \text{if} \;\;\; k \in \mathcal{D}_j. 
\end{align*} 
\item For $j=1, \ldots, q$ and $k \in \mathcal{S}_j$, compute $\bar\ell^{\bsBeta_j \bsBeta_k}$ via
\begin{align*}
\bar\ell^{\bsBeta_j \bsBeta_k} &\leftarrow \bar\ell^{\bsBeta_j \bsBeta_k} + {\mathbf {X}^k_b}^\top \text{diag}\big(\ell^{\bsEta_j\bsEta_k}_{\mathcal{D}^c}\big) {\mathbf {X}^j_b}  \quad  \;\;\; \text{if} \;\;\; k \in \mathcal{D}^c_j, \\
\bar\ell^{\bsBeta_j \bsBeta_k} & \leftarrow \bar\ell^{\bsBeta_j \bsBeta_k} + \ell^{\eta_j \eta_k}_\mathcal{D}  \quad \quad \quad \quad \quad \quad \quad \;\;\;\,
 \text{if} \;\;\; k \in \mathcal{D}_j.
\end{align*}
\end{enumerate}
%Note that steps 1 and 2 above are explicitly kept separate to mimic a software implementation where the derivatives w.r.t. $\bsEta$, which are parametrisation-specific, are evaluated in a first step, and are then used to compute those w.r.t. $\bsBeta$.
%\end{algorithmic}
\end{algorithm}

\section{Efficient Computation of the Fellner-Schall Update} \label{sec:effic_fellner}

Recall that the smoothing parameter vector, $\bsLambda$, is selected by maximising the LAML (\ref{eq:LAML}), via either the approximate FS update or its exact version (EFS). Here we provide efficient methods for computing the latter, which allow us to check whether the FS update leads to a poorer model fit and to compare the computational cost of the two methods.

As explained in Section~\ref{sec:GAM_gen}, the FS update assumes that
$c= \tr \Big(\boldsymbol{ \mathcal{H}}^{-1}{\bar\ell}^{\hat \bsBeta \hat \bsBeta\lambda_u}  \Big)$
in (\ref{eq:EFS_exact}) is zero. This simplifies computation, because $c$ is obtained by evaluating the third-order log-likelihood derivatives w.r.t. $\boldsymbol \eta$, rendering the EFS update unattractive under the logM parametrisation. However, under the MCD-based model, the third- and second-order derivatives incur the same $\mathcal{O}(d^4)$ cost due to sparsity, which, in combination with the methods described below, makes the adoption of the EFS update feasible.

In particular, the block of ${\bar\ell}^{\hat \bsBeta \hat \bsBeta\lambda_u}$ corresponding to the $j$-th and $k$-th linear predictor is
\begin{equation} \label{eq:dHess2_dlambda}
\bar\ell^{\hat \bsBeta_j \hat \bsBeta_k\lambda_u} = {\mathbf {X}^k}^\top \mathbf W^{jk}  {\mathbf {X}^j}\,\,,
\end{equation}
where $\mathbf W^{jk} = \text{diag}\big( %\lambda_u 
 \sum_{l}\ell^{\bsEta_j\bsEta_k\eta_l}\eta_l^{\lambda_u}\big)$ and $\eta_l^{\lambda_u} = {\rm d} \eta_l / {\rm d} \lambda_u$ can computed by implicit differentiation \citep[see][for details]{wood2016}. %As explained in \citet{wood2016}, the quantity \label{eq:dHess_dlambda} is also needed for optimising the LAML via quasi-Newton method, also accounting for the reparametrisation $\boldsymbol \rho = \log (\boldsymbol \lambda)$.   
Using (\ref{eq:dHess2_dlambda}), we can now write
\begin{equation}\label{eq:c_naive} 
c =  \sum_{k = 1}^q \sum_{j = 1}^q \tr \big(\tilde{\boldsymbol{\mathcal{H}}}_{jk} {\mathbf {X}^k}^\top \mathbf W^{jk}  {\mathbf {X}^j}\big)\,\,,
\end{equation}
where $\tilde{\boldsymbol{\mathcal{H}}}_{jk} = \boldsymbol{\mathcal{H}}^{-1}_{jk}$, the latter being the $jk$-th block of $\boldsymbol{\mathcal{H}}^{-1}$.  However, a more computationally efficient expression for computing $c$ is
\begin{equation}\label{eq:c_efficient} 
c =   \sum_{k = 1}^q \sum_{j = 1}^q\mathbf a_{jk}^\top  \diagv\big(\mathbf W^{jk}\big)\,\,,
\end{equation}
where $\mathbf a_{jk} = \big({\mathbf {X}^j}\tilde{\boldsymbol{\mathcal{H}}}_{jk}\odot {\mathbf {X}^k}\big) {\mathbf 1}_p$, with ${\mathbf 1}_p$ a p-dimensional vector of ones, and  $\diagv(\cdot)$ is the matrix-to-vector diagonal operator.

\begin{remark} 
Assume, for simplicity, that each model matrix has $p$ columns. Then, computing $\boldsymbol{\mathcal{H}}$ involves an $\mathcal{O}\big(np^2d^4\big)$, which is the leading cost under the FS update. Under an MCD-based model, the EFS update involves a further $\mathcal{O}\big(Unp^2d^4\big)$ cost, if $c$ is evaluated via (\ref{eq:c_naive}). However, the cost of computing $c$ can be reduced to $\mathcal{O}\big(Unpd^4\big)$ by adopting (\ref{eq:c_efficient}). This implies that the additional cost of EFS, relative to FS, is moderate unless $U \gg 
p$.
\end{remark}

To compare the FS and EFS methods, we consider the simulation setting of Section \ref{NumExp_eta}, with $n=10^4$, $d \in \{2,5,10,15,20\}$, and ten runs for each value of $d$. We include in the comparison the back-fitting (BF) methods provided by the \texttt{bamlss R} package \citep{umlauf2021jss}.  %an initialised version of the BFGS approach, where the starting values of the algorithm are given by $\hat {\boldsymbol \beta}$ and $\hat { \boldsymbol \lambda}$ of the \texttt{FS}-based fit (here denoted as \texttt{BFGS - FS}).
%We also compare in terms of computational times of the proposed model fitting approaches 
%a comparison with the frequentist version of the MCD-based model exploited by \citet{muschinski2024}  (here denoted as BAMLSS), that leverage a different optimisation algorithm based on backfitting \citep[see][]{umlauf2018bamlss}. The latter can be used as a useful benchmark from literature, being similar in spirit to the proposed modelling approach and only differing for the model fitting procedure.  
The times reported in Table \ref{tab:comparison} shows that EFS is slightly slower than FS, due to the cost of computing the $c$ term in (\ref{eq:c_efficient}) and of the third-order log-likelihood derivatives w.r.t. $\boldsymbol \eta$ (see SM \ref{app:Dllk_Deta_mcd}). Note that BF leads to model fits that are very similar to those obtained via either FS or EFS (the relative differences in log-likelihood at convergence, not shown here, are less than $10^{-4}$), which highlights the scalability of the proposed methods. %It is worth noting that we consider only the grid $d \in {2,5,10}$ due to the poor computational performance of the BF method. 

\begin{table}
%The lowest 
\centering
\begin{tabular*}{\textwidth}{@{\extracolsep{\fill}}llcccccc}
\toprule
\textbf{Value} & \textbf{Method} & \textbf{2} & \textbf{5} & \textbf{10} & \textbf{15} & \textbf{20} \\ 
\midrule
\multirow{1}{*}{Time (min)} & FS & 0.2 (0.2) & 1.2 (1.3) & 6.8 (7.3) & 27.0 (32.9) & 109.1 (155.1) \\
\toprule
\multirow{2}{*}{Rel. time} & EFS & 1.1 (1.2) & 1.1 (1.2) & 1.2 (1.3) & 1.5 (1.8) & 1.5 (1.8) \\
& BF & 5.0 (6.1) & 17.1 (28.4) & 58.5 (91.0) & - (-) & - (-) \\
\bottomrule
\end{tabular*}
\label{tab:comparison}
\caption{The first row reports the average (max) times in minutes for fitting the models ten times, with $n = 10000$ and $d\in \{2,5,10,15,20\}$, using FS, while the last two rows report the average (max) times relative to FS, for fitting the models using EFS and BF. } 
\end{table}

Finally, recall that both the EFS update and its approximate FS version aim to select the smoothing parameters by minimising the LAML. Interestingly, in the simulation study conducted here, the two updates converge to the same LAML score, within a relative difference of less than $10^{-6}$. Hence the approximation error induced by the simpler FS update is minimal, which motivates its adoption in the following section.

% and that by , by examining the relative differences in terms of LAML, we can conclude that there is no discernible difference between FS and EFS (relative differences $<|10^{-6}|$). Similarly, in terms of out-of-sample log-score (the negative log-likelihood) between FS and BF (differences $<|6\times 10^{-4}|$ relative to the generation log-score), we can confirm that the FS update represents the preferred choice for $\boldsymbol \lambda$ selection.

\section{Application to Electricity Load Modelling}\label{sec:App}

Here we compare the performance of the logM- and MCD-based models in a realistic setting.
We consider data from the electricity load forecasting track of the GEFCom2014 challenge \citep{hong2016}. The response vector consists of hourly electricity loads ($\text{load}_{ih}$), where $h = 0, \ldots, 23$, indicates the hour and $i = 1, \ldots, n$, is the index of the day, spanning the period from the 2nd of January 2005 to the 30th of November 2011. Hence, we have $d=24$ and $n=2520$. The covariates are progressive time ($t_i$), day of the year ($\text{doy}_i$), day of the week ($\text{dow}_i$), hourly loads of the previous day ($\text{load}^{24}_{ih}$), hourly temperatures ($\text{temp}_{ih})$ and exponential smoothed temperatures ($\text{temp}^S_{ih} = \alpha\text{temp}^S_{ih-1} + (1-\alpha)\text{temp}_{ih}$, with $\alpha = 0.95$).

We consider the model $\text{\textbf{load}}_{i} \sim \mathcal{N}(\bsMu_i, \bsS_i)$, where $\bsS_i$ is parametrised via either the logM or the MCD. The joint mean and covariance matrix model has $q=324$ linear predictors, which entails a challenging model selection problem. For the mean vector, we use the model
\begin{equation*}
\label{eq:meanSpec}
\eta_{ij} =  g_{1j}(\text{dow}_i) + g_{2j}(\text{load}^{24}_{ij-1}) + f^{4}_{1j}(\text{t}_{i}) +  f^{20}_{2j}(\text{doy}_{i}) + f^{15}_{3j}(\text{temp}_{ij-1}) + f^{15}_{4j}(\text{temp}^S_{ij-1})   \,, \\
\end{equation*}
for $j = 1, \dots, d$, where $g_{1j}$ and $g_{2j}$ are parametric linear effects, while $f_{1j}$ to $f_{4j}$ are smooth effects, with superscripts denoting the spline bases dimensions. To model the linear predictor controlling $\bsS_i$,  we consider a full model where
\begin{equation*}
\label{eq:CovSpec}
\eta_{ij}=f^{10}_{1j}(\text{doy}_{i}) \,, \,\,\quad \quad\,\,\, j = d+1, \dots, q,
\end{equation*} 
under either the MCD or the logM parametrisation. Having 3000 parameters controlling $\bsS_i$, the full model is over-parametrised, hence we simplify by backward stepwise effect selection. In particular, we fit the full model on data from up to 2010, and then we remove the five effects with the largest p-values. We iterate this procedure until we get to a model where $\eta_{ij}$ for $j = d+1, \dots, q$, are all constant, that is they are modelled only via intercepts.

The number of effects to be retained in the model is determined by optimising the out-of-sample predictive performance on 2011 data. To evaluate such performance, we adopt a 1-month block rolling origin forecasting procedure. In particular, having fitted each model to data up to the 31st of December 2010, we produce predictions of $\bsMu_i$ and $\bsS_i$ for the next month. We then update each model by fitting it to data up to the 31st of January 2011, and obtain a new set of predictions. By repeating the process over the whole of 2011, we obtain a full year of predictions, which we use to compute the log-score (LS), that is the negative log-likelihood of the out-of-sample data. 

%While \cite{gioia2022additive} consider a semi-automatic effect selection procedure, based on gradient boosting, for ranking the importance of several candidate effects involved in covariance matrix modelling, here we exploit a ranking strategy based on p-values for smooth components \citep{wood2013pval}. In particular, we use data from 2005 to 2010 to select the covariance matrix models, by iterating the deletion of the 5 smooth effects having the highest p-values from the \texttt{Full} to the \texttt{Fixed} model. Such an approach approach is similar to the backward elimination detailed in \citet{marra2011practical}. Then, we use the 2011 test data for selecting the total number of covariate-dependent entries of the logM- and MCD-based models. In this respect, we adopt a  1-month block rolling origin forecasting procedure starting from the 1st of January 2011 to predict the value of $\eta_{ij}$, for $j = 1, \ldots, q$, covering the whole 2011 test data. Thus, we use the out-of-sample predictions of $\boldsymbol \mu_i$ and  $\boldsymbol \Sigma_i$ to compute the out-of-sample LS metric, as defined in Section \ref{NumExp_eta}. \\

\begin{figure}
\centering
\includegraphics[scale=0.40]{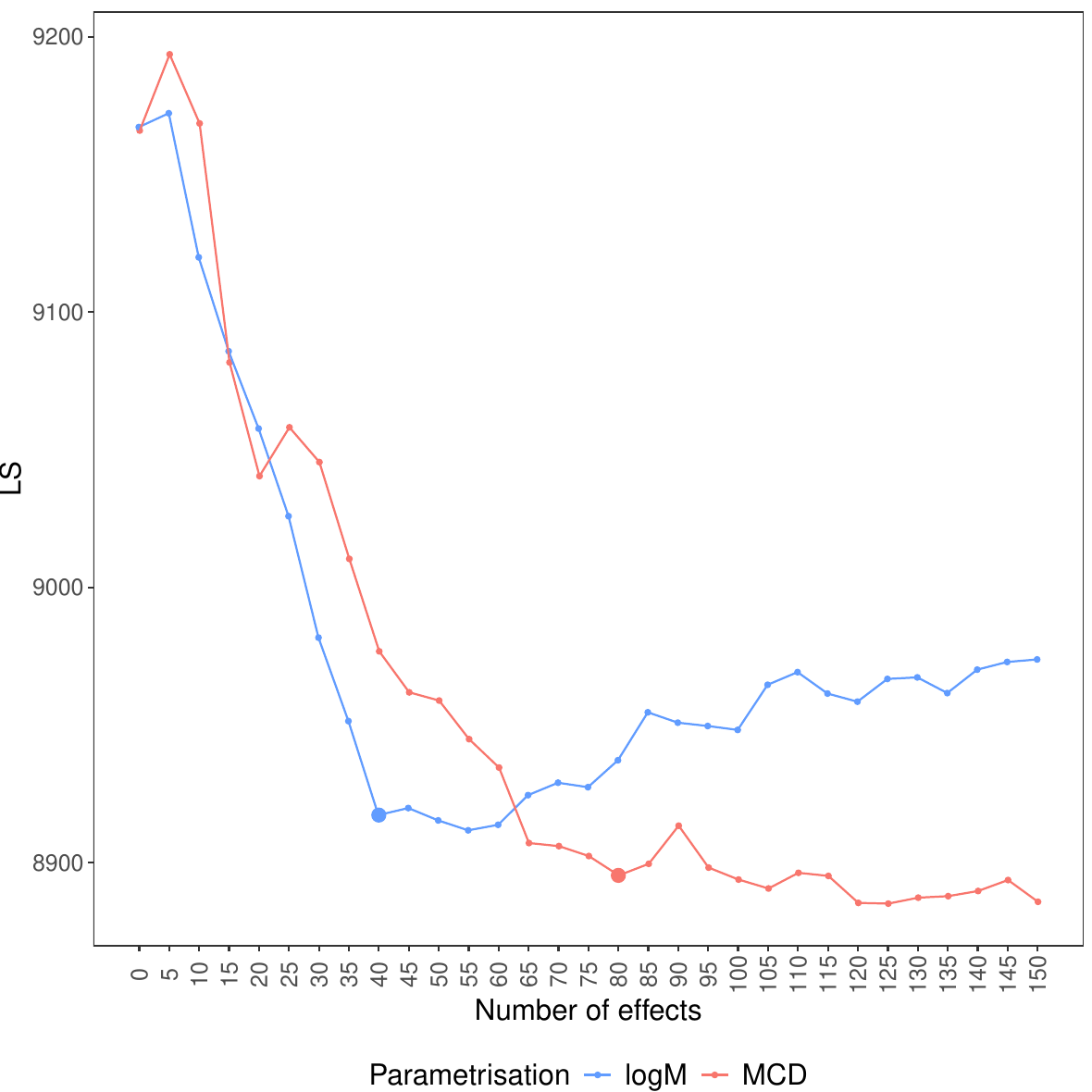}
\includegraphics[scale=0.40]{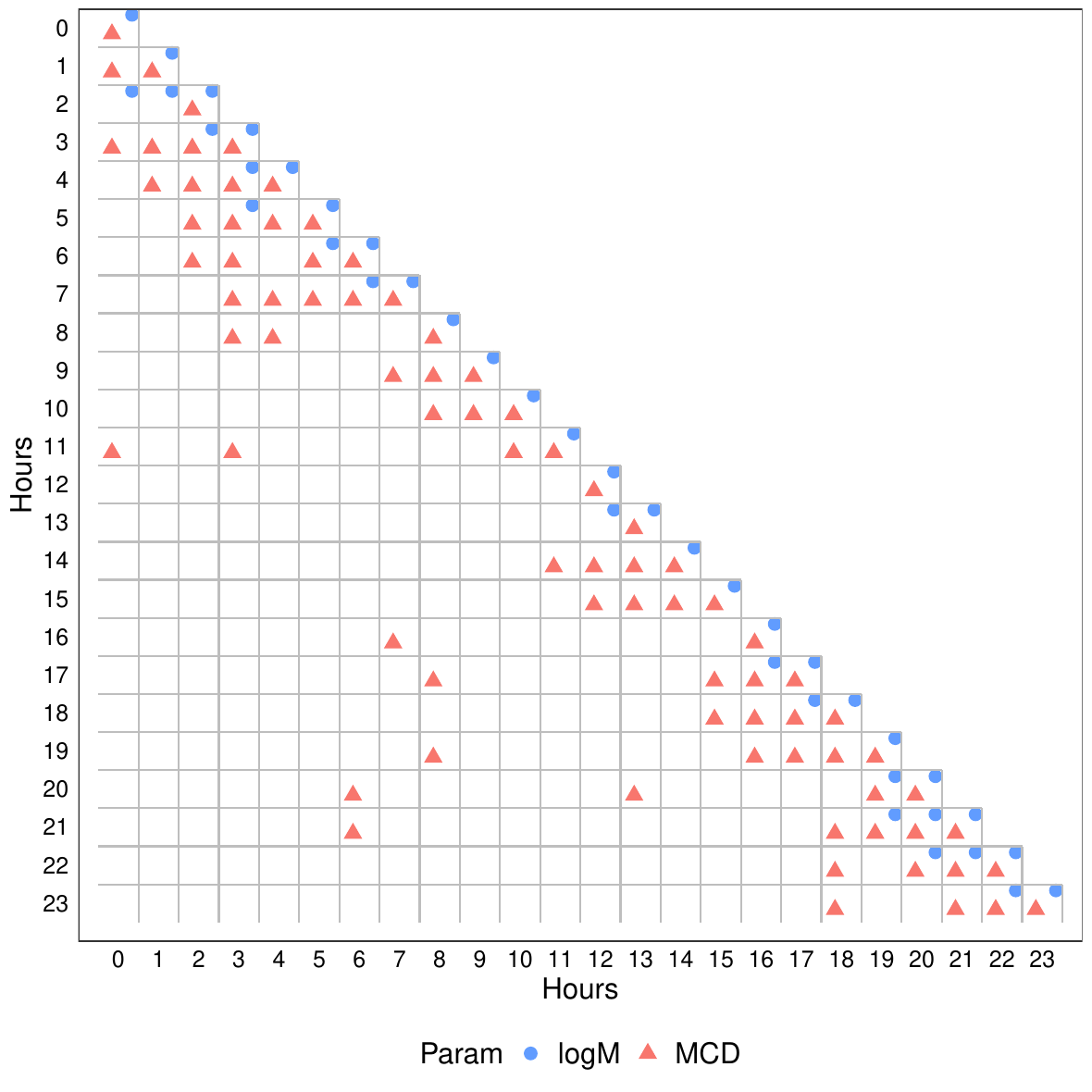}
\caption{Left: LS metric on the 2011 test data, as a function of the number effects used to model $\bsS_i$ via either the MCD or logM parametrisation. Right: elements of the parametrisations being modelled via smooth effects of $\text{doy}_i$ when using 40 (logM) and 80 (MCD) such effects. The empty cells denote linear predictors modelled only via intercepts. } 
\label{fig:perf_mod_selection_0}
\end{figure}

While the fitting times on the training data are reported in SM \ref{app:application}, the plot on the left of Figure \ref{fig:perf_mod_selection_0} shows the LS metric for models ranging from 0 to 150 effects to model $\bsS_i$. Considering the shape of the two curves, we sped up computation by not evaluating the performance of models containing more than 150 effects. Indeed, under the logM model, predictive performance has a well-defined minimum between 40 and 60 effects while, under the MCD model, the gains are diminishing for models containing more than 80 effects. Hence, the logM parametrisation leads to more parsimonious models for this data, but their predictive performance is inferior to that of more complex MCD-based models. 

On the right of Figure \ref{fig:perf_mod_selection_0}, we show which elements of the logM and MCD parametrisation matrices have been modelled via smooth effects of $\text{doy}_i$, when using respectively 40 and 80 such effects. Considering the interpretation of the MCD parametrisation presented in Section \ref{sec:MCD}, it is not surprising to see that all the diagonal elements of $\mathbf D$ and many of the elements on the first few leading sub-diagonal $\mathbf T$ are being modelled. This suggests that the residual correlation between consecutive hours, and the residual variance after having conditioned on previous hours, vary significantly with $\text{doy}_i$. Under the logM model, all diagonal elements of $\boldsymbol \Theta$ are allowed to vary with $\text{doy}_i$, with the remaining effects appearing on the first or second leading subdiagonal. Although the elements of the logM model do not have any direct interpretation in general, our understanding is that, for the data considered here, the logM model allows the variances to vary strongly with $\text{doy}_i$, while the correlation structure is kept relatively constant. See SM \ref{app:application} for more details.    

\begin{figure}
\centering
\includegraphics[scale=0.29]{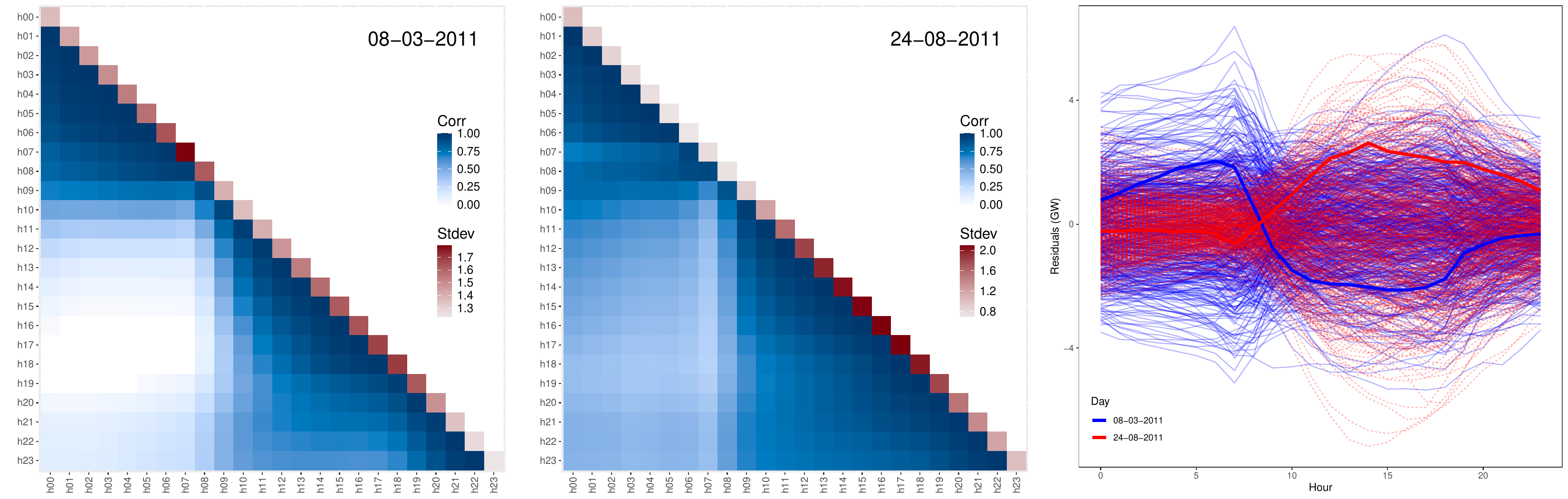}
\caption{Left and centre: Standard deviations (diagonal) and correlations (lower triangle) predicted by an MCD-based model with 80 effects, on two different dates. Right: Observed model residuals (thick lines) and 500 simulated residuals trajectories from the model.} 
\label{fig:nice_plots_2}
\end{figure}

The heatmaps in Figure \ref{fig:nice_plots_2} represent the covariance matrices predicted by the MCD-based model with 80 effects on two different days, demonstrating the importance of a dynamic covariance matrix model in this context. In particular, note that in March the residual variance is largest at 7 am and 6 pm, which correspond to the morning and evening demand ramps (the times at which demand increases at the fastest rate). In August, the variance has a wider peak around 4 pm, indicating that load prediction errors are expected to be largest at this time. Note also that the correlation structure changes with the time of year, with the correlation between morning and evening residuals being much weaker in March than in August. 
The plot on the right of Figure \ref{fig:nice_plots_2} shows the observed residuals on the two days, together with 500 residual trajectories simulated from the model. The shape of the simulated trajectories is coherent with that of the observed curve, meaning that the model could be used to generate realistic demand curves, which could serve as inputs for decision-making under uncertainty in the power industry (e.g., to inform trading strategies or stochastic power-production scheduling).

\section{Conclusion}\label{sec:Conc}

This work provides computationally efficient methods for fitting multivariate Gaussian additive models with covariate-dependent covariance matrices, with a particular focus on the widely used MCD and logM unconstrained parametrisations. Under the model fitting framework proposed here, computational efficiency requires fast evaluation of the log-likelihood derivatives. Hence, we show how this can be achieved by exploiting sparsity in the MCD case or by employing highly optimised derivative expressions under the logM model. Further, we provide computational methods for exploiting parsimonious modelling scenarios and show that doing so can greatly speed up computation, especially under the logM model and in high dimensions. We also enable efficient computation of the exact FS update to maximise the LAML. While the results provided here support the use of the original FS update, which is cheaper to compute, more theoretical work is needed to verify whether the approximation error implied by this update is negligible in general.

Future work could also focus on developing covariance matrix models that better exploit known structures in the data. For example, in the application considered here, the MCD elements can be interpreted as the parameters of an autoregressive model on the residuals. Here, we do not exploit this interpretation and choose the model via an expensive backward effect selection routine to fairly compare the MCD and logM models. But it would be interesting to integrate the computational methods proposed here with the models considered by \cite{huang2007estimation}, who exploit the MCD interpretation to control the whole covariance matrix via a few parameters.

% TODO:
% \begin{enumerate}
%     \item Mention (in the conclusions) parallel strategies over n or over the j-k blocks.
%     \item Mention  possible developments for block covariance or colletion matrix \cite{archakov2024},
%     \item \cite{kim2024} (multivariate longitidinal data - MCD); \cite{xu2012}: bivariate longitudinal setting (MCD vs logM)
%     \item \cite{hirsch2024} online GAMLSS
% %    \item could we mention Simon's recent proposal on the CV for time series; 
% \end{enumerate}

\putbib
\end{bibunit}
\newpage
\appendix
\counterwithin*{equation}{section}
\renewcommand\theequation{\thesection.\arabic{equation}}
\counterwithin*{figure}{section}
\renewcommand\thefigure{\thesection.\arabic{figure}}  
\counterwithin*{table}{section}
\renewcommand\thetable{\thesection.\arabic{table}} 

\begin{bibunit}
\section*{Supplementary Material to ``Scalable Fitting Methods for Multivariate Gaussian Additive Models with \\ Covariate-dependent Covariance Matrices''}

\label{AppA}
\setcounter{page}{1}

\section{Parametrisation Specific Quantities Needed for Model Fitting} \label{AppA1}

\subsection{Notation and Useful Quantities}  \label{AppNot}

Recall the following notation.  The operator $\diag(\cdot)$ is the vector-to-matrix diagonal operator, while $\diagv(\cdot)$ is the matrix-to-vector diagonal operator. The matrix exponential operator is denoted with $\exp(\cdot)$, the Hadamard product with $\odot$,  and $\mathbbm{1}$ is the indicator function. Further, $\mathbf X_{q\cdot}$ and $\mathbf X_{\cdot r}$ represent, respectively, the $q$-th row and the $r$-th column of a matrix $\mathbf X$. Finally, the derivatives w.r.t. $\bsEta$ are denoted with $f^{\bsEta_j}$, $f^{\bsEta_j \bsEta_k}$, and $f^{\bsEta_j\bsEta_k\bsEta_l}$, corresponding to the vectors with $i$-th elements  
%$$
%f^{\eta_{ij}} = \frac{\partial f}{\partial \eta_{ij}}\;,\; \quad \quad f^{\eta_{ij} \eta_{ik}} = \frac{\partial^2 f}{\partial \eta_{ij}\partial\eta_{ik}} \quad  \text{and} \quad
%f^{\eta_{ij} \eta_{ik}\eta_{il}} = \frac{\partial^3 f}{ \partial \eta_{ij} \partial \eta_{ik} \partial \eta_{il}}\;,
%$$ 
$$
(f^{\boldsymbol \eta_{j}})_i = \frac{\partial f}{\partial \eta_{ij}}\;,\; \quad \quad (f^{\boldsymbol \eta_{j} \boldsymbol \eta_{k}})_i = \frac{\partial^2 f}{\partial \eta_{ij}\partial\eta_{ik}} \quad  \text{and} \quad
(f^{\boldsymbol \eta_{j} \boldsymbol \eta_{k} \boldsymbol \eta_{l}})_i = \frac{\partial^3 f}{ \partial \eta_{ij} \partial \eta_{ik} \partial \eta_{il}}\;,
$$ 
where $\eta_{ij}$ refers to the element ($i,j$)  of the $n \times q$ matrix including the  linear predictors, $\boldsymbol \eta$, and $f$ is a scalar-valued function. 

Then, let $\mathbf G$ a $(d-1)\times(d-1)$ lower triangular matrix such that 
$${\rm G}_{jk}= \left[\binom{j+1}{2} - (j-k) +2d \right]\mathbbm{1}_{\{k\leq j\}},$$ with $j,k=1, \ldots, d-1$,
%${\rm G}_{jk}={\rm C}_{jk}+2d\mathbbm{1}_{\{k\leq j\}}$, where 
%$$ {\rm C}_{jk}=\begin{cases}\quad \binom{j+1}{2}\quad \quad  \quad \,\,\,k=j \\  {\rm C}_{j(k+1)}-1\quad \quad \,  k<j\\  \quad\quad 0 \quad \quad \quad \quad \, \, k>j \,\, ,\end{cases}$$ 
%
and let $\mathbf Z$ and $\mathbf W$ two $(d-1)\times(d-1)$ lower triangular matrices  such that ${\rm Z}_{jk}=k\mathbbm{1}_{\{k\leq j\}}$ and ${\rm W}_{jk}=(j+1)\mathbbm{1}_{\{k\leq j\}}$. Denoting with $\rvech(\cdot)$ the row-wise half-vectorisation operator, that is $\rvech(\mathbf Z) = (Z_{11}, Z_{21}, Z_{22}, Z_{31}, Z_{32}, Z_{33}, \dots, Z_{(d-1)(d-1)})^\top$, we define $\mathbf z=\rvech( \mathbf Z) $ and  $\mathbf w=\rvech( \mathbf W)$. In the following, the subscript $i$ is omitted to lighten the notation. 
 
\subsection{Derivatives for the logM Parametrisation} \label{app:Dllk_Deta_logm}
This section reports the formulation of the likelihood quantities needed for implementing the logM-based multivariate Gaussian model.  Some quantities, already introduced in Section \ref{sec:logM}, are recalled for clarity. Let $\boldsymbol \Theta = \log \mathbf \Sigma= \mathbf U \boldsymbol \Gamma \mathbf U^\top$, where $\mathbf U$ is the orthonormal matrix of the eigenvectors and $\boldsymbol \Gamma$ is the diagonal matrix of the log-eigenvalues,  resulting from eigen-decomposing $\boldsymbol \Sigma$, that is $\Gamma_{jj}=\gamma_j$, for  $j=1, \ldots, d$. Then, $\exp\big(-\boldsymbol  \Theta \big)=\mathbf U \mathbf L \mathbf U^\top, $ with $\mathbf L =\diag\big(e^{- \gamma_{1}},\ldots, e^{- \gamma_{d}}\big)$. The Gaussian log-density, up to an additive constant and omitting the subscript $i$, is
 \begin{equation}  \label{eqApp:logLik}
\ell=-\frac{1}{2}\tr\big(\boldsymbol \Theta \big) -\frac{1}{2}\mathbf r^\top \exp\big(-\boldsymbol \Theta \big)\mathbf r  =-\frac{1}{2} \sum_{j=d+1}^{2d}\eta_j- \frac{1}{2}\mathbf s^\top \mathbf L  \mathbf s \,\,,
\end{equation}
where $\mathbf s=\mathbf U^\top \mathbf r$ and $\mathbf r= \mathbf y -\boldsymbol \mu$. 

%The $d \times d$ symmetric matrix $\boldsymbol \Delta$ has elements 
%\begin{align*}
%\Delta_{jj} &=e^{-\gamma_j}\,\,,  \;\;\;   \quad \quad \quad \quad \quad  \quad \quad \quad  \quad \quad \quad j =1, \ldots, d,\\
%\Delta_{jk}&= \big(e^{-\gamma_k}-e^{-\gamma_j}\big)/(\gamma_j  - \gamma_k)\,\,, \quad \,\,\,\quad\;\;\;  j \neq k.
%\end{align*}

In addition, recall that $\mathbf F =  \mathbf U \diag(\mathbf s)$ and the relation $\boldsymbol \Theta= \sum_{j=1}^{d(d+1)/2}{\eta_{j+d}  \mathbf V^j}$, where the ${\mathbf V}^j$'s are basis matrices; that is, if $d=3$ they take the form %$(\mathbf V^1)_{11}=1$, $(\mathbf V^2)_{22}=1$, $(\mathbf V^3)_{33}=1$, $(\mathbf V^4)_{12}=(\mathbf V^4)_{21}=1$,  $(\mathbf V^5)_{13}=(\mathbf V^5)_{31}=1$, $(\mathbf V^6)_{23}=(\mathbf V^6)_{32}=1$ and $0$ otherwise.
\renewcommand{\arraystretch}{0.75}
$$\mathbf V^1 = \begin{pmatrix} 1 & 0 & 0 \\ 
                                             0 & 0 & 0 \\
                                             0 & 0 & 0 \\ \end{pmatrix}, \quad \mathbf V^2 = \begin{pmatrix}0 & 0 & 0 \\
                                             0 & 1 & 0 \\
                                             0 & 0 & 0 \\ \end{pmatrix}, \quad \mathbf V^3 = \begin{pmatrix}0 & 0 & 0 \\
                                             0 & 0 & 0 \\
                                             0 & 0 & 1 \\ \end{pmatrix}, $$ 
$$\mathbf V^4 = \begin{pmatrix} 0 & 1 & 0 \\
                                             1 & 0 & 0 \\
                                             0 & 0 & 0 \\ \end{pmatrix}, \quad \mathbf V^5 = \begin{pmatrix}0 & 0 & 1 \\
                                             0 & 0 & 0 \\
                                             1 & 0 & 0 \\ \end{pmatrix}, \quad \mathbf V^6 = \begin{pmatrix}0 & 0 & 0 \\
                                             0 & 0 & 1 \\
                                             0 & 1 & 0 \\ \end{pmatrix}. $$
\renewcommand{\arraystretch}{1}

\subsubsection*{Gradient}

The log-likelihood gradient $\ell^{\bsEta} = (\ell^{\eta_1}, \dots, \ell^{\eta_q})^\top = (\partial \ell/ \partial \eta_1, \dots, \partial \ell/ \partial \eta_q)^\top$ is given by 
\begin{align}
\ell^{\eta_l}&=\big[\mathbf U \mathbf L \mathbf s\big]_l \,\,,  \;\;\;\,  \quad \quad \quad \quad \quad  l = 1, \dots, d, \nonumber \\
\ell^{\eta_l}&=\frac{1}{2} [\boldsymbol \Xi]_{l-d,l-d} -\frac{1}{2}\,\, , \;\;\; \quad  \,\, \, \,  l =d+1,\ldots, 2d, \label{eq:grad_form_logm1} \\
\ell^{\eta_l}& =[\boldsymbol \Xi]_{z_{l-2d},w_{l-2d}}\,\, , \;\;\; \quad \quad \quad  l =2d+1,\ldots, q, \label{eq:grad_form_logm2}
\end{align}
where $\mathbf z$ and $\mathbf w$ are vectors of indices defined in SM \ref{AppNot},  while $
\boldsymbol \Xi =  \mathbf F \boldsymbol \Delta \mathbf F^\top
$
with $\mathbf F =  \mathbf U \diag(\mathbf s)$ and  $\boldsymbol \Delta$ is a $d \times d$ matrix such that  %$\Delta_{jj} =e^{-\gamma_j}$ and  $\Delta_{jk}= \big(e^{-\gamma_k}-e^{-\gamma_j}\big)/(\gamma_j  - \gamma_k)$.
\begin{align*}
\Delta_{jj} &=e^{-\gamma_j}\,\,,  \;\;\;   \quad \quad \quad \quad \quad  \quad \quad \quad  \quad \quad \quad j =1, \ldots, d,\\
\Delta_{jk}&= \big(e^{-\gamma_k}-e^{-\gamma_j}\big)/(\gamma_j  - \gamma_k)\,\,, \quad \,\,\,\quad\;\;\;  j \neq k.
\end{align*}

While the derivatives $\ell^{\eta_l}$, $l = 1, \ldots, d$, are trivial to obtain, those for $l = d+1, \ldots, q$, are derived from 
\begin{equation} \label{eq:naive1} 
\ell^{\eta_l}  = -\frac{1}{2} \mathbbm{1}_{\{l \leq 2d\}} - \frac{1}{2} \mathbf r^\top\Bigg\{\frac{\partial}{\partial \eta_l} \exp(-\boldsymbol \Theta)\Bigg\}\mathbf r\,\,. 
\end{equation}
Leveraging \cite{najfeld1995}, the following expression can be obtained via
\begin{equation} \label{eq:naive2}
\frac{\partial}{\partial \eta_l} \exp(-\boldsymbol \Theta) =  \mathbf U \Big (\mathbf {\overline V}^{l-d} \odot \mathbf \Delta \Big)\mathbf U^\top \,\,,
\end{equation}
where $\mathbf {\overline V}^j = -\mathbf U^\top {\mathbf V}^j \mathbf U $, $j=1, \ldots, d(d+1)/2$. 
By exploiting the property $\mathbf x^\top (\mathbf A \odot \mathbf B) \mathbf y= \tr\big\{\diag(\mathbf  x)\mathbf A \diag(\mathbf y) \mathbf B^\top \big\}$, the derivatives $\ell^{\eta_l}$, $l = d+1, \ldots, q$, take the form
\begin{align*} %\label{eq:naive1} 
\ell^{\eta_l}  &= -\frac{1}{2} \mathbbm{1}_{\{l \leq 2d\}} - \frac{1}{2} \mathbf s^\top  \Big(\mathbf {\overline V}^{l-d} \odot \mathbf \Delta\Big) \mathbf s \\
& = -\frac{1}{2} \mathbbm{1}_{\{l \leq 2d\}} + \frac{1}{2} \tr\big\{\diag(\mathbf s)\mathbf U^\top  \mathbf {V}^{l-d} \mathbf U \diag(\mathbf s) \boldsymbol \Delta \big\}\\
& = -\frac{1}{2} \mathbbm{1}_{\{l \leq 2d\}} + \frac{1}{2} \tr\Big(\mathbf F \boldsymbol \Delta \mathbf F^\top \mathbf {V}^{l-d}  \Big) \,\,.
\end{align*}
Then, the result is obtained by substituting in $\boldsymbol \Xi = \mathbf F \boldsymbol  \Delta  \mathbf F^\top$, and noticing that $\tr\big(\boldsymbol \Xi\mathbf {V}^{l-d}\big)$ is the element of $\boldsymbol \Xi$ accessed by the indices given in (\ref{eq:grad_form_logm1}) and (\ref{eq:grad_form_logm2}). 

\subsubsection*{First $d$ Rows of the Hessian Matrix} %are given by
Denoting with $\boldsymbol \Pi=\mathbf F \mathbf \Delta$,  the second-order derivatives of  (\ref{eqApp:logLik}), for $l=1,\ldots,d$,  are 
\begin{align}
\ell^{\eta_l\eta_m} &= -\big[\mathbf U \mathbf L \mathbf U^\top\big]_{lm}\,\, ,  \;\;\;  \quad \quad \quad \quad \quad \,\,\,\,\,\,\, m = l, \ldots, d,  \label{eqApp:Hess_1_1} \\
\ell^{\eta_l\eta_m} &=   -\mathbf U_{ l\cdot} (\boldsymbol \Pi_{(m-d)\cdot} \odot \mathbf U_{(m-d)\cdot})^\top\,\,, \;\;\; \,\,\,   m=d+ 1, \ldots, 2d,  \label{eqApp:Hess_1_2} \\
\ell^{\eta_l\eta_m} &= -\mathbf U_{l\cdot}   {\mathbf {\tilde u}^{m-2d}}\,\,, \;\;\;  \quad \quad \quad \quad \quad \quad \quad m=2d+ 1, \ldots, q, \label{eqApp:Hess_1_3}
\end{align}
where the $j$-th element of ${\mathbf {\tilde u}^{m-2d}}$, for $j=1, \ldots, d$, is 
\begin{equation*}  
{\tilde{\mathbf  u}^{m-2d}_j} =\Pi_{ z_{m-2d}j}{\rm U}_{w_{m-2d}j } +\Pi_{ w_{m-2d}j} {\rm U}_{ z_{m-2d}j}\,\, .
\end{equation*}
%\end{prop}

 Expression (\ref{eqApp:Hess_1_1}) is trivial to obtain, while the derivatives $\ell^{\eta_l\eta_m}$, for  $l = 1, \ldots, d$, and $m=d+1, \ldots, q$, are derived from
\begin{equation}\label{eq:naivh1}
\ell^{\eta_l\eta_m} =-\frac{1}{2}\frac{\partial}{\partial \eta_l}  \mathbf r^\top\mathbf U \big(\mathbf {\overline V}^{m-d} \odot \mathbf \Delta\big)\mathbf U^\top \mathbf r
 = \big[\mathbf U \big(\mathbf {\overline V}^{m-d} \odot \mathbf \Delta\big) \mathbf s\big]_{l}\,\,.
\end{equation}
%where $\mathbf {\overline V}^j = -\mathbf U^\top {\mathbf V}^j \mathbf U $, $j=1, \ldots, d(d+1)/2$. 
By using the property $(\mathbf A \odot \mathbf B) \mathbf y= \diagv \big(\mathbf A \diag(\mathbf y) \mathbf B^\top\big)$, it is possible to obtain the form
\begin{align*}
\ell^{\eta_l\eta_m} &=- \big[\mathbf U \diagv(\mathbf U^\top  \mathbf {V}^{m-d} \mathbf U \diag(\mathbf s) \mathbf \Delta)\big]_{l}\\
 &=- \big[\mathbf U \diagv \big(\mathbf U^\top  \mathbf {V}^{m-d} \mathbf F \mathbf \Delta \big)\big]_{l}\\
 &= - \mathbf U_{l\cdot} \diagv \big(\mathbf U^\top  \mathbf {V}^{m-d}\boldsymbol \Pi\big) \, \,.
\end{align*}
Then, for $m=d+1, \ldots, 2d$, the expression (\ref{eqApp:Hess_1_2}) is obtained by observing that 
$
\big[\mathbf U^\top \mathbf V^{m-d}\big]_{\cdot j} = \mathbf U_{(m-d)\cdot}
$,
for $j = m-d$, and $
\big[\mathbf U^\top \mathbf V^{m-d}\big]_{\cdot j} = \mathbf 0
$, for $ j\neq m-d$, which implies that
$$
\diagv \big(\mathbf U^\top \mathbf V^{m-d} \boldsymbol \Pi \big)=(\boldsymbol \Pi_{(m-d)\cdot} \odot \mathbf U_{(m-d)\cdot})^\top\;\;.
$$ 
Finally,  for $m=2d+1, \ldots, q$, to obtain (\ref{eqApp:Hess_1_3}) note that $\big[\mathbf U^\top \mathbf V^{m-d}\big]_{\cdot j}= \mathbf U_{j\cdot}$, with $j \in \{ w_{m-2d},z_{m-2d} \}$, and $\big[\mathbf U^\top \mathbf V^{m-d}\big]_{\cdot j} = \mathbf 0$, with $j\notin \{z_{m-2d}, w_{m-2d}\},$ which leads to 
$$ 
\diagv \big(\mathbf U^\top \mathbf V^{m-d} \boldsymbol \Pi\big)={\mathbf {\tilde u}^{m-2d}} \;\; . 
$$

\subsubsection*{Last $q-d$ Rows of the Hessian Matrix}
Denote with $l' = l-d$, $l'' = l-2d$, $m'=m-d$, and $m''=m-d$, and recall from Section \ref{sec:logM} the following quantities;  $\mathbf A$ is a $d\times d$ matrix obtained as $\mathbf A = \mathbf F {\tilde {\boldsymbol \Delta}}^\top$, where $\boldsymbol {\tilde \Delta}$ is such that $\tilde \Delta_{jk}=(\Delta_{jk}-\Delta_{kk})/(\gamma_j-\gamma_k)$ if $j\neq k$ and 0 otherwise; $\mathbf A'$ is a $d\times d\times d$ array given by $\mathbf A'_{\cdot\cdot j} = \mathbf F\big(\mathbf F \odot  {\overbar {\mathbf \Delta}}_{j}\big)^\top$, where $ {\overbar {\mathbf \Delta}}_{j}$ is a matrix whose $k$-th row, $k=1, \ldots,d$, is given by $\tilde {\mathbf \Delta}_{\cdot j}^\top$;  in indicial form it is 
$${\rm A}'_{rst}= \sum_{k\neq t}^{}{\rm F}_{rk}{\rm F}_{sk}\bigg(\frac{\Delta_{tk}-\Delta_{kk}}{\gamma_k-\gamma_t} \bigg)\,\,, \quad  r,s,t = 1, \ldots,d; $$
then, $\mathbf A''$ is a $d\times d\times d$ array given by $\mathbf A''_{\cdot\cdot j} = \mathbf F \big(\mathbf F \boldsymbol \Delta^*_{\cdot\cdot j}\big)^\top$, where $\Delta^*_{kk'j}=(\Delta_{k'j} - \Delta_{kk'})/(\gamma_k-\gamma_{j})$ if $j\neq k $ and $k'\neq, j, k$ and 0 otherwise; in indicial form it is 
$${\rm A}''_{rst}= \sum_{k \neq j} {\rm F}_{sk}\sum_{k'\neq j,k}{\rm F}_{rk'}\Delta^*_{tkk'}\;\;,  \quad  r,s,t = 1, \ldots,d;
$$
finally, $\mathbf K$ is the array, whose elements are of the form
$$
{\rm K}_{rst} =  {\rm F}_{rt}{\rm F}_{st} \frac {\Delta_{tt}}{2} + {\rm F}_{rt} {\rm A}_{st} + {\rm F}_{st} {\rm A}_{rt} + {\rm A}'_{rst} + {\rm A}''_{rst}\;\;,  \quad  r,s,t = 1, \ldots,d.
$$
Then,  the second-order derivatives of  (\ref{eqApp:logLik}), for $l=d+1,\ldots,2d$,  are 
\begin{align} 
\ell^{\eta_l\eta_m} &=  -\sum_{j=1}^{d} {\rm U}_{l'j}{\rm U}_{m'j} {\rm K}_{l'm'j}\,\,, \,\,\quad \quad \quad \quad \quad  \quad \quad \quad \,\,\, \,\,\,\,\, m=l, \ldots, 2d,\label{eqApp:Hess22}\\
\ell^{\eta_l\eta_m} &=  -\sum_{j=1}^{d} {\rm U}_{l'j}\big({\rm U}_{w_{m''}j}{\rm K}_{l'z_{m''}j} + {\rm U}_{z_{m''}j}{\rm K}_{l'w_{m''}j}  \big)\,\,, \,\,    \,\,\,  m=2d+1, \ldots, q, \label{eqApp:Hess23}
\end{align}
and, for $l=2d+1, \ldots, q$ and $m=l, \ldots, q$, take the form
\begin{align}
\ell^{\eta_l\eta_m} =  -\sum_{j=1}^{d} \Big\{&{\rm U}_{w_{l''}j}\big({\rm U}_{w_{m''}j}{\rm K}_{z_{l''}z_{m''}j} + {\rm U}_{z_{m''}j}{\rm K}_{z_{l''}w_{m''}j}  \big) + \nonumber \\ & {\rm U}_{z_{l''}j}\big({\rm U}_{w_{m''}j}{\rm K}_{w_{l''}z_{m''}j} + {\rm U}_{z_{m''}j}{\rm K}_{w_{l''}w_{m''}j}  \big)\Big\}\,\,, \label{eqApp:Hess33}
\end{align}

The second derivatives $\ell^{\eta_l\eta_m}$, for $l=d+1, \ldots, q,$ and $m=l, \ldots, q$, are obtained via $$\frac{\partial}{\partial \eta_l \partial \eta_m} \ell( \bsEta) = -\frac{1}{2}\mathbf r^\top \Bigg\{\frac{\partial }{\partial \eta_l \partial \eta_m} \exp(-\boldsymbol \Theta)\Bigg\}  \mathbf r\;\;, $$
whose solution require computing $\partial \exp(-\boldsymbol \Theta) / \partial \eta_l \partial \eta_m$. To simplify,  $\partial \exp(\boldsymbol \Theta)/\partial \eta_l \partial \eta_m$ is carried out because the final result is simply obtained by changing the sign of the eigenvalues.  Similarly to \citet{najfeld1995}, $\partial \exp(\boldsymbol \Theta)/\partial \eta_l \partial \eta_m$ can be obtained via directional derivatives (w.r.t. to the direction ${\mathbf V}^{l'}$ and ${\mathbf V}^{m'}$) of $\exp(\boldsymbol \Theta)$, denoted with ${\rm D}_{{\mathbf V}^{l'} {\mathbf V}^{m'}}(1, \boldsymbol \Theta)$, that are solution of the integral 

\begin{align} \label{eqApp:dirDer}
{\rm D}_{{\mathbf V}^{l'} {\mathbf V}^{m'}}(1, \boldsymbol \Theta)&= \int_{0}^{1} \exp\{(1-\tau) \boldsymbol \Theta\} {\mathbf V}^{m'}  \Bigg[ \int_{0}^{\tau} \exp\{(\tau-\omega) \boldsymbol \Theta\} {\mathbf V}^{l'} \exp(\omega \boldsymbol \Theta) d\omega \Bigg] d\tau \nonumber\\
&+ \int_{0}^{1} \exp\{(1-\tau) \boldsymbol \Theta\} {\mathbf V}^{l'}  \Bigg[ \int_{0}^{\tau} \exp\{(\tau-\omega) \boldsymbol \Theta\} {\mathbf V}^{m'} \exp(\omega \boldsymbol \Theta) d\omega \Bigg] d\tau \,\,.
\end{align}
Only one of the two integrals in (\ref{eqApp:dirDer}) must be solved, the other is simply obtained by permuting the indices $l'$ and $m'$. Instead of solving directly   (\ref{eqApp:dirDer}), an efficient formulation is carried out by considering  $\mathbf r^\top {\rm D}_{{\mathbf V}^{l'} {\mathbf V}^{m'}}(1, \boldsymbol \Theta) \mathbf r$, which allows using  the property $\mathbf x^\top (\mathbf A \odot \mathbf B) \mathbf y= \tr \big(\diag(\mathbf  x)\mathbf A \diag(\mathbf y) \mathbf B^\top\big)$.
Then, the first inner integral of (\ref{eqApp:dirDer}) takes the form $$\int_{0}^{\tau} \exp\{(\tau-\omega) \boldsymbol \Theta\} {\mathbf  V}^{l'} \exp(\omega \boldsymbol \Theta) d\omega= \mathbf U \big(\mathbf U^\top {\mathbf V}^{l'} \mathbf U \odot \mathbf \Phi(\tau)\big)\mathbf U^\top\,\,,$$ with 
$\Phi(\tau)_{jj} =\tau e^{\tau \gamma_j},$ and % \;\;, \quad \quad  \quad \quad \quad \quad \quad \quad \quad \quad \quad \; \nonumber \\
$\Phi(\tau)_{jk} = \big(e^{\tau \gamma_j} - e^{\tau \gamma_k}\big)/(\gamma_j - \gamma_k)$ 
(note that $\boldsymbol \Phi(1)=\boldsymbol \Delta$, after changing the sign of the eigenvalues), and consider that $\exp\{(1-\tau) \boldsymbol \Gamma\} \mathbf U^\top {\mathbf V}^{m'}  \mathbf U =  \boldsymbol \Psi(\tau) \odot  \mathbf U^\top {\mathbf V}^{m'}  \mathbf U$, with $\boldsymbol \Psi(\tau)$ a $d\times d$ matrix  such that $\Psi(\tau)_{jk}=e^{(1-\tau) \gamma_j}$, for $j,k=1, \ldots, d$. Denote with ${\mathbf v}_j$ the $j$-th column of ${\mathbf V}^j$, $j=1,\ldots,d$, and consider the identity matrix $\mathbf I_d=\sum_{j=1}^{d}\mathbf V^j=\sum_{j=1}^{d} {\mathbf v}_j {\mathbf {v}^\top_j}$. Thus, it is possible to obtain 
\begin{align}\label{eqApp:proof_strong}
&\mathbf r^\top \, \int_{0}^{1} \exp\{(1-\tau) \boldsymbol \Theta\} {\mathbf V}^{m'}  \Bigg[ \int_{0}^{\tau} \exp\{(\tau-\omega) \boldsymbol \Theta\} {\mathbf V}^{l'} \exp(\omega \boldsymbol \Theta) d\omega \Bigg] d\tau \,  \mathbf r \nonumber\\
= &  \int_{0}^{1} \mathbf r^\top \mathbf U \exp\{(1-\tau) \boldsymbol \Gamma\} \mathbf U^\top {\mathbf V}^{m'}  \mathbf U \big(\mathbf U^\top \mathbf V^{l'} \mathbf U \odot \mathbf \Phi(\tau)\big) {\mathbf U}^\top {\mathbf r}\, d\tau \nonumber\\
= & \int_{0}^{1} \mathbf s^\top  \big(\boldsymbol \psi(\tau) \odot  \mathbf U^\top {\mathbf V}^{m'}  \mathbf U\big) \mathbf{I}_d \big(\mathbf U^\top \mathbf V^{l'} \mathbf U \odot \mathbf \Phi(\tau)\big)\mathbf {\mathbf s}\, d\tau \nonumber\\
= & \int_{0}^{1} \sum_{j=1}^{d} \mathbf s^\top  \big(\boldsymbol \psi(\tau) \odot  \mathbf U^\top {\mathbf V}^{m'}  \mathbf U\big) {\mathbf v}_j {\mathbf {v}_j}^\top (\mathbf U^\top \mathbf V^{l'} \mathbf U \odot \mathbf \Phi(\tau))\mathbf {\mathbf s}\, d\tau \nonumber \\
= & \int_{0}^{1} \sum_{j=1}^{d} \tr\Big\{\diag(\mathbf s)  \boldsymbol \Psi(\tau) \diag\big({\mathbf v}^j\big)  \mathbf U^\top {\mathbf V}^{m'}  \mathbf U\Big\}\tr\Big\{\diag\big(\mathbf v^j\big) \mathbf U^\top {\mathbf V}^{l'}  \mathbf U  \diag({\mathbf s}) \boldsymbol \Phi(\tau) \Big\} d\tau \nonumber\\
= & \int_{0}^{1} \sum_{j=1}^{d} \tr\Big\{\mathbf F \boldsymbol \Psi(\tau) \tilde {\mathbf  U}^{j,m'}\Big\}\tr\Big\{\mathbf F \boldsymbol \Phi(\tau) \tilde {\mathbf  U}^{j,l'} \Big\} d\tau
\end{align}
where $\tilde {\mathbf  U}^{j,k}=\diag\big(\mathbf v^j\big) \mathbf U^\top \mathbf V^k$.
Then, for simplicity distinguish three cases 
\begin{itemize}
    \item[1)] For $l=d+1, \ldots, 2d$, and  $m=l, \ldots, 2d,$ 
   \begin{align*}
\tr\Big\{\mathbf F \boldsymbol \Psi(\tau) \tilde {\mathbf  U}^{j,m'} \Big\} &= {\rm U}_{m'j}\sum_{k=1}^{d}{\rm F}_{m'k} e^{(1-\tau)\gamma_k}\,\,, \\
 \tr\Big\{\mathbf F \boldsymbol \Phi(\tau) \tilde {\mathbf  U}^{j,l'} \Big\} &= {\rm U}_{l'j}\Bigg( {\rm F}_{l'j} \tau e^{\tau \gamma_j}  + \sum_{k\neq j}^{}{\rm F}_{l'k} \frac{e^{\tau \gamma_k} -e^{\tau \gamma_j}}{\gamma_k - \gamma_j}\Bigg) \,\,;
\end{align*}
    \item[2)] For $l=d+1, \ldots, 2d$, and  $m=2d+1, \ldots, q,$ 
\begin{align*}
   \tr\Big\{\mathbf F \boldsymbol \Psi(\tau) \tilde {\mathbf  U}^{j,m'} \Big\} &= {\rm U}_{w_{m''}j}\sum_{k=1}^{d}{\rm F}_{z_{m''}k} e^{(1-\tau)\gamma_k} + {\rm U}_{z_{m''}j}\sum_{k=1}^{d}{\rm F}_{w_{m''}k} e^{(1-\tau)\gamma_k}\,\,,\\
        \tr\Big\{\mathbf F \boldsymbol \Phi(\tau) \tilde {\mathbf  U}^{j,l'} \Big\}&= {\rm U}_{l'j}\Bigg( {\rm F}_{l'j} \tau e^{\tau \gamma_j}  + \sum_{k\neq j}^{}{\rm F}_{l'k} \frac{e^{\tau \gamma_k} -e^{\tau \gamma_j}}{\gamma_k - \gamma_j}\Bigg) \,\,;
\end{align*}
    \item[3)] For $l=2d+1, \ldots, q$ and  $m=l, \ldots,q$, 
        \begin{align*} \label{eqApp:trace}
\tr\Big\{\mathbf F \boldsymbol \Psi(\tau) \tilde {\mathbf  U}^{j,m'} \Big\} &= {\rm U}_{w_{m''}j}\sum_{k=1}^{d}{\rm F}_{z_{m''}k} e^{(1-\tau)\gamma_k} + {\rm U}_{z_{m''}j}\sum_{k=1}^{d}{\rm F}_{w_{m''}k} e^{(1-\tau)\gamma_k}\nonumber \\ 
              \tr\Big\{\mathbf F \boldsymbol \Phi(\tau) \tilde {\mathbf  U}^{j,l'} \Big\} &= {\rm U}_{w_{l''}j}\Bigg( {\rm F}_{z_{l''}j} \tau e^{\tau \gamma_j}  + \sum_{k\neq j}^{}{\rm F}_{z_{l''}k} \frac{e^{\tau \gamma_k} -e^{\tau \gamma_j}}{\gamma_k - \gamma_j}	\Bigg) \nonumber\\ 
             &+{\rm U}_{z_{l''}j}\Bigg( {\rm F}_{w_{l''}j} \tau e^{\tau \gamma_j}  + \sum_{k\neq j}^{}{\rm F}_{w_{l''}k} \frac{e^{\tau \gamma_k} -e^{\tau \gamma_j}}{\gamma_k - \gamma_j} \Bigg)\,\,.
        \end{align*}     
\end{itemize}
Then, replacing the above case-expressions into (\ref{eqApp:proof_strong}) and developing the integral, the expressions of the arrays $\mathbf A$, $\mathbf A'$ and $\mathbf A''$ are obtained after some algebraic arrangements, including the change of sign for the $\gamma$'s. Hence the solutions (\ref{eqApp:Hess22}), (\ref{eqApp:Hess23}), and (\ref{eqApp:Hess33}).

\subsection{Derivatives for the MCD Parametrisation} \label{app:Dllk_Deta_mcd}

The MCD parametrisation is expressed as $\boldsymbol \Sigma^{-1} = \mathbf{T}^\top\mathbf{D}^{-2}\mathbf{T}$, where $\mathbf{D}^2$ is a diagonal matrix, such that ${\rm D}^2_{jj} = e^{\eta_{j+d}}$, $j=1, \ldots, d,$    and 
\begin{equation*}\label{eq:mcd_T}
\mathbf T=\begin{pmatrix} 1 & 0 & 0 & \cdots& 0 \\
                               \eta_{2d+1} & 1 &  0 & \cdots& 0  \\

                               \eta_{2d+2} & \eta_{2d+3} & 1& \cdots   & 0 \\

                               \vdots & \vdots & \vdots & \ddots   &  \vdots&  \\

                             \eta_{q-d+2} & \eta_{q-d+3} & \cdots &  \eta_{q} & 1\\

\end{pmatrix}.
\end{equation*}
Omitting the constants that do not depend on $\bsEta$ and denoting with $r_k$ the $k$-th element $\mathbf r = \mathbf y -\boldsymbol \mu $, the $i$-th log-likelihood contribution is 
\begin{align} \label{eqApp:llik_mcd}
\ell & = -\frac{1}{2} \big\{\tr\big(\log \mathbf D^2\big)+\mathbf r^\top\mathbf{T}^\top\mathbf{D}^{-2}\mathbf{T}\mathbf r\big\} \nonumber \\
& = -\frac{1}{2}\sum_{j=1}^{d}\Bigg\{\eta_{j+d} + e^{- \eta_{j+d}}\Bigg(\sum_{k=1}^{j-1} \eta_{{\rm G}_{(j-1)k}} r_k+ r_j \Bigg)^2\Bigg\}\,\, ,
\end{align}
after using $\log |\boldsymbol \Sigma| = \tr\big(\log \mathbf{D}^2\big)=\sum_{j=1}^{d}  \eta_{j+d}$ and implicitly assuming that the sum $\sum_{k=1}^{j-1}$ should not be computed when $j = 1$ (the same convention is used in several places below). Similarly, below the sum $\sum_{j=l+1}^{d}$ will not be computed when $l=d$. Here the derivatives of $\ell$ w.r.t. ${\bsEta}$ up to the third order are obtained, both in compact matrix form and indicial form, the latter being more useful for efficient numerical implementation.  Finally, the following $d\times d$ matrices are introduced, being largely used in the matrix form expression of the likelihood derivatives reported below, that is $\mathbf{Q}^l$, for $l=1, \ldots, d$, and  $\mathbf{P}^l $, for $ l=1, \ldots, d(d-1)/2$,  such that $\big[\mathbf{Q}^l\big]_{ll}=e^{-\eta_{l+d}}$ and $\big[\mathbf{P}^l\big]_{{z}_{l}{w}_{l}}=1$, while all other elements are equal to zero. 

\subsubsection*{Score Vector}

The elements of $\ell^{\bsEta} = (\ell^{\eta_1}, \dots, \ell^{\eta_q})^\top = (\partial \ell/ \partial \eta_1, \dots, \partial \ell/ \partial \eta_q)^\top$ are 
\begin{align*}
\ell^{\eta_l} & =\big[ \mathbf{T}^\top\mathbf{D}^{-2}\mathbf{T}\mathbf{r} \big]_l   \\  
&=e^{- \eta_{d+l}}\Bigg(\sum_{k=1}^{l-1} \eta_{ {\rm G}_{(l-1)k}}  r_k +  r_l\Bigg) + \sum_{j=l+1}^{d}e^{-\eta_{j+d}}\Bigg(\sum_{k=1}^{j-1}  \eta_{ {\rm G}_{(j-1)k}} r_k+ r_j \Bigg) \eta_{ {\rm G}_{(j-1)l}} \,\, ,
\end{align*}
for $l=1, \ldots, d$,  %
\begin{align*}
\ell^{\eta_l}  & =  \frac{1}{2}\mathbf{r}^\top \mathbf{T}^\top\mathbf{Q}^{l-d}\mathbf{T}\mathbf{r} -\frac{1}{2} \\  
& =\frac{1}{2}e^{-\eta_l}\Bigg(\sum_{k=1}^{l-d-1} \eta_{ {\rm G}_{(l-d-1)k}} r_k+ r_{l-d} \Bigg)^2-\frac{1}{2}\,\, ,
\end{align*}
for $l=d+1, \ldots, 2d$, and 
\begin{align*}
\ell^{\eta_l} &= -\mathbf{r}^\top \mathbf{P}^{l-2d}\mathbf{D}^{-2}\mathbf{T}\mathbf{r}     \\
&= -e^{ \eta_{w_{l-2d}+d}}\Bigg(\sum_{k=1}^{w_{l-2d}-1} \eta_{ {\rm G}_{( w_{l-2d}-1)k}} r_k+ r_{ w_{l-2d}} \Bigg) r_{ z_{l-2d}}\,\,,  
\end{align*}
for $l=2d+1, \ldots, q$. 

\subsubsection*{Hessian Matrix} 
The elements forming the upper triangle of $\ell^{\bsEta\bsEta}$ (here $\ell^{\eta_l\eta_m} = \partial^2 \ell / \partial \eta_l \partial \eta_m$), are 
\begin{align*}
\ell^{\eta_l\eta_m} &=-\big [\mathbf{T}^\top\mathbf{D}^{-2}\mathbf{T} \big ]_{lm} \\
&=-\Bigg\{e^{-{\eta}_{l+d}} + \sum_{k=l+1}^{d}e^{-{\eta}_{k+d}} \big({\eta}_{ {\rm G}_{(k-1)l}}  \big)^2\Bigg\}\mathbbm{1}_{\{m=l\}} \\
&\hspace{5mm}-\Bigg(e^{-{\eta}_{m+d}} {\eta}_{ {\rm G}_{(m-1)l}}+\sum_{k=m+1}^{d}e^{-{\eta}_{k+d}} {\eta}_{ {\rm G}_{(k-1)l}}  {\eta}_{ {\rm G}_{(k-1)m}} \Bigg) \mathbbm{1}_{\{m>l\}}\,\,,
\end{align*}
for $l = 1, \ldots, d$ and $m=l, \ldots, d$,
\begin{align*}
\ell^{\eta_l\eta_m}&=-\big[ \mathbf{T}^\top\mathbf{Q}^{m-d}\mathbf{T}\mathbf{r} \big]_l  \\ 
&=-e^{-{\eta}_{m} }\Bigg\{\Bigg(\sum_{k=1}^{l-1}  \eta_{ G_{(l-1)k}} r_k+ r_{l} \Bigg) \mathbbm{1}_{\{m-d=l\}} \\ &\hspace{5mm}+\Bigg(\sum_{k=1}^{m-d-1}  \eta_{ G_{(m-d-1)k}} r_k+ r_{m-d} \Bigg) \eta_{ G_{(m-d-1)l}}  \mathbbm{1}_{\{m-d>l\}}\Bigg\}\,\, ,
\end{align*}
for $l=1, \ldots, d$ and $m=d+1, \ldots, 2d$,  
\begin{align*}
\ell^{\eta_l\eta_m}&=\big [ \mathbf{P}^{m-2d}\mathbf{D}^{-2}\mathbf{T}\mathbf{r}  + \mathbf{T}^\top\mathbf{D}^{-2}\big(\mathbf{P}^{m-2d}\big)^\top \mathbf{r}  \big ]_l\,\, \\  
&=e^{- \eta_{w_{m-2d}+d}}\Bigg\{  r_{ z_{m-2d}}\Big(\mathbbm{1}_{\{ w_{m-2d}=l\}}+ \eta_{{\rm G}_{( w_{m-2d}-1)l}}\mathbbm{1}_{\{ w_{m-2d}>l\}}\Big) \\
&\hspace{5mm}+\Bigg(\sum_{k=1}^{ w_{m-2d}-1}  \eta_{ {\rm G}_{( w_{m-2d}-1)k}} r_k+r_{ w_{m-2d}} \Bigg)\mathbbm{1}_{\{ z_{m-2d}=l\}}\Bigg\}\,\, , 
\end{align*}
for $l=1, \ldots, d$ and $m=2d+1, \ldots, q$, 
\begin{align*}
\ell^{\eta_l\eta_m}&=-\frac{1}{2}\mathbf{r}^\top \mathbf{T}^\top\mathbf{Q}^{l-d}\mathbf{T}\mathbf{r}  \\
&=-\frac{1}{2}e^{- \eta_{l}}\Bigg(\sum_{k=1}^{l-d-1} \eta_{{\rm G}_{(l-d-1)k}}  r_{k}+ r_{l-d} \Bigg)^2\mathbbm{1}_{\{ m=l\}}\,\, ,  
\end{align*}
for $l = d+1, \ldots, 2d$ and $m=l, \ldots, 2d$,
\begin{align*}
\ell^{\eta_l\eta_m}&=\mathbf{r}^\top \mathbf{P}^{m-2d}\mathbf{Q}^{l-d}\mathbf{T}\mathbf{r}   \\
&=e^{- \eta_{l}}\Bigg(\sum_{k=1}^{l-d-1} \eta_{ {\rm G}_{(l-d-1)k}}  r_{k}+ r_{l-d} \Bigg)r_{ z_{m-2d}}\mathbbm{1}_{\{ w_{m-2d}=l-d\}}\,\, ,
\end{align*}
for $l=d+1, \ldots, 2d$ and $m=2d+1, \ldots, q$, and finally
\begin{align*}
\ell^{\eta_l\eta_m}&=-\mathbf{r}^\top \mathbf{P}^{l-2d}\mathbf{D}^{-2}\big(\mathbf{P}^{m-2d}\big)^\top\mathbf{r} \\
&=-e^{-{\eta}_{ w_{l-2d}+d}}  r_{ z_{l-2d}} r_{ z_{m-2d}} \mathbbm{1}_{\{ w_{m-2d}=w_{l-2d}\}} \,\, ,
\end{align*}
for $l = 2d+1, \ldots, q$ and $m = l, \ldots, q$.

\subsubsection*{Third Derivatives Array}

The elements forming the third derivatives array $\ell^{\bsEta\bsEta\bsEta}$ (here $\ell^{\eta_k\eta_l\eta_m} = \partial^3 \ell / \partial \eta_k \partial \eta_l \partial \eta_m$), are 
$$\ell^{\eta_k\eta_l\eta_m}  = 0 \;\;,$$
for $k=1,\ldots,d,\quad l= k,\ldots,d, \quad m = l,\ldots,d,$
\begin{align*}
\ell^{\eta_k\eta_l\eta_m}  &= \big[{ {\mathbf T}}^\top {{\mathbf {Q}}^{m-d}} {\rm {\mathbf T}}\big]_{kl} \\
&= e^{-\eta_m}\Big\{\mathbbm{1}_{\{k=l=m-d\}}+  \eta_{{\rm { G}}_{(l-1)k}} \mathbbm{1}_{\{l=m-d\}}\mathbbm{1}_{\{l>k\}} +  \eta_{{\rm { G}}_{(m-d-1)k}}\eta_{{\rm { G}}_{(m-d-1)l}} \mathbbm{1}_{\{m-d > l \geq k\}} \Big \}\,\, ,
\end{align*}
for $k=1,\ldots,d,\quad l= k,\ldots,d,\quad m = d+1,\ldots,2d,$
\begin{align*}
\ell^{\eta_k\eta_l\eta_m}  &= -\big[{ {\mathbf P}}^{m-2d} {\rm {\mathbf {D}}^{-2}} {\rm {\mathbf T}} + {\rm {\mathbf T}}^\top {\rm {\mathbf {D}}^{-2}} \big({\mathbf P}^{m-2d}\big)^\top\big]_{kl} \\
&= -e^{-\eta_{w_{m-2d}+d}}\Big\{\mathbbm{1}_{\{k=z_{m-2d}\}}\mathbbm{1}_{\{l=w_{m-2d}\}} \mathbbm{1}_{\{ l > k\}} + 2\eta_{m} \mathbbm{1}_{\{k=l=z_{m-2d}\}} +\\
&\hspace{2.7cm} \big(\eta_{{\rm { G}}_{(w_{m-2d}-1)l}} \mathbbm{1}_{\{k=z_{m-2d}\}} + \eta_{{\rm { G}}_{(w_{m-2d}-1)k}}\mathbbm{1}_{\{l=z_{m-2d}\}}\big) \mathbbm{1}_{\{w_{m-2d}> l > k\}} \Big\} \;\;,
\end{align*}
for $k=1,\ldots,d,\quad l= k,\ldots,d,\quad m=2d+1,\ldots,q,$
\begin{align*}
\ell^{\eta_k\eta_l\eta_m}  &= -\big[{\rm {\mathbf T}}^\top {{\mathbf {Q}}^{l-d}} {\rm {\mathbf T}} \mathbf{r}\big]_{k} \\
&= e^{-\eta_{l}}\Bigg\{\Bigg( \sum_{j=1}^{k-1} \eta_{{\rm {G}}_{(k-1)j}} r_{j}+r_{l-d} \Bigg)\mathbbm{1}_{\{k=l-d\}}  + \\
&\hspace{1.6cm}\Bigg( \sum_{j=1}^{l-d-1} \eta_{{\rm {G}}_{(l-d-1)j}} r_{j}+r_{l-d} \Bigg)\eta_{{\rm {G}}_{(l-d-1)k}}\mathbbm{1}_{\{l-d > k\}}  \Bigg\}\mathbbm{1}_{\{l=m\}}\;\;,
\end{align*}
for $k=1,\ldots,d,\quad l= d+1,\ldots,2d,\quad m = l,\ldots,2d,$
\begin{align*}
\ell^{\eta_k\eta_l\eta_m}  &= -\big[{\rm {\mathbf P}}^{m-2d} {{\mathbf {Q}}^{l-d}} {\rm {\mathbf T}} \mathbf{r} + {\rm {\mathbf T}}^\top {{\mathbf {Q}}^{l-d}} \big({{\mathbf P}}^{m-2d}\big)^\top\mathbf{r} \big]_{k} \\
&=-e^{-\eta_{l}}\Bigg\{r_{z_{m-2d}}(\mathbbm{1}_{\{l-d = k\}} + \eta_{{\rm {G}}_{(l-d-1)k}}\mathbbm{1}_{\{l-d > k\}}) +\\
& \hspace{1.7cm} \Bigg( \sum_{j=1}^{l-d-1} \eta_{{\rm {G}}_{(l-d-1)j}} r_{j}+r_{l-d} \Bigg)\mathbbm{1}_{\{k=z_{m-2d}\}} \Bigg\}\mathbbm{1}_{\{l-d=w_{m-2d}\}}\;\;,
\end{align*}
for $k=1,\ldots,d,\quad l= d+1,\ldots,2d,\quad m=2d+1,\ldots,q,$

\begin{align*}
\ell^{\eta_k\eta_l\eta_m}  &= -\big[{\rm {\mathbf P}}^{l-2d} {\rm {\mathbf {D}}^{-2}} \big({\rm {\mathbf P}}^{m-2d}\big)^\top \mathbf{r} + {\rm {\mathbf P}}^{m-2d} {\rm {\mathbf {D}}^{-2}} \big({\rm {\mathbf P}}^{l-2d}\big)^\top \mathbf{r} \big]_{k} \\
&=e^{-\eta_{w_{l-2d}+d}}\Big\{r_{z_{m-2d}}\mathbbm{1}_{\{z_{l-2d}=k\}}+ r_{z_{l-2d}}\mathbbm{1}_{\{z_{m-2d}=k\}}\Big \} \mathbbm{1}_{\{w_{l-2d}=w_{m-2d}\}}\,\, ,
\end{align*}
for $k=1,\ldots,d,\quad l= 2d+1,\ldots,q,\quad m = l,\ldots,q,$
\begin{align*}
\ell^{\eta_k\eta_l\eta_m}  &= \frac{1}{2}\mathbf {r}^\top {\rm {\mathbf T}}^\top {{\mathbf {Q}}^{k-d}} {\rm {\mathbf T}} \mathbf{r} \\
&= \frac{1}{2}e^{-\eta_k}\Bigg\{ \sum_{j=1}^{k-d-1} \eta_{{\rm {G}}_{(k-d-1)j}} r_{j}+r_{k-d}\Bigg \}^2 \mathbbm{1}_{\{k=l=m\}}\,\, ,
\end{align*}
for $k=d+1,\ldots,2d,\quad l=k,\ldots,2d,\quad m = l,\ldots,2d,$
\begin{align*}
\ell^{\eta_k\eta_l\eta_m}  &= -\mathbf {r}^\top { {\mathbf P}^{m-2d}} {{\mathbf {Q}}^{k-d}} {\rm {\mathbf T}} \mathbf{r} \\
&= e^{-\eta_k}\Bigg\{ \sum_{j=1}^{k-d-1} \eta_{{\rm { G}}_{(k-d-1)j}} r_{j}+r_{k-d}\Bigg \} r_{z_{m-2d}}\mathbbm{1}_{\{k-d=l-d=w_{m-2d}\}} \,\, ,
\end{align*}
for $k=d+1,\ldots,2d,\quad l=k,\ldots,2d,\quad m=2d+1,\ldots,q,$
\begin{align*}
\ell^{\eta_k\eta_l\eta_m} &= \mathbf {r}^\top { {\mathbf P}^{l-2d}} {{\mathbf {Q}}^{k-d}} \big({\mathbf P}^{m-2d}\big)^\top \mathbf{r} \\
&= e^{-\eta_k}r_{z_{l-2d}}r_{z_{m-2d}}\mathbbm{1}_{\{k-d=w_{l-2d}=w_{m-2d}\}} \,\,,
\end{align*}
for $k=d+1,\ldots,2d,\quad l=2d+1,\ldots,q,\quad m=l,\ldots,q,$
$$\ell^{\eta_k\eta_l\eta_m} = 0\,\,,$$
for $k=2d+1,\ldots,q,\quad l=k,\ldots,q,\quad m=l,\ldots,q$.

\subsubsection{Sparsity of Derivatives for the MCD Parametrisation} \label{app:sparsity}
\begin{figure}
\centering
\includegraphics[scale=0.585]{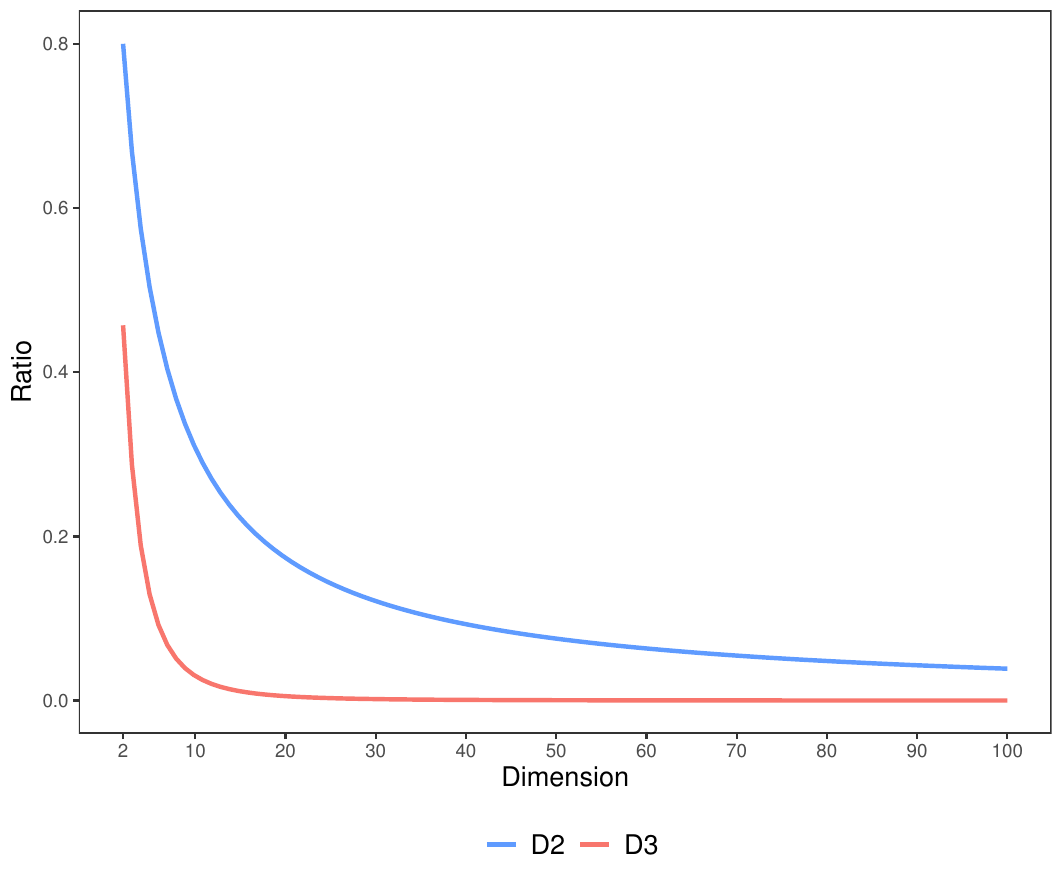}
\caption{Ratio between the total number of non-zero elements and the total number of the Hessian matrix (red) and the third derivatives array (blue) elements. }
\label{fig:sparsity}
\end{figure}

The sparsity of the second- and third-order derivatives is directly related to the dimension of the outcome $d$. Consider the following notation. Let $\tilde {\boldsymbol \eta}_j$, $j=1,2,3$, the subset of $\boldsymbol \eta$ corresponding to the  predictors involved in modelling the vector $\boldsymbol \mu$ ($j=1$), the matrix $\log \mathbf D^2$ ($j=2$) and the $\mathbf T$ matrix ($j=3$). Then, $\tilde {\boldsymbol \eta}_1 = (\eta_1, \ldots, \eta_d)$, $\tilde {\boldsymbol \eta}_2 = (\eta_{d+1}, \ldots, \eta_{2d})$, and $\tilde {\boldsymbol \eta}_3 = (\eta_{2d+1}, \ldots, \eta_{q})$. Hence, for the Hessian matrix, ${\rm E}_{jk}$ denotes the number of non-redundant elements forming the block $\ell^{\tilde \bsEta_j \tilde \bsEta_k}$, $j,k=1,2,3$,  and $e_{jk} \leq {\rm E}_{jk}$ denotes the number of non-zero elements of such blocks.   For instance, ${\rm E}_{11}$ is the number of non-redundant elements forming the upper triangular block of $\ell^{\tilde {\boldsymbol \eta}_j \tilde {\boldsymbol \eta}_k}$, $j,k=1$, and, thus, $e_{11}$ is the number of non-redundant elements different from zero forming such triangular block, while  $E_{12}$ is the number of non-redundant elements forming the square block $\ell^{\tilde {\boldsymbol \eta}_j \tilde {\boldsymbol \eta}_k}$, $j=1$, $k=2$, and $e_{12}$ is the number of non-redundant elements different from zero forming such square block, and so on. Similarly, for the third derivatives array,  we denote with ${\rm E}_{jkl}$ the number of non-redundant elements forming the block  $\ell^{\tilde {\boldsymbol \eta}_j \tilde {\boldsymbol \eta}_k \tilde {\boldsymbol \eta}_l}$, $j,k,l=1,2,3$, and with $e_{jkl} \leq E_{jkl}$ the number of non-zero elements of such blocks. It is easy to obtain the total number of elements for each block and the number of elements different from zero within them. To obtain such  closed forms the   identity  $\sum_{m=k}^{n} \binom{m}{k}=\binom{n+1}{k+1}$ is massively used. 
 %(see Charles H. Jones (1994), GENERALIZED HOCKEY STICK IDENTITIES AND iV-DIMENSIONAL BLOCKWALKING).
 In Table \ref{tab:sparsityd2d3}, the number of non-redundant elements different from zero and the total number of non-redundant elements for each block is reported for the Hessian matrix and the third derivative array, respectively, while the last row considers all the blocks.   
 Instead, in Figure \ref{fig:sparsity} we report the ratio between the total number of elements different from zero and the total number of elements, by varying the dimension of the outcome $d$.

\begin{table}[htbp]\footnotesize
\centering
\renewcommand\arraystretch{1.25}
\begin{tabular}{c|c|c|c|c}%\footnotesize%|cc} 
  \hline
$\ell^{\tilde \bsEta_j \tilde \bsEta_k}$ &\\ \hline
%$\ell^{\tilde \bsEta_j \tilde \bsEta_k}$ & & \\
$(j,k)$ & $e_{jk}$ & $E_{jk}$ & $\mathcal{O}(e_{jk}/E_{jk})$   \\%&  &  \\ 
  \hline
(1,1) &$\frac{d(d+1)}{2}$ &$\frac{d(d+1)}{2}$ & 1 \\%& &  \\ 
(1,2) &$\frac{d(d+1)}{2}$ &$d^2$ & 1 \\%& &  \\ 
(1,3) &$d(d-1)\left[1 + \frac{(d-2)
}{3}\mathbbm{1}_{\{d>2\}}\right]$  &$\frac{d^2(d-1)}{2}$ & $d^{-1} \mathbbm{1}_{\{d=2\}} + 1 \mathbbm{1}_{\{d>2\}}$ \\%& &  \\ 
(2,2) &$d$  &$\frac{d(d+1)}{2}$ &$d^{-1}$\\% &  & \\
(2,3) &$\frac{d(d-1)}{2}$  &$\frac{d^2(d-1)}{2}$ &$d^{-1}$\\%&  & \\
(3,3) &$\frac{(d+1)d(d-1)}{6}$ &$\frac{d(d-1)(d^2-d+2)}{8}$ &$d^{-1}$ \\%&  & \\
Total &$\frac{d\left[(d^2+15d+2)+ 2(d-1)(d-2)\mathbbm{1}_{\{d>2\}}\right]}{6} $ &$\frac{d(d+1)(d+2)(d+3)}{8}$ &  $d^ {-1}$\\%& & \\
%$d(d^2+15d+2)/6 + d(d-1)(d-2)/3\mathbbm{1}_{\{d>2\}} $
\hline
\\
  \hline
  $\ell^{\tilde \bsEta_j \tilde \bsEta_k \tilde \bsEta_l}$ & \\ \hline
%$\ell^{\tilde \bsEta_j \tilde \bsEta_k \tilde \bsEta_l}$ & & \\
$(j,k,l)$& $e_{jkl}$& $E_{jkl}$  & $\mathcal{O}(e_{jkl}/E_{jkl})$ \\ \hline%&  &  \\\
  \hline
(1,1,1) &$0 $ &$\frac{d(d+1)(d+2)}{6} $  & 0\\%& &  \\ 
(1,1,2) &$\frac{d(d+1)(d+2)}{6} $ &$\frac{d^2(d+1)}{2} $ &1 \\% & &  \\ 
(1,1,3) &$\frac{(d-1)d(d+1)}{3} $ &$\frac{d^2(d-1)(d+1)}{4} $ &$d^{-1}$ \\% & &  \\ 
(1,2,2) &$\frac{d(d+1)}{2} $ &$\frac{d^2(d+1)}{2}$ & $d^{-1}$\\% & &  \\ 
(1,2,3) &$\frac{(d-1)d(d+1)}{3} $ &$\frac{d^3(d-1)}{2} $ & \textbf{$d^{-1}$} \\% & &  \\ 
(1,3,3) &$\frac{(d-1)d(2d-1)}{6} $ &$\frac{d^2(d-1)[d(d-1)+2]}{8} $ &  \textbf{$d^{-2}$}\\% & &  
(2,2,2) &$d $ &$\frac{d(d+1)(d+2)}{6} $ &  \textbf{$d^{-2}$}\\ %\\ & &  \\ 
(2,2,3) &$\frac{d(d-1)}{2}$ &$\frac{d^2(d-1)(d+1)}{4} $ &  \textbf{$d^{-2}$}\\% & &  \\ 
(2,3,3) &$\frac{d(d+1)(d-1)}{6} $ &$\frac{d^2(d-1)[d(d-1)+2]}{8} $ & \textbf{$d^{-2}$} \\% & &  \\ 
(3,3,3) &$0 $ &$\frac{d(d-1)[d(d-1)+2][d(d-1)+4]}{48} $ & 0 \\% & &  \\ \hline
Total &$\frac{d(4d^2+3d+2)}{3}$ &$\frac{d(d+3)(d^4+6d^3+15d^2+18d+8)}{48}$ &$d^{-3}$\\% & & \\
\hline
\end{tabular}
\caption{Number of non-redundant elements different from zero ($e_{jk}$ and $e_{jkl}$) and total number of elements ($E_{jk}$ and $E_{jkl}$), for each block and overall for the Hessian matrix ($\ell^{\tilde \bsEta_j \tilde \bsEta_k}$) and the third derivative array ($\ell^{\tilde \bsEta_j \tilde \bsEta_k \tilde \bsEta_l}$), as function of the dimension $d$. The last column reports the order of $e_{jk}/E_{jk}$ and of $e_{jkl}/E_{jkl}$.}
\label{tab:sparsityd2d3}
\end{table}

\newpage
\label{AppB}

\section{Further Details and Results} \label{AppB1}

\subsection{Computer System and Software Details} 
The log-likelihood derivatives w.r.t. $\bsEta$ and $\bsBeta$, as well as auxiliary parametrisation-specific quantities, are written using the \verb|C++| language and interfaced with the \verb|R| \citep{Rsoftware} statistical software by means of \verb|Rcpp| \citep{RcppPackage} and \verb|RcppArmadillo| \citep{eddelbuettel2014rcpparmadillo}. Further, methods for building and fitting the multivariate Gaussian additive models are implemented by the  \verb|SCM| \verb|R| package, which is available at \verb|https://github.com/VinGioia90/SCM| and leverage the model fitting routines of the \verb|mgcv| package \citep{wood2011fast, wood2016}. At the time of writing, being not yet on CRAN, a developer package version of \verb|mgcv| is used. The code for reproducing the results in this article is available at \verb|https://github.com/VinGioia90/SACM|. All the computations are done by using a 12-core Intel Xeon Gold 6130 2.10GHz CPU and 256GBytes of RAM. 

\subsection{Simulation Results} \label{Sec:sim1}
\begin{figure}[htbp]
\centering
\includegraphics[scale=0.54]{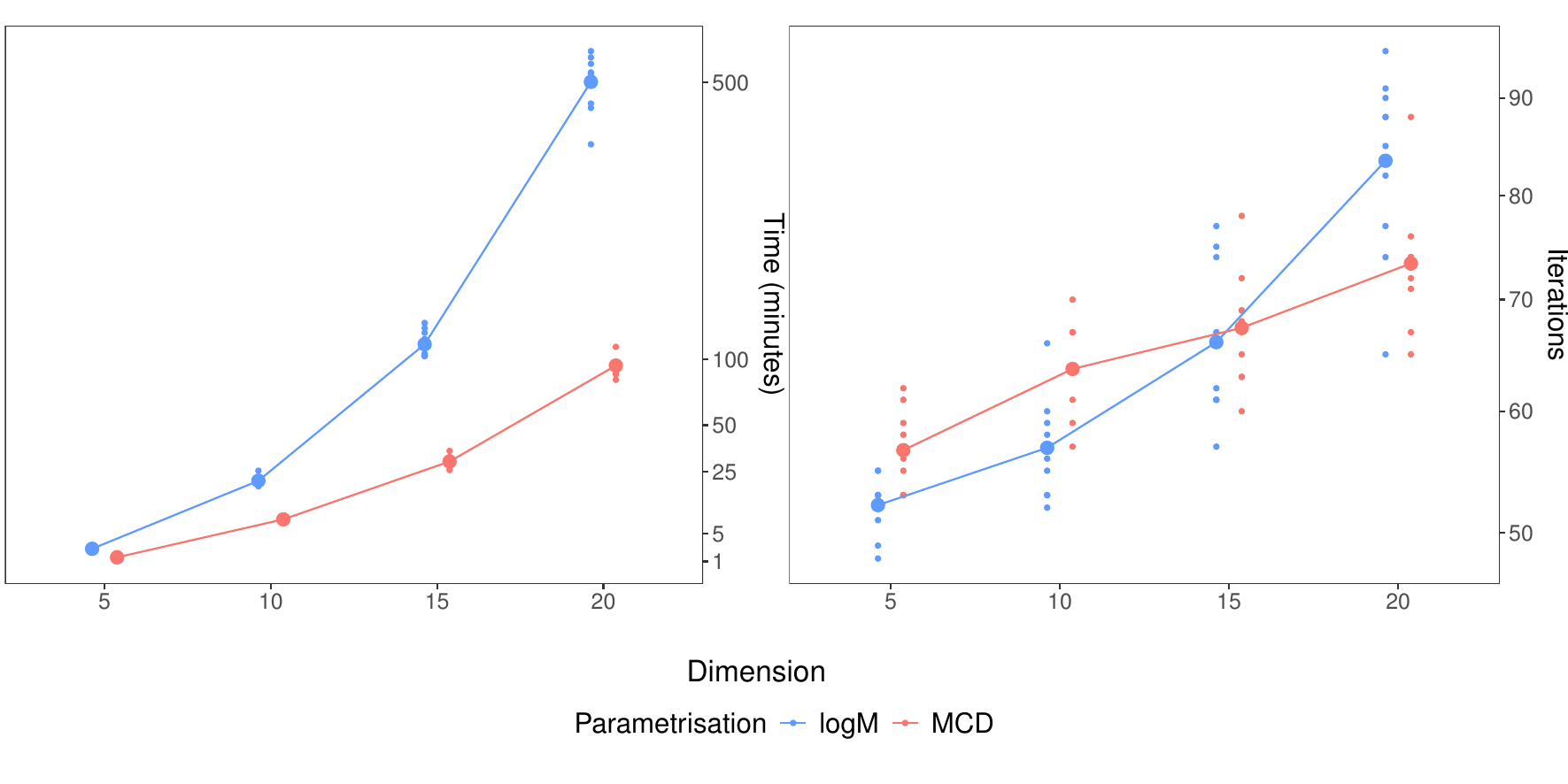}
\caption{Model fitting results by generating $\boldsymbol \Sigma$ via logM parametrisation. Left: fitting
times (in minutes); Right: inner Newton-Raphson iterations. The connected bigger dots represent the mean values (times and iterations), while smaller dots depict the ten observed values. Blue refers to the logM-based model and red to the MCD one. The $y$-axes are on a square root scale.} 
\label{fig:logM_vs_MCD_logM_generation} 
\end{figure}

Figure \ref{fig:logM_vs_MCD_logM_generation} shows the model fitting times and the total number of Newton's iterations used to obtain $\hat{\boldsymbol \beta}$, for both the logM- and MCD-based models, when true covariance matrix $\bsS_i$ is related to $\boldsymbol \Theta_i$ via the logM model. 
\begin{figure}
\centering
\includegraphics[scale=0.55]{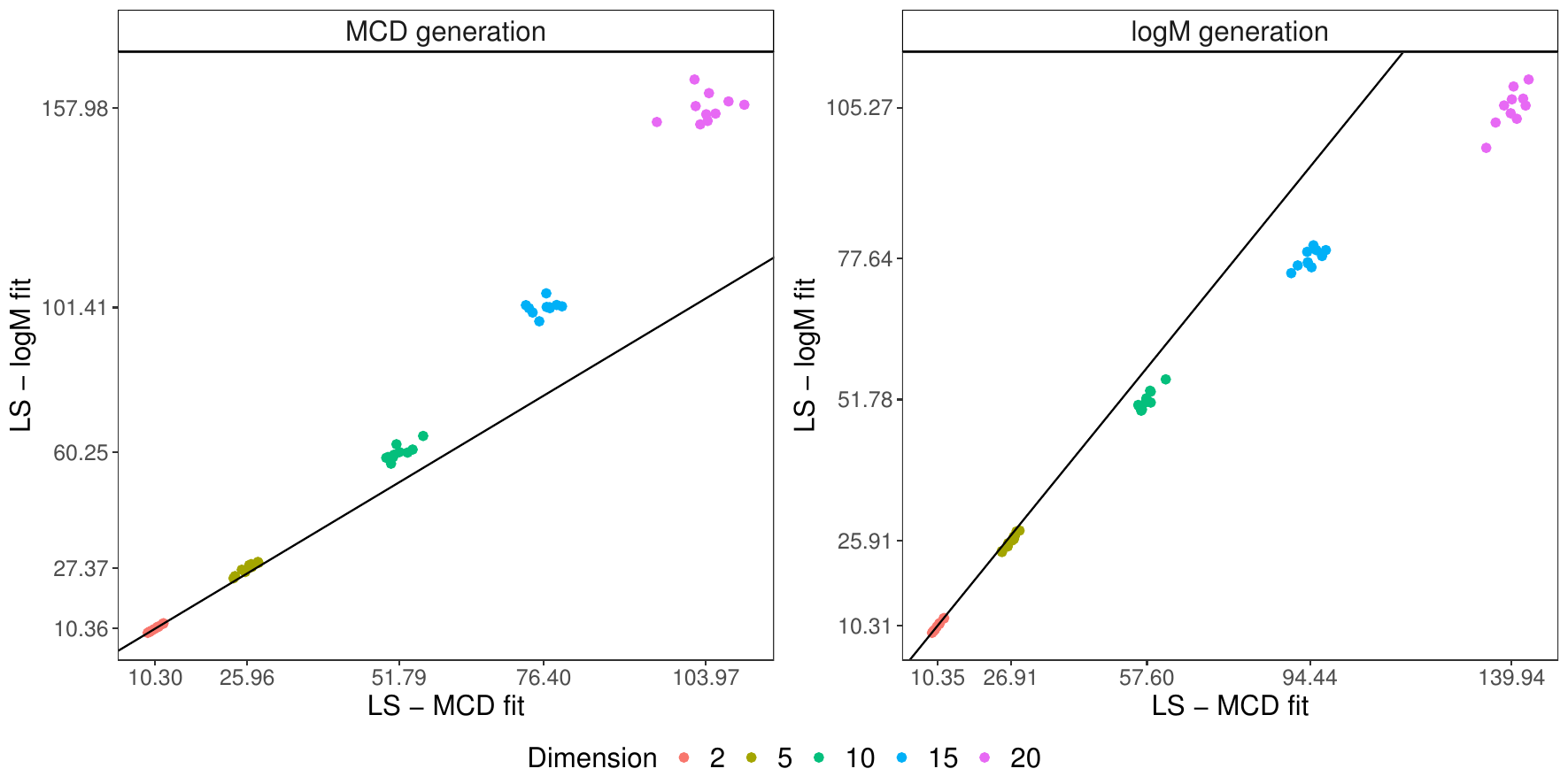}
\caption{Log-score of the fitted MCD model ($x$-axis) and logM model ($y$-axis) when the true covariance matrix of the response is related to $\boldsymbol \Theta_i$ via the MCD (left) or logM (right) parametrisation. The log-score values are expressed in thousands.}
\label{fig:logS_sim1}
\end{figure}
It is interesting to compare the MCD and logM models also in terms of goodness-of-fit. Hence, here we consider each model's log-score (LS), that is the negative log-likelihood, as a performance metric. In agreement with intuition, Figure \ref{fig:logS_sim1} shows that the MCD (logM) model leads to a better fit when the true covariance matrix $\bsS_i$ is related to $\boldsymbol \Theta_i$ via the MCD (logM) model. That is, fitting the true covariance model leads to a better fit, and the goodness-of-fit gains increase with $d$.

\subsection{Further Simulation Results under Parsimonious Models} \label{Sec:sim2}

\begin{figure}
\centering
\includegraphics[scale=0.55]{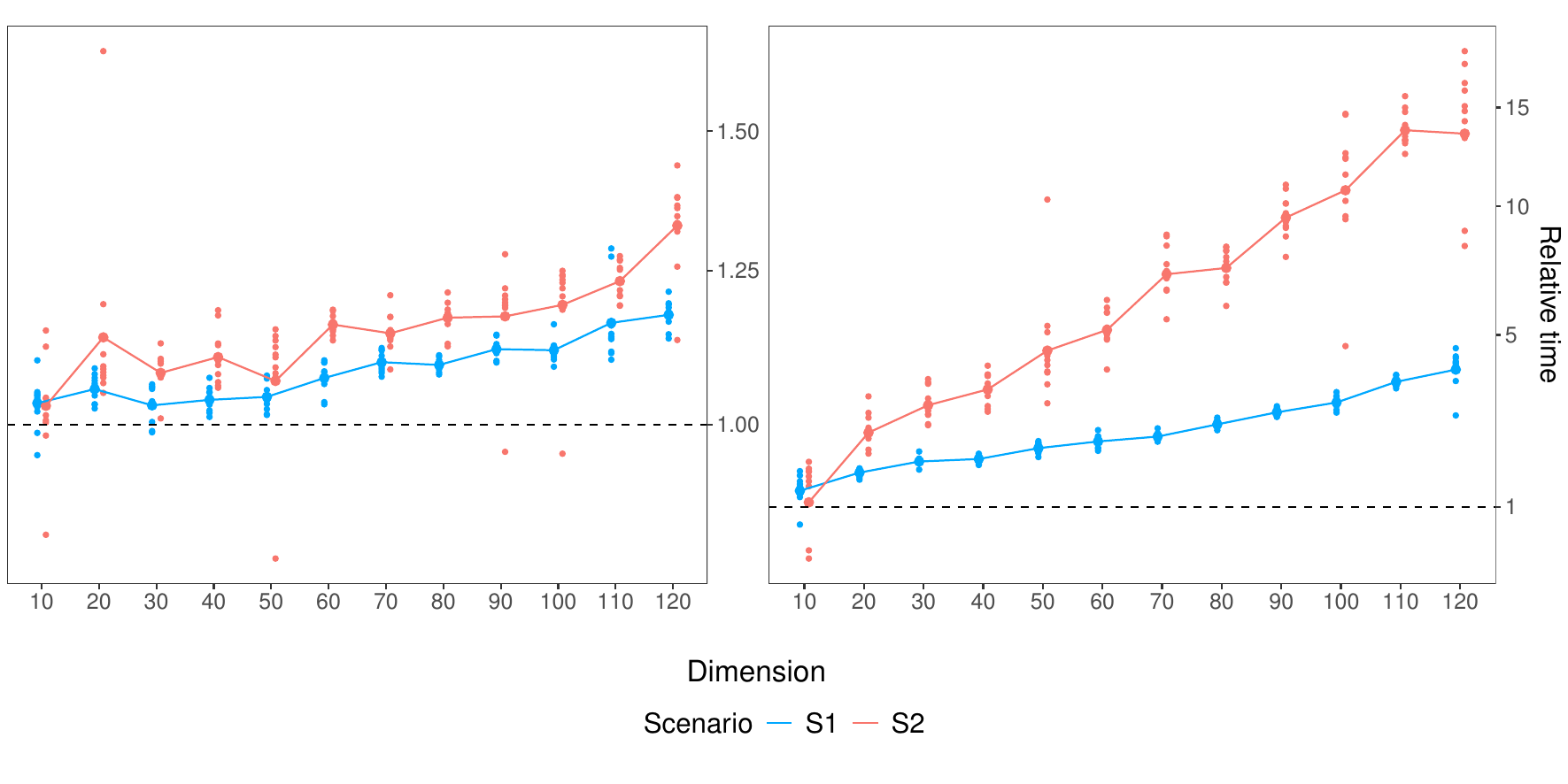}
\caption{Parsimonious modelling setting results under MCD-based models. Ratios between the time needed to compute $\bar\ell^{\bsBeta \bsBeta}$ under the \texttt{Standard} and the \texttt{Parsimonious} approach, under the simulation setting considered in Section \ref{sec:parsimonious} (left) and under a modified scenario where the linear predictors controlling the mean vector are also fixed to intercepts, that is $\eta_{ij} =  \beta_{1j}$ for $j=1, \ldots, d$ (right). The connected dots represent the ratio of the mean times, while the small dots are the ratios of the ten observed times. The y-axes are on square root scale.}
\label{fig:RelTime_MCD}
\end{figure}

Here we present further results using the simulation setting described in Section \ref{sec:parsimonious}. In particular, the plot on the left-hand side of Figure \ref{fig:RelTime_MCD} shows the time needed to compute $\bar\ell^{\bsBeta \bsBeta}$ under the \texttt{Standard} approach, relative to that required under the \texttt{Parsimonious} methods for the MCD-based model. Clearly, the gains obtained by using the latter are very marginal.

To understand why this is the case consider that, under a logM model where each linear predictor is controlled via $p$ parameters (hence not a parsimonious setting),  computing $\bar \ell^{\bsBeta_j \bsBeta_k}$, for each $j$ and $k$, requires  $\mathcal{O}\big(nd^4p^2\big)$ operations. Hence, when $p^2 > d$, the Hessian w.r.t. $\bsBeta$ determines the leading cost of computation, as computing $\ell^{\bsEta_j \bsEta_k}$ involves $\mathcal{O}\big(nd^5\big)$ operations. Under the MCD-based model, both derivatives cost $\mathcal{O}\big(d\big)$ less, due to sparsity, but their relative cost is unchanged. However, the number of blocks required to keep memory usage constant as $d$ increases is $\mathcal{O}\big(d^3\big)$, rather than $\mathcal{O}\big(d^4\big)$, hence the computational slowdown due to small blocks is less severe than under the logM model.

\begin{figure}
\centering
\includegraphics[scale=0.55]{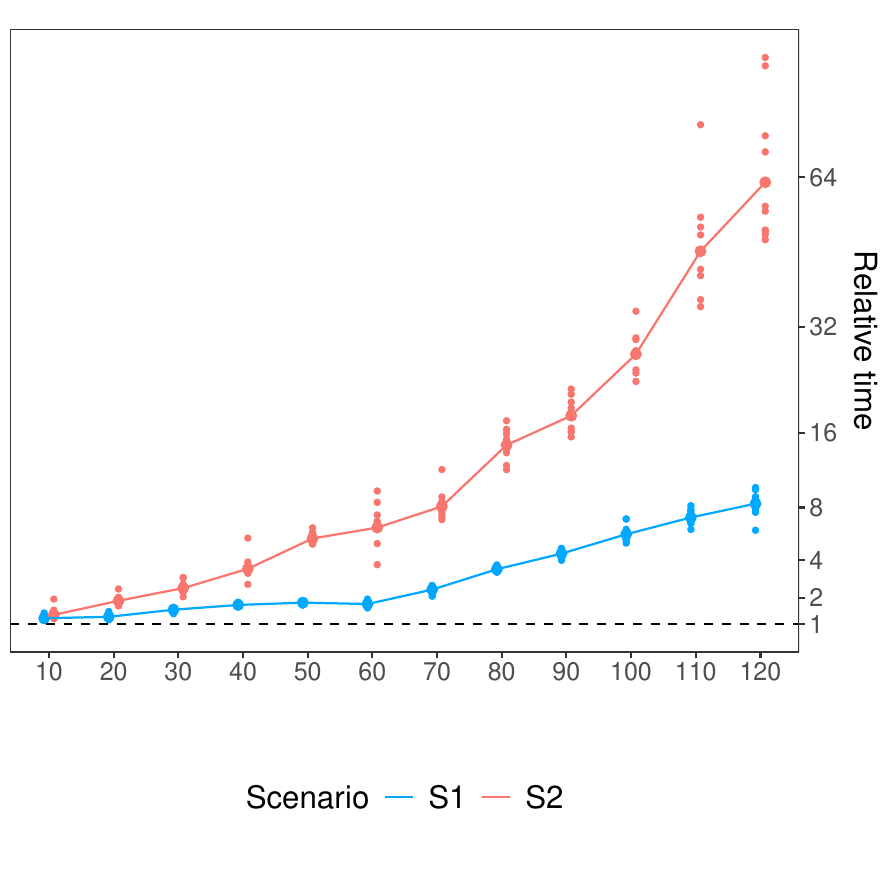}
\caption{Parsimonious model setting results under the logM-based model when considering the modified scenario $\eta_{ij} =  \beta_{1j}$ for $j=1, \ldots, d$. Ratio between the time needed to compute $\bar\ell^{\bsBeta \bsBeta}$ under the \texttt{Standard} and the \texttt{Parsimonious} approach.  The connected dots represent the ratio of the mean times, while the small dots are the ratios of the ten observed times. The y-axis is on square root scale.}
\label{fig:RelTime_logM_nomean}
\end{figure}

To assess the impact of sparsity on this issue, it is useful to consider a simulation setting in which the linear predictors controlling the mean vector are fixed to their intercepts, i.e. $\eta_{ij} = \beta_{1j}$ for $j = 1, \ldots, d$. The corresponding relative computational times are shown on the right-hand side of Figure \ref{fig:RelTime_MCD}. In this scenario, the \texttt{Parsimonious} methods scale better with $d$, particularly in the second, more parsimonious, case. Note that fixing the mean vector to intercepts does not significantly affect the results under the logM model, as demonstrated by Figure \ref{fig:RelTime_logM_nomean}, which is consistent with the right-hand plot of Figure \ref{fig:hessian_beta}.

In summary, under the MCD model, the sparsity of the second-order log-likelihood derivatives with respect to $\eta_{ij}$ for $j = d+1, \ldots, q$ means that storing these derivatives requires little memory compared with the dense logM case. Consequently, adopting \texttt{Parsimonious} computational methods does not yield substantial gains, as most of the memory is allocated to storing the (dense) second-order derivatives associated with the mean vector, $\eta_{ij}$ for $j = 1, \ldots, p$. This implies that \texttt{Parsimonious} computational methods are effective for MCD models only when many of the linear predictors controlling the mean vector are fixed to their intercepts. By contrast, under the logM parametrisation, the second-order derivatives with respect to $\eta_{ij}$ are non-zero for $j = 1, \dots, q$, so that \texttt{Parsimonious} computational methods lead to a speed-up regardless of whether the unmodelled linear predictors (i.e. those fixed to intercepts) control the mean vector or the covariance matrix.

\subsection{Further Details on the Application}\label{app:application} 

\begin{figure}
\centering
\includegraphics[scale=0.5]{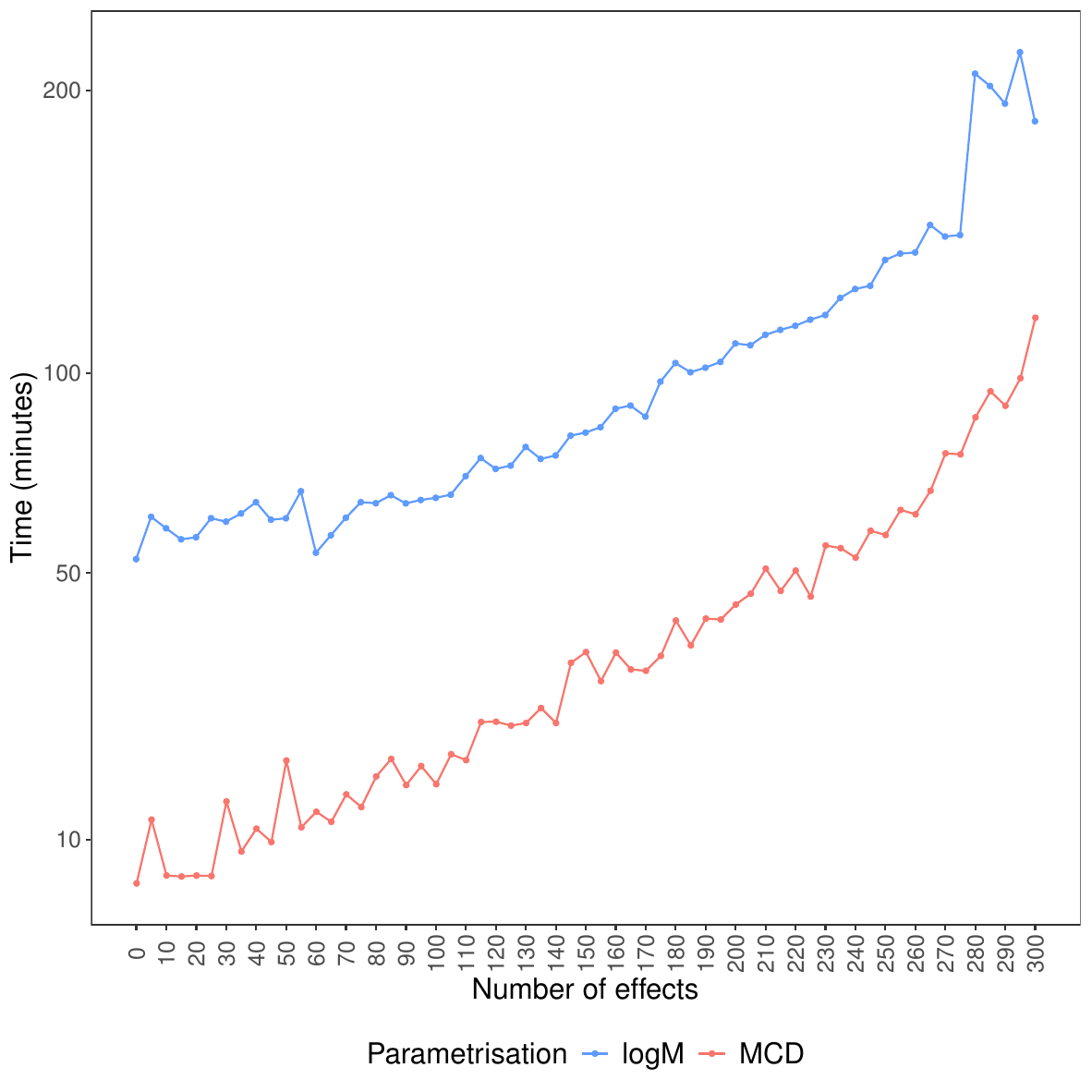}
\caption{Computational times (in minutes) for fitting the models on the 2005-2010 train data, by varying the number of covariate-dependent components involved in modelling $\eta_{ij}$, $j =  d+1, \ldots, q$. The plot is on a square root scale.  }
\label{fig:time_application}
\end{figure}

Figure \ref{fig:time_application} illustrates the time required to fit the models described in Section \ref{sec:App}. Under both parametrisations, model fitting was performed using the \texttt{FS} update for selecting the smoothing parameter and the computational methods detailed in Section \ref{sec:blocked_D_beta}, which are designed to exploit parsimonious modelling settings. Note that the MCD models are considerably faster than the logM models when only a few effects are used to model $\eta_{ij}$, for $j = d+1, \ldots, q$. This is because computing the second-order log-likelihood derivatives with respect to $\boldsymbol \eta$ is much less computationally expensive under the MCD model, and this cost becomes dominant in highly parsimonious scenarios.

\begin{figure}
\centering
\includegraphics[scale=0.23]{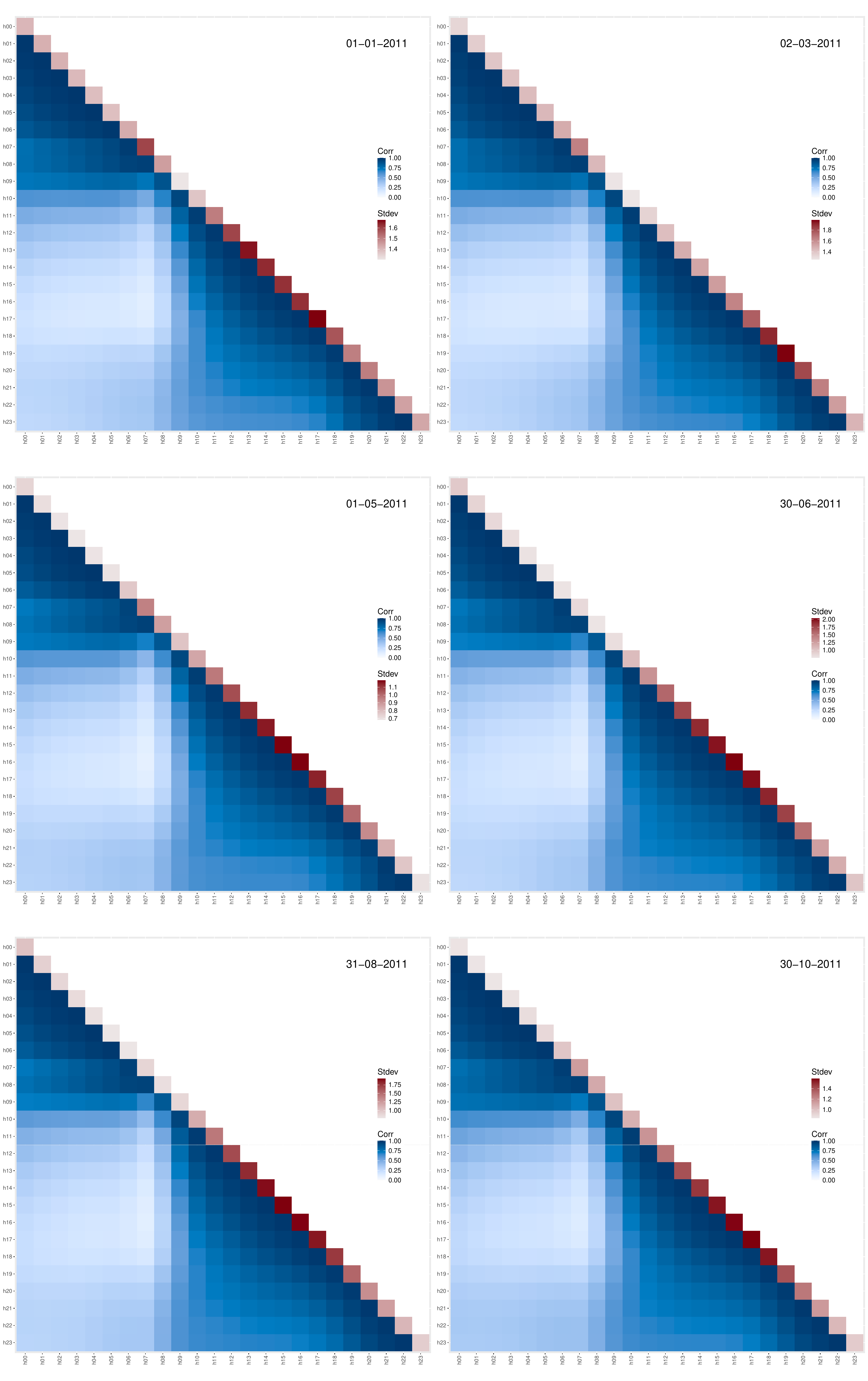}
\caption{Standard deviations (diagonal) and correlations (lower triangle) predicted by an logM-based model with 40 effects, on six different dates.}
\label{fig:Heats_logM}
\end{figure}

\begin{figure}
\centering
\includegraphics[scale=0.23]{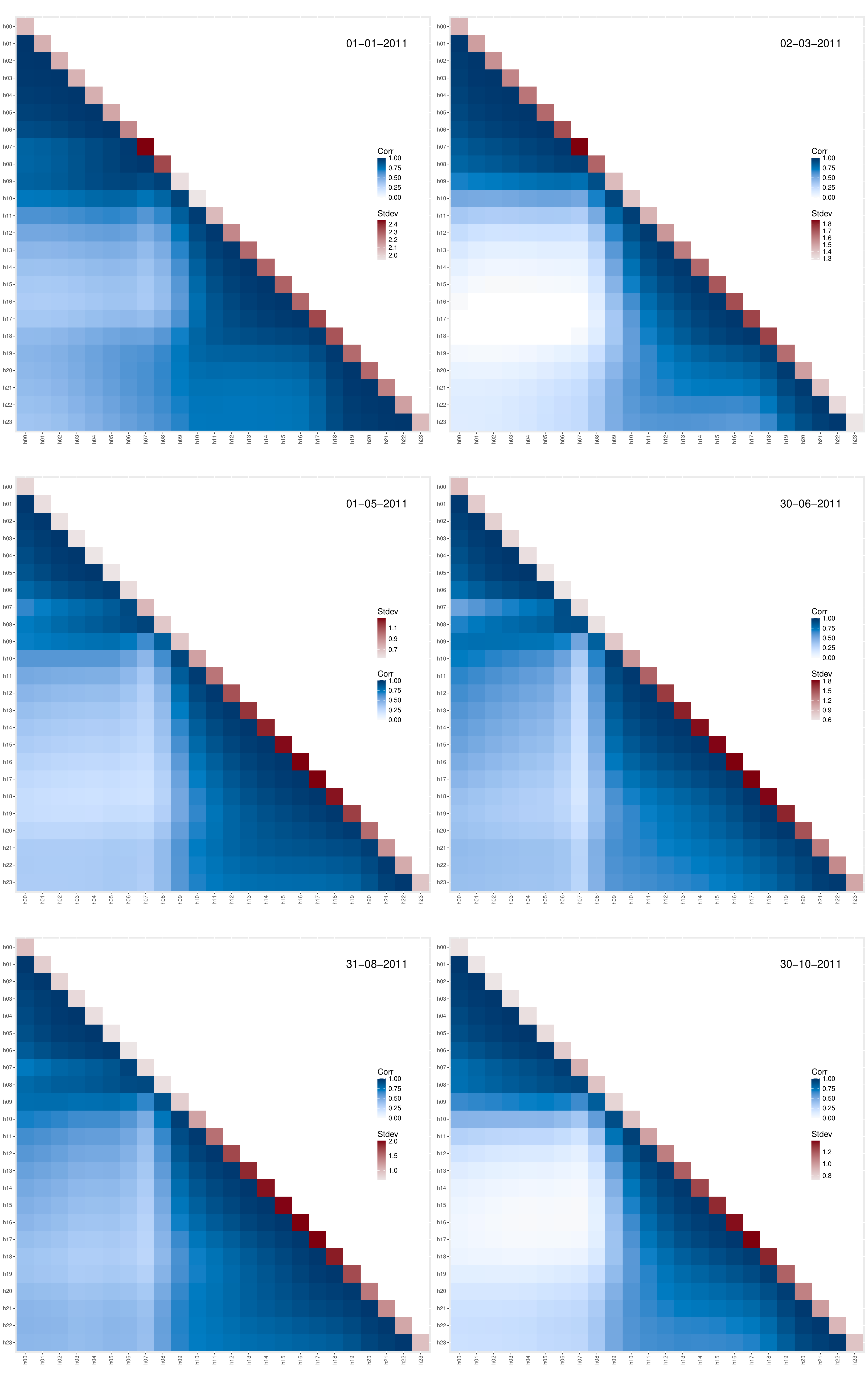}
\caption{Standard deviations (diagonal) and correlations (lower triangle) predicted by an MCD-based model with 80 effects, on six different dates.}
\label{fig:Heats_MCD}
\end{figure}

Recall from Section \ref{sec:App} that backward effect selection leads to a logM model where almost all the effects fall on the main diagonal or on the first subdiagonal. The reason for this is not immediately obvious because, in contrast with the MCD model, the element of the logM parametrisation to not have, to our best knowledge, a clear interpretation. However, Figure \ref{fig:Heats_logM} suggests that the logM elements that are being modelled as functions of the time of year ($\text{doy}_i$) mostly control the conditional variances. In fact, the plots show that the conditional variances of electricity load vary strongly with $\text{doy}_i$, while the correlation structure is roughly constant through the year. In contrast, the selected MCD model allows both the variance and the correlation to vary through the year, as shown by \ref{fig:Heats_MCD}.

\newpage

\putbib
\end{bibunit}

\end{document}